\documentclass[10pt,a4paper,twoside]{article}
\usepackage{cite}
\usepackage{epsfig}
\usepackage[innercaption]{sidecap}
\usepackage{eurosym}
\usepackage{color}

\begin{document}

\setlength{\oddsidemargin}{2.3cm}
\setlength{\evensidemargin}{1.7cm}

\pagestyle{empty}
\begin{center}
\color{blue}
{\huge \bf  Proposal}\\[3mm]
{\large  for the }\\[3mm]
{\huge  \bf  Wide Angle Shower Apparatus (WASA)}\\[2mm]
{\huge \bf at COSY-J\"ulich}\\[6mm]
{\huge \bf ''WASA at COSY''}\\[2cm]
\color{black}

{\large J\"ulich, October 5, 2004}\\


\end{center}

\pagebreak

\begin{tabular}{ll}
Corresponding authors:&\\
Bo H\"oistad & (bo.hoistad@tsl.uu.se)\\
James Ritman & (j.ritman@fz-juelich.de)\\
\end{tabular}

\vfill

\begin{tabular}{l}
\parbox{\textwidth}{
  {\bf Preprint version.} The figures in here have 
  reduced size and, thus, 
  reduced quality. The original proposal can be accessed at 
  http://www.fz-juelich.de/ikp/wasa.\\
}
\end{tabular} 

\begin{tabular}{ll}
Released & October 5, 2004 \\
Revised  & October 11, 2004 \\
\end{tabular}

\cleardoublepage

\pagestyle{plain}
\setcounter{page}{1}
\pagenumbering{roman}

\newcounter{institute}
\newcommand{\instno}{\refstepcounter{institute}\theinstitute}
\newcommand{\newinst}[1]{$^\mathrm{\instno\label{#1}}$}
\newcommand{\inst}[1]{$^\mathrm{\ref{#1}}$}

\begin{center}
\noindent
H.-H.~Adam\inst{muenster}, 
M.~Bashkanov\inst{tuebingen}, 
U.~Bechstedt\inst{ikp}, 
J.~Bisplinghoff\inst{hiskp},
D.~Bogoslovski\inst{jinr},
A.~Bondar\inst{budker},
B.~Borasoy\inst{hiskp},
M.~B\"uscher\inst{ikp},
K.-Th.~Brinkmann\inst{dresden},
P.~Brylski\inst{ip-cracow}
% ...

\vspace{5mm}
\noindent
{\footnotesize
\newinst{muenster}%
Institut f\"ur Kernphysik, Westf\"alische Wilhelms-Universit\"at M\"unster, Germany\\
\newinst{tuebingen}%
Physikalisches Institut, Universit\"at T\"ubingen, Germany\\
\newinst{ikp}%
Institut f\"ur Kernphysik, Forschungszentrum J\"ulich, Germany\\
\newinst{hiskp}%
Helmholtz-Institut f\"ur Strahlen- und Kernphysik, 
Rheinische Friedrich-Wilhelms-Universit\"at Bonn, Germany\\
\newinst{jinr}%
Joint Institute for Nuclear Research, Dubna, Russia\\
\newinst{budker}%
Budker Institute of Nuclear Physics, Novosibirsk, Russia\\
\newinst{dresden}%
Institut f\"ur Kern- und Teilchenphysik, Technische Universit\"at Dresden, Germany\\
\newinst{ip-cracow}%
Institute of Physics, Jagiellonian University, Cracow, Poland\\
\newinst{tsl}%
The Svedberg Laboratory, Uppsala University, Sweden\\
\newinst{drs-upp}%
Department of Radiation Sciences, Uppsala University, Sweden\\
\newinst{itep}%
Institute of Theoretical and Experimental Physics, Moscow, Russia\\
\newinst{zel}%
Zentralinstitut f\"ur Elektronik, Forschungszentrum J\"ulich, Germany\\
\newinst{inr}%
Institute for Nuclear Research, Russian Academy of Sciences, Moscow, Russia\\
\newinst{mepi}%
Moscow Engineering Physics Institute, Moscow, Russia\\
\newinst{koeln}%
Institut f\"ur Kernphysik, Universit\"at zu K\"oln, Germany\\
\newinst{inp-cracow}%
Institute of Nuclear Physics, Polish Academy of Sciences, Cracow, Poland\\
\newinst{sofia}%
Faculty of Physics, St. Kliment Ohridski University of Sofia, Bulgaria\\
\newinst{tu-kosice}%
Technical University Kosice, Kosice, Slovakia\\
\newinst{fairfax}%
Physics Department, George Mason University, Fairfax (VA), USA\\
\newinst{uni-kosice}%
University of P.J.~Saf\'{a}rik, Kosice, Slovakia\\
\newinst{hamburg}%
Institut f\"ur Experimentalphysik, Universit\"at Hamburg, Germany\\
\newinst{katowice}%
Institute of Physics, University of Silesia, Katowice, Poland\\
\newinst{soltan}%
The Andrzej Soltan Institute of Nuclear Studies, \'Swierk, Warsaw and {\L}\'od\'z, Poland\\
\newinst{iep-warsaw}%
Institute of Experimental Physics, Warsaw University, Warsaw, Poland\\
}
\end{center}

\cleardoublepage

\section*{Executive Summary}

The document at hand describes the physics case, the key
experiments proposed and the technical solution for the project ``WASA at
COSY'', which concerns the transfer of the WASA pellet-target and
detector system from CELSIUS (TSL, Uppsala, Sweden) to COSY (FZJ,
J\"ulich, Germany) and its installation at an internal target position.

WASA at COSY will provide unique scientific possibilities for
research in hadron physics with hadronic probes, since it combines:
\begin{itemize}
\item COSY with its (phase-space cooled, polarized) proton and deuteron
  beams with energies sufficient to cover the strange quark sector
  including $\phi$-mesons
\item WASA, a close to $4\pi$ detector for both photons and charged 
  particles
\end{itemize}

Since both COSY and WASA are operational devices, the 
time needed to install and commission it is short: it is
anticipated that the physics program will start at the end of 2006.\\

\noindent
The physics that will be investigated comprises:
\begin{description}
\item {\bf Symmetries and symmetry breaking:} \\ 
  {\em The fundamental theories of physics are based on symmetry, yet:
  our world is filled with asymmetry} (T.D. Lee).

  Finding and studying symmetries and symmetry violations in hadronic 
  reactions will help
  to better understand the strong interaction.
  Hadron beams are particularly well suited for such investigations due to
  the spin- and isospin filter mechanism. In addition, it is even conceivable 
  to search for physics beyond the Standard Model.

\item{\bf Hadron structure and interactions:} \\
   {\em That [intermediate distance] scale is the richest
   phenomenologically, and is certainly the crux region to understand
   \ldots what QCD is really about. And at the heart of the subject is
   the hadron spectrum, in particular the spectrum built from light
   quarks. (\ldots) Without question, there is a great need \ldots for
   a new round of experiments, especially utilizing hadron beams}
   (J.D. Bjorken).

  Finding and further
  investigating specific hadronic bound systems will --- together with
  corresponding progress in theory --- provide a more fundamental
  insight into how nature makes hadrons.

\end{description}

The proposal concentrates on the study of 
symmetry breaking as the primary objective, in particular in $\eta$ 
and $\eta{}'$-decays and
meson production. In addition, the potential of 
further studies is discussed.   

An experienced and enthusiastic community of more than one hundred scientists
from all over Europe and abroad has started to work for ''WASA at COSY''
to become a reality and is looking forward to exploit its scientific
potential.

\cleardoublepage

{\bf
\tableofcontents
}

\cleardoublepage

\pagenumbering{arabic}
\setcounter{page}{1}

\section{Introduction}

Hadron physics is concerned with one of the most difficult but
also most fascinating problems in contemporary physics.
It is commonly accepted that the underlying theory, 
Quantum Chromo Dynamics (QCD), is correct. However, very little is 
known about the solution of this non--linear quantum field theory in 
the regime of small
and moderate energy scales pertinent to the matter surrounding us.
QCD is characterized by two particular phenomena,
confinement and chiral symmetry breaking. QCD is formulated in terms
of colored quarks and gluons. These fundamental particles have
never been observed as free states but only in composite systems,
so called hadrons, i.e. they are confined within colorless mesons,
baryons and, possibly, exotics. 
Furthermore, in the sector of the light up, down and strange
quarks, QCD possesses an approximate chiral symmetry, which is
not observed in nature but rather broken spontaneously with the
appearance of almost massless Goldstone bosons. These can be
identified with the lowest octet of pseudoscalar mesons. Such broken
symmetries also play an important role in phenomenon of superconductivity 
or the Higgs mechanism, which is believed to lead to the creation masses of the
quarks and leptons. The consequences of the dynamical and also the 
explicit chiral symmetry violation through the current quark masses 
in QCD can be systematically
explored
with
effective field theory. Presumably
related to these phenomena is the generation of hadron masses in 
QCD, e.g. the
proton and the neutron are much heavier than the current quarks they
are made of. In the energy regime considered here, the interaction
between the quarks and gluons is highly non-linear and also
non-perturbative, excluding the use of standard perturbative methods
as applied e.g. in Quantum Electro Dynamics or QCD at high energies,
where asymptotic freedom is the dominant feature of the
theory. A deeper understanding of the structure and dynamics of
hadrons, in particular their excitation spectrum,  as derived from 
QCD would explore
one  of the last {\it terrae} {\it incognitae} of the Standard Model.

\noindent
To understand complex systems such as hadrons, a variety of
complementary experimental and theoretical studies are mandatory.
The WASA detector at COSY offers a unique possibility to
deepen our understanding of aspects connected with QCD in the
non-perturbative regime through a precise study of symmetry breakings
and very specific investigations of hadron structure. 
The $\eta$ and $\eta'$ decays, that vanish in the limit of equal quark masses,
allow us to explore the explicit isospin symmetry 
breaking
in QCD.

It is well established that  isospin violating quark mass 
effects are often masked by electromagnetic corrections and thus
such studies require
the detection of neutral decay particles.
A textbook example for this is the threshold 
photoproduction of neutral pions off protons which has so clearly 
revealed the chiral loops inherent
to the spontaneous chiral symmetry breaking of QCD.
Through the same quark mass term of QCD, such quark mass effects also
appear in systems of a few nucleons, so that  $dd$ collisions 
producing $^4$He plus neutral Goldstone bosons at COSY energies 
offer yet another tool to explore this area - also because only now
theoretical tools are becoming available that allow a model-independent
analysis of these processes. Furthermore, precision measuremetns of 
rare  $\eta$ and $\eta'$ 
decays can be used to get new
limits on the breaking of the fundamental $C$, $P$ and $T$ symmetries, or
combinations thereof. In addition, reactions involving the $\eta '$
might also offer insight into the elusive glue  through the
anomaly. 

\noindent
The spontaneous chiral symmetry breaking is believed
to be at the heart of the mass generation in QCD. Still, this issue cannot
be separated from the question: what kind of hadrons are really generated
in QCD? In the course of the last year, the paradigm of the constituent quark model,
i.e. the hadrons are either quark-antiquark or three quark states, has been
challenged by many experiments. However, a definite picture concerning the
nature of e.g. the light scalar mesons, the $a_0 (980)$, $f_0 (980)$, etc., 
the $\Lambda (1405)$ or the
pentaquark, the $\Theta^+ (1540)$, has not yet emerged. Evidently, if one
does not understand the nature of these states and others, a true comprehension
of confinement can not be achieved. WASA at COSY is capable to contribute
significantly to test the models which are offered to explain
these states through the precise measurements of decay chains or the
coupling to other hadrons. It should
be stressed that all these proposed experiments 
will be {\em precision measurements}, because only accurate data
will allow to test and constrain the theory. 
With the advent of effective field theory methods, supplemented
by coupled channel analyses, the theoretical tools of the required precision
are available to analyze the data in a model-independent way, reflecting the 
change of paradigm in the foundations in this area of theoretical physics,
that in the past has been dominated by model building. 
Moreover, lattice
gauge theory offers prospects to confront certain observables, in particular 
spectroscopic issues, with ab initio calculations.
A very close contact between 
experiment and theory is foreseen and indispensable to achieve progress in
the field of hadron physics.

\noindent
In this document it is proposed to perform high precision
measurements
of several hadronic reactions to confront the theoretical
predictions
sketched above.  In particular, investigations of symmetry breaking
in the decay of the $\eta$ and $\eta'$ mesons and isospin violation
in the
reaction $\vec{d}d\to\alpha\pi^0$  are seen to be very
promising for this purpose.   The combination of the detector system
WASA
and the COSY accelerator facility perfectly matches the requirements
needed to measure these reactions.  The beam energy range, the phase
space cooling and the availability of (polarized) proton and
deuteron beams at COSY are essential aspects for high luminosity
measurements with low background of these processes.  The WASA
detector provides nearly full solid angle coverage for both charged
and neutral particles, which is needed to for kinematically complete
measurements of the multiparticle final states.  Furthermore, the
usage of frozen pellets of liquid hydrogen and deuterium as the
target minimizes background from secondary reactions but at the same
time allows high luminosity conditions.  Since the developement and
construction of a detector system fulfilling these requirements
requires a number of years, the transfer of an existing detector
like WASA to COSY for these measurements will greatly reduce the
lead-time until physics output can be expected.

\label{WASAintro}

\noindent
To summarize: The challenge of hadron physics is not only 
to gain an understanding
of QCD in the non--perturbative regime, e.g. to explain the mechanism
underlying confinement, but also the possibility of unhinging the foundations
of the Standard Model at low energies. The  understanding of the structure 
of matter is intimately linked to the structure and dynamics of 
hadrons. Thus, hadron physics plays a central role in the whole building 
of physics. This is the area of research where WASA at COSY
can contribute significantly as will be demonstrated in the following
sections.

\clearpage


\section{Physics case}
\label{sec:physics}

\subsection{Primary objective:\\ Symmetries and symmetry breaking}
\label{sec:symmetries}

The degrees of freedom within the Standard Model are leptons and
quarks for the matter fields and gluons, photons and the weak gauge bosons. 
However, at low and intermediate energies
only the leptons and photons play a role as dynamical degrees of
freedom: $W^\pm$ and $Z^0$ are too heavy and quarks and gluons are
trapped inside composite objects (hadrons, glueballs, and possibly
exotics). The mechanisms behind this process are confinement and
spontaneous breakdown of chiral symmetry. However, through its
symmetries the underlying theory is still visible. 
If a symmetry is spontaneously broken, there is no obvious
connection between the excitation spectrum and the symmetry.
The currents are still conserved and, as a consequence, the
interactions of the hadrons at low energies are largely constrained.
The same holds true in the presence of a small explicit symmetry
breaking: in nature the chiral symmetry is broken by the non-vanishing
quark masses.  We will discuss in the following
examples, where a study of symmetries and symmetry breaking patterns
in hadronic systems allows to get insights on symmetries and symmetry
breaking patterns of QCD.

\subsubsection{\boldmath Decays of $\eta$ and $\eta^\prime$ mesons}
\label{subsubetap}

Isospin symmetry forbidden decays $\eta \left(\eta^\prime\right) 
\rightarrow 3\pi$ are sensitive to isospin symmetry breaking (ISB) due to 
the light quark mass difference $\Delta \mbox{m} = 
\mbox{m}_d - \mbox{m}_u$, and provide an approach complementary to ISB 
estimates from pseudoscalar meson masses and production processes (see
section~\ref{subsubalphapi} and \cite{SWeinberg77,JGasser82,miller}).
ISB effects in these decay channels can be associated directly with 
$\Delta \mbox{m}$, since electromagnetic effects are expected to be 
small~\cite{Sutherland:1966pl,Baur:1995gc}.
Moreover, the ratio of symmetry breaking and allowed $\eta^\prime$ decays 
to $3\pi$ and $\eta\,2\pi$, respectively, is directly proportional to the 
square of $\Delta \mbox{m}$ (see Ref.~\cite{Gross:1979ur} and 
section~\ref{subsubetapday1}):

Mixing between isospin eigenstates $\mid\tilde{\pi}^0>$ and 
$\mid\tilde{\eta}>$ occurs due to the non--vanishing matrix element of the 
quark mass contribution to the QCD Hamiltonian density ${\cal H}_m = 
\mbox{m}_u u \bar{u} + \mbox{m}_d d \bar{d} + \mbox{m}_s s \bar{s}$
\begin{equation}
 <\tilde{\pi}^0\mid {\cal H}_m \mid\tilde{\eta}>  =  
 \frac{1}{\sqrt{6}} \left< \left( u \bar{u} - d \bar{d} \right) \mid 
 {\cal H}_m \mid 
  \left( u \bar{u} + d \bar{d} - s \bar{s} \right) \right> = 
  - \frac{1}{\sqrt{6}} \Delta\mbox{m}  \;,
\label{eq_pi_eta_delta_m}
\end{equation}
where SU(3) meson wave functions and a pseudoscalar singlet--octet mixing 
angle $\Theta_{ps} \approx -19.5^\circ$ have been employed~\cite{Nefkens:2002sa}.
The mixing angle $\Theta_{\pi\eta}$ for the physical states $\pi^0$ and 
$\eta$ is related to $\Delta \mbox{m}$ by~\cite{Gross:1979ur,Kaiser:2000, Feldmann:1998}:
\begin{equation}
\left( \begin{array}{l}
  \pi^0 \\  \eta 
\end{array} \right) =
\left( \begin{array}{rr}
  \cos{\Theta_{\pi\eta}} & \sin{\Theta_{\pi\eta}}\\
 -\sin{\Theta_{\pi\eta}} & \cos{\Theta_{\pi\eta}}
\end{array} \right)\;
\left( \begin{array}{l}
  \tilde{\pi}^0 \\ \tilde{\eta}
\end{array} \right) \; ; \;
\sin{\Theta_{\pi\eta}} = 
 \frac{\sqrt{3}\,\Delta \mbox{m}}
      {4\left(\mbox{m}_s - \hat{\mbox{m}} \right)}
\label{eq_pi_eta_mix}
\end{equation}
with the average light quark mass $\hat{\mbox{m}} = 
(\mbox{m}_u + \mbox{m}_d)/2$.

Measuring the ratios of isospin symmetry breaking $\eta^\prime \rightarrow 
3\pi$ decays with respect to the allowed $\eta\,2\pi$ modes
\begin{equation}
{\cal R}_1 = 
 \frac{\Gamma\left( \eta^\prime \rightarrow \pi^0 \pi^0 \pi^0 \right)}
      {\Gamma\left( \eta^\prime \rightarrow \eta \pi^0 \pi^0 \right)} \; ;\;
{\cal R}_2 = 
 \frac{\Gamma\left( \eta^\prime \rightarrow \pi^0 \pi^+ \pi^- \right)}
      {\Gamma\left( \eta^\prime \rightarrow \eta \pi^+ \pi^- \right)} \; ,
\label{eq_pi_eta_ratio}
\end{equation}
the $\pi$--$\eta$ mixing angle in this approach can be determined in 
$\eta^\prime$ decays from 
\begin{equation}
{\cal R}_i = \mbox{P}_i \sin^2{\Theta_{\pi\eta}}, 
\end{equation}
with $\mbox{P}_i$ denoting phase space factors~\cite{Gross:1979ur}.

The energy dependence of hadronic decays of $\eta$ and $\eta^\prime$ to 
pseudoscalars, i.e.\ $\eta\left(\eta^\prime\right) \rightarrow 3\,\pi$ and 
$\eta^\prime \rightarrow \eta\pi\pi$, is contained in the Dalitz plot 
distribution of the decay products.
Some of the published results obtained for slope parameters of the Dalitz 
plots for $\eta \rightarrow 3\pi$~\cite{Gormley:1970qz,Layter:1973ti,
Alde:1984wj,Amsler:1995sy,Amsler:1997up,Abele:1998yi,Abele:1998yj,
Tippens:2001fm} are not in agreement with each other and with theoretical 
predictions~\cite{Kambor:1995yc,Beisert:2003zs}.
In particular, the predicted slope parameter for the neutral decay is 
excluded by the most recent published data~\cite{Tippens:2001fm}.
Preliminary results on both charged and neutral $3\pi$ decay modes from a 
high statistics measurement at the KLOE facility have recently been 
reported~\cite{DiMicco:2003cd}.

The energy dependence of hadronic decays $\eta^\prime \rightarrow 
\eta\,2\pi$ and $\eta^\prime \rightarrow 3\pi$ is discussed in the ChPT 
framework in more detail in~\cite{Beisert:2002ad,Beisert:2003zs}.
The presently available data with highest statistics on the $\eta\,2\pi$ 
decay mode are still inconclusive, whether Dalitz plot distributions 
deviate from phase space.
For the $\eta^\prime \rightarrow 3\pi$ decay, no experimental data on the 
Dalitz plot parameters are presently available.

Hadronic decay modes also constrain the interpretation of the scalar meson 
nonet:
Isovector $a_0(980)$ exchange is found to dominate $\eta^\prime \rightarrow 
\eta\,2\pi$~\cite{Fariborz:1999gr,Beisert:2003zs}, and the experimental 
data are used to fix unknown scalar--pseudoscalar--pseudoscalar 
couplings~\cite{Fariborz:1999gr}.
In comparison to $\eta \rightarrow 3\pi$ decays, exchange diagrams 
involving $f_0(980)$ and $a_0(980)$ are expected to have a stronger 
influence in $\eta^\prime$ decays to the $3\pi$ channel, since the exchange 
particles are much closer to the mass shell~\cite{Abdel-Rehim:2002an}.

The axial U(1) anomaly prevents the pseudoscalar singlet $\eta_0$ from being 
a 
Goldstone boson.
The dominant contribution to the mass of the singlet state arises from the 
divergence of the singlet axial current, that acquires an additional term 
with the gluonic field strength tensor and, consequently, does not vanish 
in the chiral limit. 
Since the physical states $\eta$ and $\eta^\prime$ are mixtures of the 
octet and singlet fields, $\eta_8$ and $\eta_0$, respectively, the 
$\eta^\prime$ cannot be disentangled from the $\eta$ and a proper study of 
$\eta^\prime$ physics requires to account for the $\eta$ as well.
Precision data on decays of $\eta$ and $\eta^\prime$ allow to probe a 
variety of aspects of low--energy hadron dynamics.
Experimental results can be compared with model--independent calculations 
of chiral perturbation theory (ChPT) with a clear theoretical connection to 
QCD.
Two-photon decays in the $\eta$--$\eta^\prime$ system allows ChPT predictions
to be tested, 
particularly its extension to the U(3) framework including the axial U(1) 
anomaly. 

Radiative decays $\eta\left(\eta^\prime\right) \rightarrow 2\gamma$ and 
$\eta\left(\eta^\prime\right) \rightarrow \pi^+ \pi^- \gamma$ test 
parameters of the QCD triangle and box anomalies.
Two--photon decays $\eta\left(\eta^\prime\right) \rightarrow 
\gamma \gamma$ provide insight into the axial $\mbox{U}_A(1)$ anomaly of 
QCD, since they are determined by the pseudoscalar singlet and octet 
couplings to the divergences of the relevant axial--vector currents and the 
singlet--octet mixing angle $\Theta_{ps}$ (see 
section~\ref{subsubetapmedium} and~\cite{Ball:1996zv}).
The latter can be constrained from radiative decays of vector ($V$) and 
pseudoscalar ($P$) mesons in $V\,(P) \rightarrow P\,(V) \gamma$ processes, 
i.e.\ from $\eta^\prime$ by the $\rho \gamma$ and $\omega \gamma$ decay 
modes.
A model framework to conclude about the glue content of the $\eta^\prime$ 
is discussed in~\cite{Kou:1999tt}.
The two--photon decays of $\eta$ and $\eta^\prime$ are furthermore well 
suited to confirm the number of colors in the low--energy hadronic sector 
to be $\mbox{N}_c = 3$~\cite{Baer:2001,Borasoy:2004ua}.
Within a few weeks, the available statistics for radiative $\eta^\prime$ 
decays could be increased by two orders of magnitude at the WASA facility 
(Table~\ref{tab_etap_rates}).

Theoretically, the anomalous behavior of the effective action under chiral 
transformations is reproduced by the Wess--Zumino--Witten term of the 
effective Lagrangian~\cite{Wess:1971yu,Witten:1983tw} which yields the 
major contribution to the decays $\eta\left(\eta^\prime\right) \rightarrow 
\gamma \gamma$.
The $\pi^+ \pi^- \gamma$ mode is dominated by $\eta^\prime \rightarrow 
\rho \gamma$.
However, the amplitude of the non--resonant $\eta^\prime \rightarrow 
\pi^+ \pi^- \gamma$ decay is determined both by parameters of the AVV 
(axialvector--vector--vector) triangle anomaly, that can be derived from 
radiative decays as discussed above, and by the AAAV box 
anomaly~\cite{Wess:1971yu,Chanowitz:1975jm,Chanowitz:1980ma,
Benayoun:2003we,Borasoy:2003yb,Borasoy:2004qj}.
On the basis of the presently available statistics, non--resonant 
contributions have not been unequivocally extracted (see 
Ref.~\cite{Acciarri:1998yx,Abele:1997yi} and section~\ref{subsubetapday1}).

Dalitz decays $\eta\left(\eta^\prime\right) \rightarrow \gamma \gamma^* 
\rightarrow \gamma l^+ l^-$ with $l= e,\,\mu$, where a time--like virtual 
photon converts to a lepton pair, probe the transition form factor, i.e.\ 
the electromagnetic properties of $\eta$ and $\eta^\prime$ in terms of the 
spatial distribution of meson charge with the $l^+l^-$ invariant mass 
corresponding to the four--momentum squared of the virtual photon 
(theoretical approaches for $\eta$ decays are discussed 
in~\cite{Stepaniak:2002ad,Borasoy:2003yb}).
The available data for $\eta^\prime \rightarrow \gamma l^+ l^-$ consist of 
$33 \pm 7$ $\mu^+ \mu^- \gamma$ 
events~\cite{Dzhelyadin:1980ki,Landsberg:1986fd}. 
Thus, a detailed analysis of the leptonic mass spectrum is presently not 
possible due to the lack of statistics.
Significant progress could be made using the WASA facility at COSY 
(section~\ref{subsubetapmedium}).

The decays $\eta\left(\eta^\prime\right) \rightarrow 
l^+ l^- l^+ l^-$ into lepton pairs address decays via two off--shell 
photons and indicate whether double vector meson dominance is realized in 
nature.
The coupling to two virtual photons is substantial for the real part of the 
decay amplitude to a single lepton pair, but also an important issue in 
kaon decays, and for the hadronic contribution to the anomalous magnetic 
moment of the muon~\cite{Hayakawa:1997rq,Bijnens:1997ac}.
The corresponding $\eta \gamma^* \gamma^*$ form factor has neither been 
measured in the time--like nor in the space--like region (see 
section~\ref{subsubetapday1}), a recent theoretical investigation is 
presented in~\cite{Borasoy:2003yb}.

The dominant mechanism in the leptonic decay $\eta \rightarrow e^+ e^-$, 
which is forbidden to proceed via a single photon intermediate state, is a 
fourth order electromagnetic process with two virtual 
photons~\cite{Bergstrom:1982zq,Landsberg:1986fd}.
The decay is additionally suppressed by helicity factors $\mbox{m}_e / 
\mbox{m}_{\eta}$ at the $\gamma e^+ e^-$ vertices.
A very low branching ratio 
$5 \cdot 10^{-9}$~\cite{Savage:1992ac,Ametller:1993we}, predicted by the 
Standard Model, makes the decay sensitive to contributions from 
non--conventional effects that would significantly increase the branching 
ratio (section~\ref{subsubetapmedium}).

\renewcommand{\arraystretch}{1.1}
\begin{table}[h]
\caption{Counting rate estimates including detector efficiencies for 
$\eta$ ($\eta^\prime$) decays with WASA at COSY based on a luminosity of 
$10^{32}\,\mbox{cm}^{-2}\,\mbox{s}^{-1}$, with a $20\,\mu\mbox{b}$ 
($300\,\mbox{nb}$) cross section in $pp \rightarrow 
pp \eta\left(\eta^\prime\right)$ at $2.250\,\mbox{GeV/c}$ 
($3.350\,\mbox{GeV/c}$) beam momentum.}
\vspace{1ex}
{\begin{tabular}{@{}llr@{.}l@{$\,\pm\,$}r@{.}lrr@{\hspace{0.5ex}}l@{}} 
 \hline
 & Decay & \multicolumn{4}{c}{Branching fraction} & 
   Existing & Counting & \\
 & mode & \multicolumn{4}{c}{$\Gamma_i / \Gamma_{tot}$} & data & rate & \\ 
 & & \multicolumn{4}{c}{\cite{PDBook}} & [events] & 
   [evts/day] & \\ \hline\hline
$\eta$  & $e^+ e^- \pi^+ \pi^-$ & 4 & 
 \multicolumn{3}{l}{$\mbox{}^{+14.0}_{-2.7}$ $\cdot 10^{-4}$} & 5 & 
 7000 & \footnotemark\\ 
 {\it (semi)--leptonic} & $e^+ e^- e^+ e^-$ & $<\,6$ & 
   \multicolumn{3}{l}{\hspace{-2ex} 9 $\cdot 10^{-5}$} & - & 
   450 & \footnotemark\\ 
 & $e^+ e^-$ & $<\,7$ & \multicolumn{3}{l}{\hspace{-2ex} 7 $\cdot 10^{-5}$} 
   & - & $1/6$ &  \footnotemark\\
 & $\pi^0 e^+ e^-$ & $<\,4$ & \multicolumn{3}{l}{$\cdot 10^{-5}$} & - & 
   $1/15 - 1/2$ &  \footnotemark\\
\hline\hline
$\eta^\prime$  & $\pi^+ \pi^- \eta$ & 44 & 3 & 1 & 5 $\%$ & 
 8200 & 18000 & \\ 
 {\it hadronic} & $\pi^0 \pi^0 \eta$ & 20 & 9 & 1 & 2 $\%$ & 5400 & 14500 & \\ 
 & $3\,\pi^0$ & 1 & 56 & 0 & 26 $\cdot 10^{-3}$ & 130 & 145 & \\ 
 & $\pi^+ \pi^- \pi^0$ & $<\,5$ & \multicolumn{3}{l}{$\%$} & - & 85 & 
   \footnotemark \\ 
\hline
$\eta^\prime$  & $\rho^0 \gamma$ & 29 & 5 & 1 & 0 $\%$ & 
 9550 & 44000 & \footnotemark \\ 
 {\it radiative} & $\omega \gamma$ & 3 & 03 & 0 & 31 $\%$ & 160 & 1200 & \\ 
 & $\gamma \gamma$ & 2 & 12 & 0 & 14 $\%$ & 2767 & 17100 & \\ 
\hline
$\eta^\prime$  & $\mu^+ \mu^- \gamma$ & 1 & 04 & 0 & 
 26 $\cdot 10^{-4}$ & 33 & 15 & \\ 
 {\it semi--leptonic} & $e^+ e^- \gamma$ & $<\,9$ & \multicolumn{3}{l}{$\cdot 10^{-4}$} & 
  - & 45 & \footnotemark \\ \hline\hline
\end{tabular}}
\label{tab_etap_rates}
\end{table}

The observation of CP or C violating rare decays might hint at effects from 
new physics beyond the Standard Model: 
The single conversion decay $\eta \rightarrow \pi^+ \pi^- e^+ e^-$ allows 
to search for CP violation in flavor--conserving processes beyond the 
Cabbibo--Kobayashi--Maskawa (CKM) 
mechanism~\cite{Cabibbo:1963yz,Kobayashi:1973fv}, which are not constrained 
by limits on the neutron electric dipole moment.  
A recently proposed CP violating mechanism from an interference between 
magnetic and 
electric decay amplitudes can induce a sizable linear photon polarization 
in the $\eta \rightarrow \pi^+ \pi^- \gamma$ 
decay~\cite{Gao:2002gq,Geng:2002ua}.
The photon polarization is accessible in the associated conversion decay 
$\eta \rightarrow \pi^+ \pi^- e^+ e^-$ as an asymmetry in the angular 
distribution between the $\pi^+ \pi^-$ and $e^+ e^-$ production planes. 
Similarly, this method was used to measure CP violating effects in the 
$K^0$ system in the decay $K_L \rightarrow 
\pi^+ \pi^- e^+ e^-$~\cite{Lai:2003ad}, with an asymmetry reported as large 
as $14\,\%$.
The corresponding asymmetry for $\eta \rightarrow \pi^+ \pi^- e^+ e^-$ is 
expected to be at a level of $10^{-3}$--$10^{-2}$ (see 
section~\ref{subsubetapday1} and~\cite{Gao:2002gq,Geng:2002ua}).

\addtocounter{footnote}{-6}
\footnotetext{assuming the theoretical estimate of $3\cdot 10^{-4}$ for the
branching ratio~\cite{Jarlskog:1967np}}
\stepcounter{footnote}
\footnotetext{assuming $\Gamma(\eta \rightarrow e^+ e^- e^+ e^-) / 
 \Gamma_{tot} = 2.52 \cdot 10^{-5}$~\cite{Jarlskog:1967np}}
\stepcounter{footnote}
\footnotetext{Standard Model estimate $\Gamma(\eta \rightarrow e^+ e^-) / 
 \Gamma_{tot} = 5 \cdot 10^{-9}$ from $\eta \rightarrow 
 \mu^+ \mu^-$~\cite{Savage:1992ac,Ametller:1993we}}
\stepcounter{footnote}
\footnotetext{Standard Model estimates $\Gamma(\eta \rightarrow 
 \pi^0 e^+ e^-) / \Gamma_{tot} = 
 0.2-1.3 \cdot 10^{-8}$~\cite{Cheng:1967pr,Ng:1993sc,Jarlskog:2002zz}}
\stepcounter{footnote}
\footnotetext{assuming $\Gamma(\eta^\prime \rightarrow \pi^+ \pi^- \pi^0) 
 \approx \Gamma(\eta^\prime \rightarrow 3\,\pi^0)$, 
 see~\cite{Beisert:2003zs}} 
\stepcounter{footnote}
\footnotetext{including $\eta^\prime \rightarrow  \pi^+ \pi^- \gamma$} 
\stepcounter{footnote}
\footnotetext{assuming $\Gamma(\eta^\prime \rightarrow  e^+ e^- \gamma) / 
 \Gamma_{tot} \approx 3 \cdot 10^{-4}$~\cite{Briere:1999bp}} 
\stepcounter{footnote}

A variety of decays, e.g.\ $\eta\left(\eta^\prime\right) \rightarrow 
\pi^0 \pi^0 \gamma$, $\eta\left(\eta^\prime\right) \rightarrow 
3 \pi^0 \gamma$, $\eta\left(\eta^\prime\right) \rightarrow 3 \gamma$, 
and $\eta^\prime \rightarrow e^+ e^- \pi^0 \left(\eta\right)$ or $\eta 
\rightarrow e^+ e^- \pi^0$ probe C--invariance, for which only moderate 
experimental limits exist.
For the latter decays to the $e^+ e^- \pi^0$ mode the dominant 
Standard Model --- i.e.\ C conserving --- mechanism is an intermediate 
state with two photons $\gamma^* \gamma^* \pi^0$ resulting in a branching 
ratio of 
$0.2 - 1.3 \cdot 10^{-8}$~\cite{Cheng:1967pr,Ng:1993sc,Jarlskog:2002zz} for 
the $\eta$ decay.
If C invariance is violated, the intermediate $\gamma^* \pi^0$ state with 
one virtual photon can contribute, increasing the branching ratio with 
respect to the Standard Model prediction (section~\ref{subsubetapmedium}). 

The move of the WASA facility to COSY offers the possibility to 
extend the programs both on $\eta$ decays at CELSIUS and on $\eta^\prime$ 
production dynamics at COSY.
Until the end of the present data taking period at KLOE in the middle of 
2005, an 
integral luminosity of $2000\,\mbox{pb}^{-1}$ is anticipated (presently 
$500\,\mbox{pb}^{-1}$), corresponding to $8 \cdot 10^7$ and $4 \cdot 10^5$ 
fully reconstructed $\eta$ and $\eta^\prime$ events, respectively, from 
$\phi \rightarrow \eta\left(\eta^\prime\right) \gamma$ 
events~\cite{Kluge:2004pc}.
The KLOE data will significantly improve the presently available data set.
However, the total anticipated statistics from KLOE can be reached 
with WASA at COSY in a period of 1 to 10 days, according to the rate 
estimates given in table~\ref{tab_etap_rates}, which take into account 
acceptance and reconstruction efficiency of the existing experimental setup 
of the WASA facility. Considering these future developments we
concentrate on specific rare and not-so-rare decays.

Together with the decay data, new and important information will be gained 
on the $\eta'$--nucleon interaction. Because of the special features 
of the $\eta'$ outlined above, this is a system of interest by itself. 
In addition, it is well known that final state interactions that occur 
subsequent to short range processes are universal. As a consequence the 
$\eta'$--nucleon interaction is the same in $pp\to pp\eta'$ as in 
$B\to \eta'pX$. Thus, the measurement of the former reaction will provide
necessary information to interpret the latter one
 that will eventually be measured at the B factories 
and might shed light on the dynamics of charmless 
B decays~\cite{Piccinini:2001ja}.

\subsubsection{\boldmath Isospin violation in $\vec{d}d\,\to\,\alpha\pi^0$}
\label{sec:dd2alphapi0}
\label{subsubalphapi}

Within the standard model there are only two sources of isospin
violation, namely the electro-magnetic interaction and the differences
in the masses of the lightest quarks~\cite{SWeinberg77,JGasser82}.
Especially in situations where we are able to disentangle these two
sources, the observation of isospin violation in hadronic reactions is a
direct window to quark mass ratios~\cite{JGasser82,miller} --- quark masses
themselves are not directly observable and additional information is
necessary to assign a scale to these fundamental parameters of the standard
model (see e.g.~\cite{Leutwyler:1996qg}).
  
Already in 1977 Weinberg predicted a huge effect (up to 30\%
difference in the scattering lengths for $p\pi^0$ and $n\pi^0$) of
isospin violation in $\pi^0 N$ scattering~\cite{SWeinberg77}. Also the impact
of soft photons was studied
systematically~\cite{Meissner:1997ii,GMueller99,NFettes01,Gasser:2002am}.
Since scattering experiments with neutral pions are not feasible, in
Ref.~\cite{jouni} it was suggested to use $NN$ induced pion production
instead\footnote{The $\pi^0p$ scattering length might also be measurable
in polarized neutral pion photoproduction at
threshold~\cite{AMBernstein98}.}.
The authors demonstrated that a charge symmetry breaking
(CSB)\footnote{Charge symmetry is fulfilled, if the amplitudes are
invariant under a 180 degree rotation in isospin space.
CSB is thus a subclass of isospin breaking effects.}
$\pi N$ seagull term (the Weinberg term), required by chiral symmetry,
should be even more relevant for $A_{fb}(pn\to d\pi^0)$ --- the
forward-backward asymmetry in $pn\to{}d\pi^0$ --- than $\pi -\eta$ mixing,
previously believed to completely dominate this CSB
observable~\cite{Niskanen:1998yi}.

\begin{SCfigure}[1.0][tb]
\resizebox{0.5\textwidth}{!}{\includegraphics{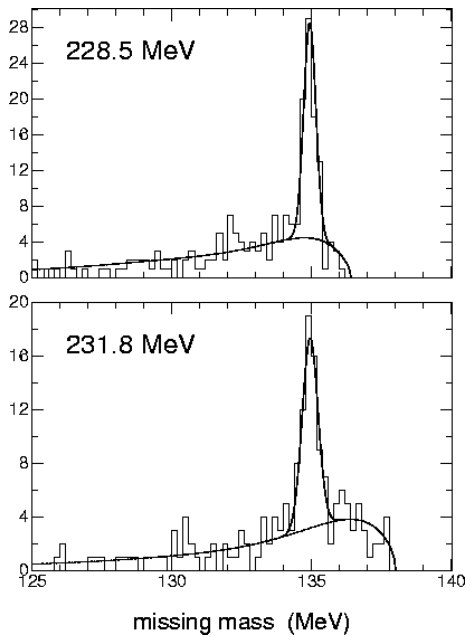}}
\caption{Experimental missing-mass spectra ($\mathrm{dd}\to\alpha$X) from 
       Ref.~\cite{Stephenson:2003dv} revealing a clean pion peak. The
       smooth curves show a reproduction of the data using a Gaussian
       peak and a continuum. The corresponding cross sections are
       $\sigma = (12.7 \pm 2.2) \mathrm{pb}$ and 
       $\sigma = (15.1 \pm 3.1) \mathrm{pb}$ 
       for the lower and higher energy, respectively. The
       data are consistent with pure $s$-waves
       contributions. \label{fig:iucf} }
\end{SCfigure}

Interest in CSB in hadronic reactions was revived recently with the
successful completion of experiments for $A_{fb}(pn\to
d\pi^0)$~\cite{Opper:2003sb} as well as $dd\to \alpha \pi^0$
(\cite{Stephenson:2003dv}, see Fig.~\ref{fig:iucf}) close to the pion
threshold. A  large collaboration has formed to perform
the corresponding calculations within effective field theory. First
results are presented in~\cite{Gardestig:2004hs}.  The studies show
that the relative importance of the various charge symmetry breaking
effects is very different for the two reactions and a consistent
investigation of both should help to further disentangle the leading
CSB matrix elements.

However, open questions will remain. It is the main result of
Ref.~\cite{Gardestig:2004hs} that, within a plane wave calculation,
the Weinberg term is suppressed due to selection rules in the reaction
$dd\to \alpha \pi^0$. Thus, the lowest order that gives finite
contributions is next-to-next-to leading order (NNLO).
At this order there are several diagrams contributing and only a NNNLO
contribution will really prove that there is a convergence in the series
for this particular reaction.
However, there is one more important test for the approach that will
be used for the analysis of the recent CSB experiments, namely, once
the parameters are fixed the $p$-waves in $dd\to \alpha \pi^0$ can be
predicted parameter free to leading and next-to-leading order. The
IUCF experiment~\cite{Stephenson:2003dv} was consistent with purely
$s$-waves contributing. Thus, the same experiment at somewhat higher
energies (but still well below the $\Delta$ region, e.g. at $Q\approx 60$
MeV) is urgently called for.

The Bose symmetry of the initial wave function strongly limits the
number of allowed partial waves. As a consequence, for $s$-wave as
well as $p$-wave production only a single partial wave is allowed,
namely $^3P_0\to s$ and $^5D_1\to p$. Thus, they do not interfere in
any unpolarized measurement. It would therefore be very important to
measure the unpolarized total cross section as well as 
polarized differential observables. In addition, the analyzing
powers are given by the imaginary parts of interference terms and thus
would provide an important test of the initial state interaction,
the ingredient to the calculations that is controlled to the smallest
extent. In a plane wave treatment the analyzing powers would vanish
identically.

In a further step, CSB effects involving the $\Delta$ isobar and
purely nucleonic ones can be disentangled taking data in the $\Delta$
regime as well ($Q\approx 160$ MeV). This will not only be an additional
important cross check of the analysis carried out previously, but
gives in addition quite direct access to matrix elements difficult to
get at from other experiments.  However, in order to study the
different partial waves polarized differential observables are as
important as for the lower energy.

It is important to stress that CSB studies for few-nucleon systems can be
based on recent developments in effective field
theories~\cite{EEpelbaum04,EEpelbaum02a,EEpelbaum02b,EEpelbaum04a,JLFriar}, thus allowing
for a model-independent analysis of the data.

\clearpage

\subsection{Additional objectives:\\ Exotic and cryptoexotic hadron resonances}
The study of hadron resonances and their decays is one important
approach to obtain information about quark dynamics in the
non-perturbative regime of QCD.  The experiments performed at the
proton-antiproton storage ring LEAR (Crystal Barrel)
 and at the new generation of experiments in
electron-positron colliders (BaBar, Belle, BES, CLEO) have found
new mesons which challenge the traditional interpretation of mesons as
quark-antiquark states\cite{Amsler:2004ps,Belle03}. In particular, new
meson candidates with the quantum numbers $J^{PC}=1^{-+}$ cannot be
generated by $q\bar{q}$ states\cite{Abele98,Kuhn04}. Because of this feature 
these states
are called exotics.
With the recent evidence for the existence of the so called pentaquark
$\Theta^+$ a first baryon candidate was added to this list.
 In addition, in the hadron spectrum there are
several states with quantum numbers consistent with the quark model.
However, some of their other properties are not. These states are called
cryptoexotics. Both mesons, like the well known $a_0$
and $f_0$, as well as baryons like the $\Lambda (1405)$, are discussed
in the literature as candidates for this class of states. However, it
is clear that additional experimental information is needed to verify
or falsify these conjectures.
Hadronic probes provide an experimental tool complementary to 
electromagnetic excitations because of different spin- and isospin
selectivities. 
In the following we elaborate why WASA is particularly
suited to investigate those exotic and cryptoexotic resonances.

In the {\it meson sector} the controversial states that are within the
COSY energy range are the scalar--isoscalar $f_0(980)$ and the
scalar--isovector $a_0(980)$. Although 
QCD can in principle be treated explicitly in the low
Mo-men\-tum-transfer regime using lattice
techniques~\cite{Kunihiro:2003xe}, those are not yet in the
position to make quantitative statements about light scalar states
($J^P{=}0^+$). Alternatively, QCD inspired models, which use effective
degrees of freedom, are to be used. The constituent quark model is one
of the most successful in this respect (see e.g.\
Ref.~\cite{qmod1,qmod2,qmod3,qmod4,qmod5,qmod6}). This approach treats 
the lightest scalar
resonances $a_0/f_0$(980) as conventional $q\bar{q}$ states. However,
they have also been identified with $K\bar{K}$
molecules~\cite{KK_1,KK_2,KK_3,KK_4,KK_5} or compact $qq$-$\bar{q}\bar{q}$ states
\cite{4q_1,4q_2,4q_3}. It has even been suggested that at masses below 1.0
GeV/c$^2$ a full nonet of 4-quark states might exist
\cite{Close:2002zu}. 

As in the reaction $dd\to \alpha \pi^0$, with WASA at COSY we can use the isospin 
selectivity in the $dd\to \alpha X$ transition, to model independently
access
 the isospin-violating 
$a_0$-$f_0$ mixing: the total cross section for the reaction
$dd\to\alpha(\pi^0\eta)$ was shown to be directly proportional to the
square of the mixing matrix element with the influence of other
isospin violating effects being suppressed by one order of magnitude
\cite{Hanhart:2003pg}. 
There is the additional possibility to extract the phase of the mixing matrix 
element from
the forward-backward asymmetry in
$pn\to da_0^0/f_0$.  Both measurements will reveal independent
 new information about the structure
of the light scalar mesons, in particular about their $K\bar K$-meson
content.

In the {\it baryon sector}, several facilities have obtained evidence for a
new baryon resonance, the $\Theta^+$, which has achieved a three-star
status in the Particle Data Group~\cite{PDBook}. If ultimately
confirmed with convincing statistics, this resonance will have a large
impact on the field.  The $\Theta^+$ has positive strangeness which can
only be generated by an antistrange quark as a building block of the
resonance and therefore cannot be accommodated by a three quark
structure.  The width of the $\Theta^+$ is unexpectedly small: while
direct determinations of the width can only give upper limits of the
order of 10~MeV/c$^2$ due to limited energy resolution, indirect analysis
point to a width of less than
1~MeV/c$^2$~\cite{Arndt03,Haid03,cahn04,sib04}.
Jaffe and Jain have recently pointed out the limitations such a small 
width imposes on the possibilities to generate the $\Theta^+$  dynamically in the
K N channel or as a Castillejo-Dalitz-Dyson pole\cite{Jaffe:2004at}.

Theorists have pointed out the possible existence of pentaquark states
with narrow widths due to the formation of colored clusters about
thirty years ago~\cite{Hogaasen:1978jw}. Recent theoretical work has
focussed on the diquark degree of freedom~\cite{Jaffe03}.  Though the
clustering of colored diquarks provides a qualitative explanation of
the small width of the $\Theta^+$, explicit dynamical calculation
require fine-tuning of the model~\cite{Stech04}. Combining the colored
diquark degree of freedom with chiral symmetry, however, produces
hadrons which are stable and do not couple to the kaon-nucleon degree
of freedom in the chiral limit~\cite{Beane,Ioffe,Melikhov}.

The production of the $\Theta^+$ via the reaction
$ p d \rightarrow p \Lambda \Theta^+$
allows access to a pure isospin $I=0$ combination $\Lambda \Theta^+$
in the final state. This allows to obtain information 
complementary to the proton-proton reaction which produces the
$\Theta^+$ starting from a pure isospin $I=1$ state.

It is well established that if there is one exotic there should be
many and it is the corresponding multiplet structure that encodes
essential information on the underlying substructures.

The $N^*(1710)$ has been assumed to be a member of the 
pentaquark antidecuplet and predictions for the branching into
various decay channels have been published\cite{Diakonov}
awaiting experimental tests.

The Roper resonance, $N(1440)$, has been studied already with WASA at
CELSIUS.  The Roper is not a manifestly exotic state, but its broad
width and small mass have caused many experimental and theoretical
investigations of its structure. In the Jaffe-Wilczek diquark scenario,
the Roper may be related to the $\Theta^+$.

In addition,
the $\Lambda(1405)$ has been 
a candidate for a non-triplet quark structure and thus for a crypto 
exotic for a long time.
 Recent work within unitarized chiral perturbation theory
suggests that the $\Lambda(1405)$ is made of two overlapping
resonances with different flavor structure. This calls for detailed
investigations of the decay channels as a function of the excitation
energy.

Recently, the ANKE collaboration has found evidence for a hyperon
$Y(1475)$ (see Fig.\ref{fig:Y1475} in section~\ref{sec:hyperons}) 
which is not predicted 
by any known quark model. The PDG gives a one-star status to a surmised
$\Sigma(1480)$.

\subsubsection{\boldmath Mixing of the scalar mesons $a_0/f_0$(980)}
\label{sec:a0_f0}

Both, the (isospin $I{=}1$) $a_0$- and the ($I{=}0$) $f_0$-resonances
can decay into $K\bar K$, whereas in the non-strange sector the decays
are into different final states according to their isospin,
$a_0^\pm{\to} (\pi^\pm\eta)_{I=1}$, $a_0^0{\to} (\pi^0\eta)_{I=1}$ and
$f_0{\to} (\pi^0\pi^0)_{I=0}$ or $(\pi^+\pi^-)_{I=0}$.  Thus, only
the non-strange decay channels have defined isospin and allow one to
(model independently) discriminate between the two mesons. It is also only by measuring the non-strange
decay channels that isospin violating effects can be investigated.  
Such measurements
can be carried out with WASA at COSY for $\pi^0$- or $\eta$-meson
identification, while the strange decay channels $a_0/f_0{\to} K_SK_S$
might be measured in parallel. Measurements of the $K\bar K$ final
state with at least one charged kaon have already been performed at
COSY using magnetic spectrometers. The main results of these
experiments and their implications on the proposed measurements are
described below.

\begin{figure}[tb]
  \centering 
  \resizebox{7.5cm}{!}{\includegraphics[scale=1]{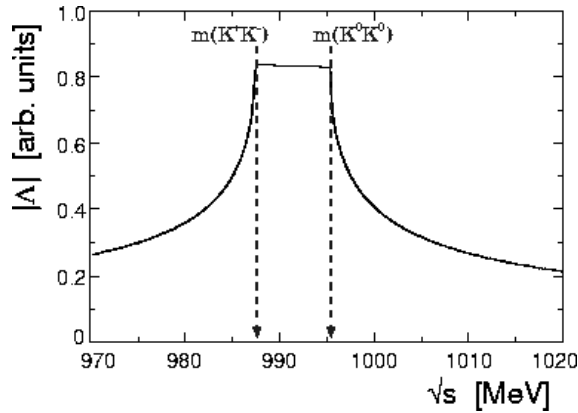}}
  \caption{Modulus of the leading term of the mixing amplitude
  $\Lambda$ defined in Eq.\ (\protect\ref{eq:llam}). The two kinks occur
  at the $K^+ K^-$ (at 987.35 MeV) and the $\bar K^0 K^0$ (995.34 MeV)
  threshold respectively.}
  \label{fig:mixsdep}
\end{figure}

In 1979 Achasov and collaborators~\cite{Achasov:1979xc} pointed out
that between the charged and neutral kaon thresholds the leading term
to the $a_0$-$f_0$ mixing amplitude is dominated by the unitary cut of
the intermediate two-kaon system. It was demonstrated, that the
leading piece of the $a_0$-$f_0$ mixing amplitude can be written as
\begin{equation}
\Lambda = \langle f_0 |T| a_0\rangle = ig_{f_0K\bar K}g_{a_0K\bar
    K}\sqrt{s}\left( p_{K^0}-p_{K^+} \right) \ + {\cal
    O}\left(\frac{p_{K^0}^2-p_{K^+}^2}{s}\right) \ ,
\label{eq:llam}
\end{equation}
where the effective coupling constants are defined through
$\Gamma_{xK\bar K}=g_{xK\bar K}^2p_K$.  This $\sqrt{s}$ dependence of
$\Lambda$ is depicted in Fig.~\ref{fig:mixsdep}. Here, electromagnetic
effects between the kaons were neglected as in~\cite{Achasov:1979xc}.

As demonstrated in a recent compilation~\cite{Baru:2003qq}, the values
for $g_{a_0K\bar K}$ and $g_{f_0K\bar K}$ range from 0.224 to 0.834
and from 1.31 to 2.84, respectively --- depending on the Flatt\'e
parameterization of measured $\pi\pi$ and $\pi\eta$ spectra. It has
also been shown in Ref.~\cite{Baru:2003qq} that the coupling of a
physical particle to mesons carries information about the nature of
that particle, as was shown by Weinberg for the deuteron case
\cite{weinberg}. Based on the existing data, the authors of
Ref.~\cite{Baru:2003qq} conclude that both the $a_0$ and $f_0$ have
significant mesonic ($K\bar K$) components, however, more quantitative
statements require a better knowledge of $g_{a_0/f_0K\bar K}$, which
could be obtained from an accurate measurement of $\Lambda$.

In addition, since $\Lambda$ (as defined in eq.~\ref{eq:llam}) is essentially 
the overlap of the
wavefunctions of $a_0$ and $f_0$, it is expected that also that part
of
the mixing amplitude
 that does not stem from kaons will reveal new insights into the
structure of the light scalar mesons.

A $pp\to dX$ reaction must lead to $a_0^+$ ($I{=}1$) production, a
$pn\to dX$ interaction is not isospin selective and both the $a_0^0$
and $f_0$ may be produced ($I{=}0,\, 1$), whereas the $dd\to\alpha X$
reaction --- neglecting the small isospin violating contributions which 
are the final
goal of the proposed experimental program --- is a filter for the
$f_0$ ($I{=}0$) resonance, since the initial deuterons and the
$\alpha$ particle in the final state have isospin $I{=}0$ (``isospin
filter''). Since at COSY it is possible to manipulate the initial
isospin one can thus selectively produce the $a_0$ or $f_0$ resonances
and can identify observables that vanish in the absence of
isospin violation~\cite{miller,ANKE_WS}. The idea behind the proposed 
experiments is
the same as behind recent measurements of isospin violation in 
the reactions
$np{\to} d\pi^0$~\cite{Opper:2003sb} and $dd{\to}
\alpha\pi^0$~\cite{Stephenson:2003dv}. However, the interpretation of the
signal from the scalar mesons is much simpler as compared to the pion
case. Since the $a_0$ and the $f_0$ are rather narrow overlapping
resonances, the $a_0$-$f_0$ mixing in the final state is enhanced by
more than an order of magnitude compared to isospin violation in the production
operator (i.e.\ ``direct'' isospin violating $dd{\to}\alpha a_0$ production) and
should, e.g., give the dominant contribution to the isospin violating effect 
via the
reaction chain $dd{\to} \alpha f_0(I{=}0) {\to}
\alpha a_0^0(I{=}1) {\to} \alpha (\pi^0\eta)$~\cite{Hanhart:2003pg}. 

The $dd{\to} \alpha (\pi^0\eta)$ reaction seems to be most promising
for the extraction of isospin violating effects. Any observation of $\pi^0\eta$
production in this particular channel is a direct indication of isospin
violation,
however, the cross section $\mathrm{d}\sigma/\mathrm{d}m$ will be
given by the product of the mixing amplitude $\Lambda(m)$ and the
$dd{\to}\alpha f_0$ production operator. It is therefore compulsory to
determine the latter in an independent measurement in order to extract
the mixing amplitude. A corresponding experiment aiming at the
measurement of the $dd\to\alpha (K^+K^-)_{I{=}l{=}0}$ cross section is
foreseen for winter 2004/05 at ANKE~\cite{dd_proposal}.  These data,
together with the information on the $dd{\to} \alpha (\pi^0\eta)$
reaction from WASA will allow one to determine $\Lambda$ 
independently.

In analogy to the measurement of isospin violation in the 
reaction $np{\to}
d\pi^0$, it has been predicted that the measurement of angular
asymmetries (i.e.\ forward-backward asymmetry in the $da_0$ c.m.s.) 
also allows to extract the $a_0$-$f_0$ mixing amplitude
\cite{Grishina:2001zj,Kudryavtsev:2001ee,Kudryavtsev:2002uu}. It was 
stressed in Ref.\ \cite{Kudryavtsev:2001ee} that --- in contrast to
the $np{\to} d\pi^0$ experiment where the forward-backward asymmetry
was found to be as small as 0.17\% \cite{Opper:2003sb} --- the
reaction $pn{\to} d\pi^0\eta$ is subject to a kinematical enhancement.
As a consequence, the effect is predicted to be significantly larger
in the $a_0$/$f_0$ case. The numbers range from some
10\%~\cite{Kudryavtsev:2001ee} to a few factors
\cite{Grishina:2001zj} and, thus, should easily be observable. It has 
been pointed out in Ref.~\cite{Kudryavtsev:2002uu} that the analyzing
power of the reaction $\vec p n{\to} d \pi^0 \eta$ also carries
information about the $a_0$-$f_0$ mixing amplitude. This quantity can
be measured at COSY as well.

\begin{figure}[tb]
  \resizebox{7.0cm}{6.5cm}
          {\includegraphics[scale=1]{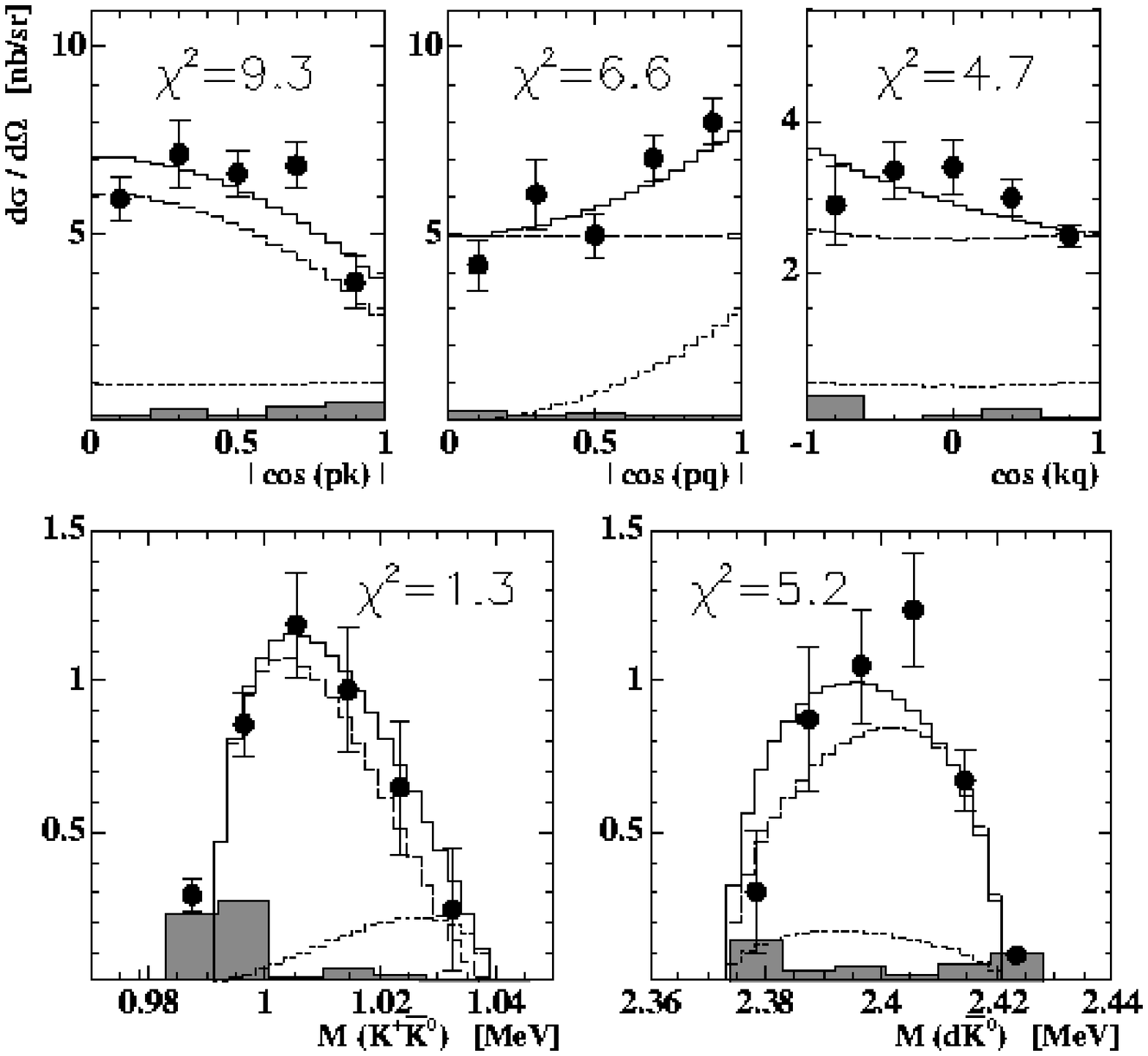}}
  \resizebox{5cm}{!}{\includegraphics[scale=1]{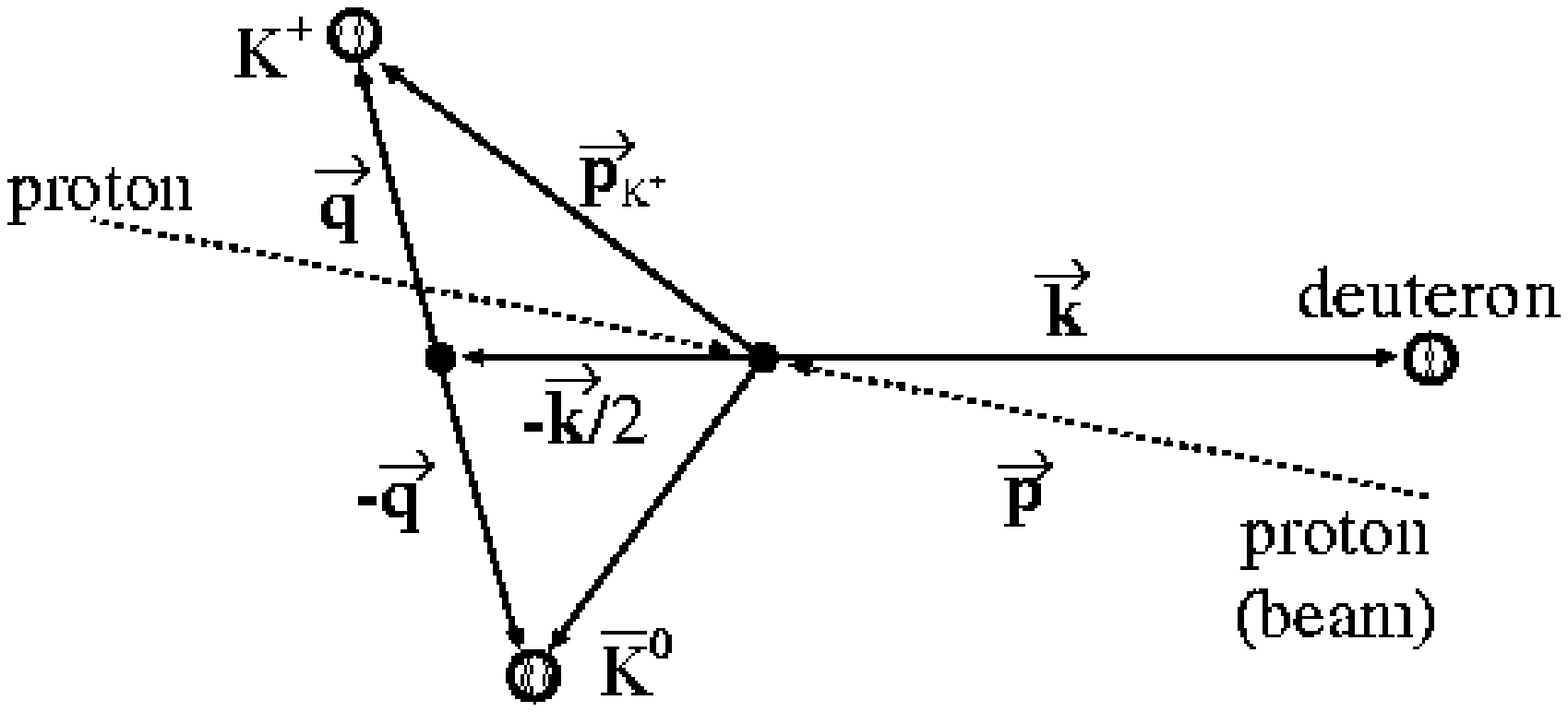}}
  \caption{ANKE data for the reaction $p(3.46\, \mathrm{GeV/c})p\to
  dK^+\bar{K}^0$~\cite{Kleber:2003kx}. The shaded areas correspond to the
  systematic uncertainties of the acceptance correction. The dashed
  (dotted) line corresponds to $K^+\bar{K}^0$-production in a relative
  $S$-($P$-) wave and the solid line is the sum of both
  contributions. For definition of the vectors $p$, $q$ and $k$ in the
  cms of the reaction $pp {\to} dK^+\bar{K^0}$ see right hand part of
  the figure.  Angular distributions with respect to the beam
  direction $\vec{p}$ have to be symmetric around $90^\circ$ since the
  two protons in the entrance channel are indistinguishable.
}
  \label{fig:pp2dKKbar}
\end{figure}

An experimental program has already been started at COSY which aims at
exclusive data on $a_0/f_0$ production close to the $K\bar{K}$
threshold from $pp$~\cite{Quentmeier:2001ec,Moskal:2002jd,a+_proposal},
$pn$ \cite{a0f0_proposal}, $pd$~\cite{a0f0_proposal,momo-KK} and $dd$
\cite{css2002,dd_proposal} interactions. The reactions $pp {\to} 
ppK^+K^-$ and $pd {\to} ^3\mathrm{He}\, K^+K^-$ have been measured at
COSY-11 \cite{Quentmeier:2001ec,Moskal:2002jd} and
MOMO~\cite{momo-KK}, respectively, at excitation energies up to
$Q=56$~MeV above the $K\bar K$ threshold. However, mainly due to the
lack of precise angular distributions, the contribution of the
$a_0/f_0$ to $K\bar K$ production remains unclear for these reactions.
At ANKE, the reaction $pp {\to} dK^+\bar{K^0}$ has been measured
exclusively (by reconstructing the $\bar{K^0}$ from the measured
$dK^+$ missing mass) at beam momenta of $p{=}3.46$ and 3.65 GeV/c
($Q{=}46$ and 103 MeV).  The differential spectra for the lower beam
momentum are shown in Fig.\
\ref{fig:pp2dKKbar}~\cite{Kleber:2003kx}. The background of
misidentified events at ANKE is less than 10\% which is crucial for
the partial-wave analysis. This analysis reveals that the
$K^+\bar{K}^0$ pairs are mainly (83\%) produced in a relative $S$-wave
(dashed line in Fig.\ \ref{fig:pp2dKKbar}), presumably via the $a_0^+$
channel~\cite{Kleber:2003kx}. Based on model calculations for the $pp
{\to} dK^+\bar{K^0}$ reaction the authors of
Ref.~\cite{Grishina:2004rd} draw the conclusion that for $Q{<}100$ MeV
$K\bar K$ pair production proceeds dominantly via the
$a_0^+$-resonance.
Limits on the $\overline{K^0}d$ scattering length have been obtained in
Ref.~\cite{ASibirtsev04}.
For a further discussion of scalar mesons in these reactions see
e.g. Ref.~\cite{EOset96}.

Data on the reaction $p(3.46\, \mathrm{GeV/c}) p\to d \pi^+X$ have
been obtained at ANKE in parallel to the kaon data.  In contrast to
the latter, where the spectra are almost background free, the $pp\to
d\pi^+\eta$ signal is on top of a huge multi-pion
background~\cite{a+_pieta}. This demonstrates the need for a photon
detector supplying active $\pi^0$- or $\eta$-meson
identification. With such a detector one may expect data of a quality
comparable to what is shown in Fig.\ \ref{fig:pp2dKKbar} also for the
non-strange decay channels.

\subsubsection{Hyperon resonances}
\label{sec:hyperons}

Compared to the spectrum of nucleon resonances, the excitation modes
of hyperons ($\Lambda$,$\Sigma$) are much less known.
The lowest excitations listed in Ref.~\cite{PDG} are the $\Lambda$(1405)
$J^P=\frac{1}{2}^-$ isospin singlet and the $\Sigma$(1385)
$J^P=\frac{3}{2}^+$ isospin triplet states, well accessible in
proton-nucleon collisions at COSY. The study of the $\Lambda$(1405)
state is particularly interesting since this resonance has still not
been understood in its nature, its spin and isospin assignment is
based on indirect arguments only. In the quark model the
$\Lambda$(1405) is interpreted as a state with orbital excitation.
However, three well established quark models with entirely different
residual interaction (gluon exchange~\cite{Isgur:1978xj}, meson
exchange~\cite{Glozman:1999vd}, and t'Hooft instanton induced
interaction~\cite{Loring:2001kx}) have difficulties in reproducing the
mass of the $\Lambda$(1405).  All
models~\cite{Isgur:1978xj,Glozman:1999vd,Loring:2001kx} find a
degeneracy of the computed $\Lambda$(1405) with the
$J^P=\frac{3}{2}^-$ $\Lambda$(1520) resonance in contrast to the
observation.  A similar problem is also manifest in very recent
lattice QCD calculations where an extrapolation to the physical pion
mass results in a too high value for the $\Lambda$(1405)
mass~\cite{Melnitchouk:2002eg}.

On the other hand, since long time the $\Lambda$(1405) has been
interpreted as a $\bar{K}N$ quasi-bound
state~\cite{RCArnold62,RHDalitz67,Jones:1977yk}, later on substantiated by
the consideration of chiral symmetry~\cite{Kaiser:1995eg,Kaiser:1996js}.
Recent work in the framework of unitarized chiral effective field
theories has pointed out that the structure of the $\Lambda$(1405) is
extremely sensitive to the breaking of the $SU(3)$
symmetry~\cite{Oset:1997it,JAOller01,Oset:2001cn,Jido:2003cb}.
Ref.~\cite{Jido:2003cb} finds two poles of the scattering matrix close to
the nominal $\Lambda$(1405) resonance, one of which couples more
strongly to $\pi\Sigma$ states and the other one mostly to
$\bar{K}N$ states.  As an experimentally observable consequence,
peak structures with different invariant mass distributions are
expected in photon and hadron induced reactions that populate the
$\Lambda$(1405) resonance~\cite{Jido:2003cb}.

The experimental information available on the $\Lambda$(1405) and
$\Sigma$(1385) resonances is primarily based on studies of $K^-$
proton collisions in bubble
chambers~\cite{Thomas:1973uh,Hemingway:1984pz,MAguilar81,MBaubillier84}.
In this entrance channel the resonances are only
observable through the decay of higher-lying $\Lambda^{\star}$ and
$\Sigma^{\star}$ resonances.

\begin{figure}[tb]
\begin{center}
\epsfig{file=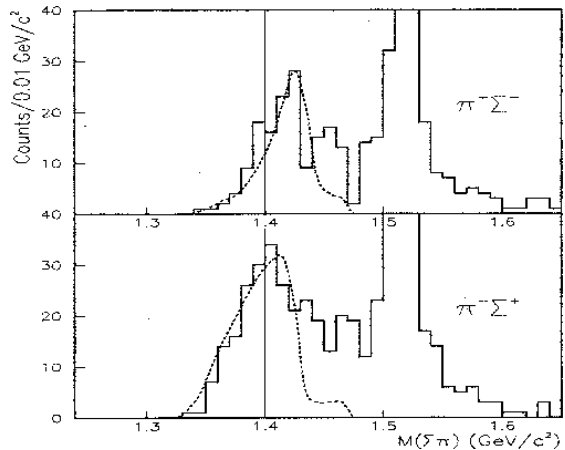, width=7.5cm}
\caption{Invariant mass spectra for $\pi^+\Sigma^-$ and $\pi^-\Sigma^+$ 
   as observed in the $p(\gamma{},K^+\pi)\Sigma$ reaction at
   SPring-8/LEPS~\cite{Ahn:2003mv}.}
\label{fig:SPring-8}
\end{center}
\end{figure}

Recently photon-induced hyperon resonance production with a statistics
of $\approx$100 events in the $\Lambda$(1405)/$\Sigma$(1385) region was
observed at SPring-8~\cite{Ahn:2003mv}, showing different spectral
shapes in $\Sigma^+\pi^-$ and $\Sigma^-\pi^+$ final states. Based on
arguments given in Ref.~\cite{Nacher:1998mi}, this was interpreted as
an indication for a meson-baryon nature of the
$\Lambda$(1405)~\cite{Ahn:2003mv} (see Fig.~\ref{fig:SPring-8}).

In view of the limited quality of the existing data and the lack of
proton induced studies, it is important to investigate
the nature of the $\Lambda$(1405) resonance in proton-proton collisions by
measuring its spectral shape with good resolution and statistics.
This will allow recent theoretical predictions based on the chiral
unitarity approach to be tested.  In particular, a comparison of the
$\Sigma^+\pi^-$ and $\Sigma^-\pi^+$ spectral distribution will reveal
a possible resonant behavior of the the $I=1$ amplitude in this mass
region, whereas the $\Sigma^0\pi^0$ distribution provides the true
shape of the $I=0$ resonance according to Ref.~\cite{Nacher:1998mi}.
A comparison with the shape of the spectral distribution measured in
photo-induced reactions is predicted to reveal the two-pole structure
of the resonance~\cite{Jido:2003cb}.

\begin{figure}[tb]
\begin{center}
\epsfig{file=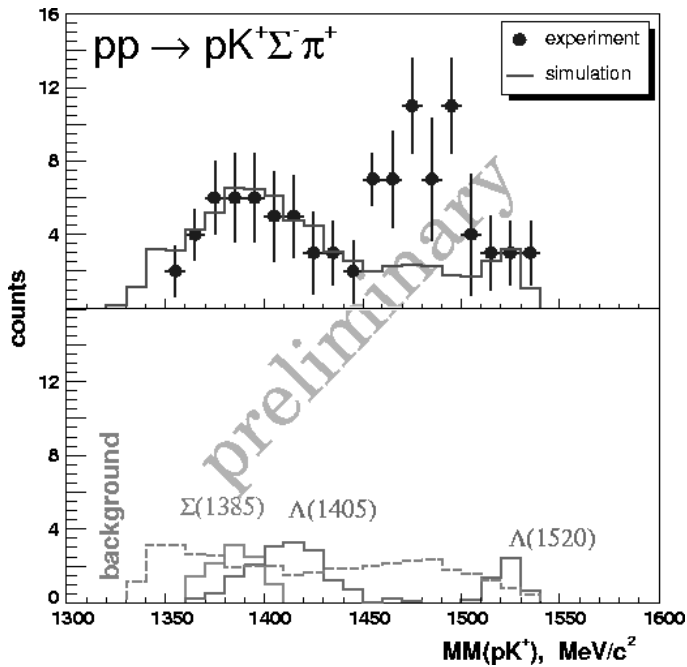, width=9cm}
\caption{$K^+p$ missing mass spectrum for the reaction 
   $p(\mathrm{3.60\ GeV/c})p \rightarrow pK^+\,\Sigma^-\pi^+$ measured
   at ANKE in comparison with a simulation (upper panel).
   The simulated $K^+p$ missing mass spectrum decomposes into background
   from misidentified events and non-resonant $\Sigma^-\pi^+$ production,
   as well as the known hyperons $\Sigma(1385)$, $\Lambda(1405)$ and
   $\Lambda(1520)$ decaying into $\Sigma^-\pi^+$ (lower panel).}
\label{fig:Y1475}
\end{center}
\end{figure}

An experimental study of the $\Lambda$(1405) resonance also delivers
information on the $\Sigma$(1385) resonance as a by-product, since due
to their similar pole masses and widths the production of these two
resonances can only be disentangled by the complete identification of
the final states populated in their decays ($\Lambda{\rm{}(1405)}
\rightarrow\Sigma\pi$ (100\%); $\Sigma{\rm{}(1385)} \rightarrow 
\Lambda\pi$ (88$\pm$2\%), $\Sigma\pi$ (12$\pm$2\%)). This involves 
the detection of both charged and neutral pions, and thus a large
acceptance detector with  charged and neutral particle detection
capability like the WASA detector. In particular, photon detection is
required for the observation of the decay channel
$\Lambda(1405) \rightarrow\Sigma^0\pi^0\rightarrow
(\Lambda\gamma)(\gamma\gamma)$.  The identification of this final
state is particularly important since it is not populated in the decay
of the $\Sigma^0(1385)$ resonance and thus provides direct
access to the spectral shape of the $\Lambda$(1405) resonance.
The $\Sigma^0$(1385) contribution in the charged $\Sigma\pi$ final states
($\Sigma^+\pi^-$,$\Sigma^-\pi^+$) can be deduced from a measurement of the
$\Lambda\pi^0$ final state which is not populated in the $\Lambda$(1405)
decay.

Recent data from ANKE on the reaction
$pp\rightarrow pK^+\,Y^*\rightarrow pK^+\,\Sigma(1190)^-\pi^+$
measured at a beam momentum of 3.60 GeV/c are consistent with a
hyperon resonance $Y^*$ with a mass of 1475~MeV/c$^2$ and a width of
45~MeV/c$^2$ (see \cite{koptev02} and Fig.~\ref{fig:Y1475}).
This state is also seen with comparable strength in the charge-conjugate
decay channel $Y^*\rightarrow \Sigma(1190)^+\pi^-$ and its production cross
section is of the same order as for the $\Lambda(1405)$.
Within the above mentioned quark
models~\cite{Isgur:1978xj,Glozman:1999vd,Loring:2001kx} there is no
room for an additional $\Lambda$ or $\Sigma$ hyperon below $\approx$1600
MeV/c$^2$.
However, a $\Sigma(1480)$ with a $4q1\bar q$ structure has been predicted
in Ref.~\cite{Hogaasen:1978jw} --- as the $I=1$ partner of the
$\Lambda(1405)$ in a common multiplet.
Further studies are needed in order to confirm the conjecture that the
observed enhancement in the $K^+p$ missing mass spectrum originates from a
new hyperon resonance.
If this interpretation will be supported after further analysis or
new measurements, the next step is to determine the isospin of the new
hyperon state.
This is not possible from the ANKE data but requires a comparison of the
possible decay channels $\Sigma^{\pm}\pi^{\mp}$, $\Sigma^0\pi^0$, and
$\Lambda\pi^0$.
\label{Y1475ref}

As an additional note, it should be mentioned that the different nature of
the $\Lambda$(1405) and $\Sigma$(1385) states is also expected to manifest
itself in different in-medium properties of these hyperon resonances.
In particular, if the meson-baryon picture of the $\Lambda$(1405) is
correct, its medium properties are intimately related to the behavior of
antikaons ($K^-$,$\bar{K}^0$) in nuclear matter, which is a still unsolved
problem, of importance in both hadron physics and astrophysics.
A comparative study of $\Lambda$(1405) and $\Sigma$(1385) resonance
production on nuclear targets allows to investigate the propagation of
these hyperon resonances inside the nuclear medium.
A better knowledge of the in-medium properties of the $\Lambda$(1405)
(absorption cross section, modification of the spectral shape) will also
help to understand the dynamics of antikaons in nuclear matter.

\subsubsection{Pentaquarks}
\label{sec:pentaquark}

Early in 2003, the LEPS Collaboration at the SPring-8 facility in
Japan observed a sharp resonance, $\Theta^{+}$, at $(1.54\pm 0.01)$~GeV/c$^2$
with a width smaller than 25~MeV/c$^2$ and a statistical significance of
$4.6 \sigma$ in the reaction $\gamma n\rightarrow
K^{+}K^{-}n$~\cite{Nakano:2003qx}. This resonance decays into
$K^{+}n$, hence carries strangeness $S= +1$. Later, many other groups
have claimed the observation of this 
state~\cite{pos_1,pos_2,pos_3,pos_4,pos_5,pos_6,pos_7,pos_8,pos_9,pos_10,pos_11}.
In particular, evidence for the $\Theta^+$ has also been seen at the
COSY-TOF experiment in the reaction
$pp\rightarrow\Sigma^+pK_s$~\cite{MAbdel-Bary04} in the $pK_s$ invariant
mass spectrum.
While in some experiments detecting a $pK_s$ in the final state the
strangeness of the resonance in this system is undefined, in this experiment
it is uniquely determined that the observed $pK_s$ system originates from
a $S=+1$ state since the full final state with an associated $\Sigma^+$ is
identified.

Since all known baryons with $B= +1$ carry negative or zero strangeness,
such a resonance must have a minimum quark content $uudd\bar{s}$, a
structure which is clearly beyond the conventional quark model. This
experimental discovery triggered a lot of theoretical activity and up
to now over 250 papers appeared trying to interpret this exotic
state~\cite{review}.

There is preliminary evidence that the $\Theta^{+}$ is an iso-scalar
because no enhancement was observed in the $pK^{+}$ invariant mass
distribution~\cite{pos_3,pos_5,pos_6,pos_11}. All the other quantum numbers 
including
its angular momentum and parity remain undetermined. Most of
theoretical work postulated its angular momentum to be one half,
because of its low mass, but the possibility of $J = 3/2$ still cannot
be excluded completely.

It is important to point out that many other experimental groups
reported negative results~\cite{neg_1,neg_2,neg_3}. A long list of experiments
yielding negative results (still unpublished at the time of the writing) can be found
in Ref.~\cite{morenegative}.

With both positive and null evidence for the $\Theta^{+}$ from a
variety of experiments, it is difficult to conclude whether the
$\Theta^{+}$ exists or not. Furthermore, the theoretical difficulties
to explain a possible narrow width, perhaps as small as 1~MeV/c$^2$, suggest
that if the $\Theta^{+}$ exists, it is very unusual indeed. Guidance
from lattice gauge theory is not helpful, as some calculations show
evidence for a negative parity resonance, while another indicates positive
parity. Furthermore, one does not see a resonance in either parity.  So the 
confirmation of the $\Theta^{+}$ is of highest importance for which 
the COSY-TOF collaboration will start an extensive experimental
investigation in October 2004. 

One should note that the $\Theta^+$ is a member of a multiplet. This calls
for the existence for additional unusual states.
WASA at COSY with its very good mass resolution and large acceptance
can make a decisive contribution to this field by providing high
statistics data on the differential cross sections of exotic baryon
production in elementary reactions.
While there are several elementary reactions with the $\Theta^{+}$ in the
final state within the COSY energy range, the
$pd\rightarrow p\Lambda\Theta^{+}$ reaction is of special interest.
Currently this reaction is investigated by the CELSIUS/WASA collaboration
close to threshold.
This reaction provides access to the pure isospin zero $\Theta^+$
production channel via the subsystem $pn\to \Theta^+ \Lambda$ and thus
allows to obtain further and independent insight into the physics of the
$\Theta^+$ pentaquark as compared to elementary transitions like
$pp\to \Theta^+ \Sigma^+$ (which is pure isospin one).
For example, one can obtain additional and complementary information on the
parity of the $\Theta^+$ \cite{CHanhart04} but explore also the $\Theta^+$
production mechanism, which is so far completely unknown.

\clearpage

\section{Key experiments}
\label{keyintro}

  In this chapter several key experiments are presented to address the
  issues discussed in the previous chapter.  After the commissioning
  phase of WASA at COSY, in which the performance of the detector will
  be demonstrated, the intention is to carry out the experiments
  discussed below.  The proposed immediate program consists of
  experiments that are feasible shortly after the initial
  commissioning of WASA at COSY, and will produce physics results
  soon. The medium term plans require either extended data taking or
  improvements of the experimental conditions depending on the
  experience gained in the initial experiments.  The order of the
  experiments listed below presents a reasonable balance between the
  physics interest and timeliness of the experimental feasibility.

\subsection{\boldmath Not-so-rare $\eta^\prime$ decays}
\label{sec:day1_eta}
\label{subsubetapday1}

\paragraph{Isospin symmetry breaking in $\eta^\prime$ decays.}
The experiment aims at a precise determination of the ratios of isospin 
symmetry breaking decays $\eta^\prime \rightarrow 3\,\pi$ with respect to 
the isospin allowed $\eta\,2\pi$ modes as defined in 
relation~(\ref{eq_pi_eta_ratio}) of section~\ref{subsubetap} to extract 
the mixing angle 
$\Theta_{\pi\eta}$, which is proportional to the light quark mass 
difference $\Delta\,\mbox{m}$ (eq.~\ref{eq_pi_eta_mix}).
While only an upper limit is known for $\eta^\prime \rightarrow 
\pi^0 \pi^+ \pi^-$~\cite{PDBook}, in the neutral decay modes in eq. 
(\ref{eq_pi_eta_ratio}) the ratio ${\cal R}_1$ has been measured with a 
statistics of $\approx 130$ events, leading to $\sin{\Theta_{\pi\eta}} = 
0.023 \pm 0.002$~\cite{Binon:1984fe}.
However, there is considerable uncertainty in the literature concerning 
$\sin{\Theta_{\pi\eta}}$:
a value of $\sin{\Theta_{\pi\eta}} = 0.010$ was extracted using PCAC 
relations for meson masses and removing electromagnetic 
contributions~\cite{Gross:1979ur}, in agreement with the estimate 
in~\cite{Gasser:1985gg}.
Further results given in the literature cover a range between 0.013 and 
0.014~\cite{Piekarewicz:1993ad,Chan:1994kg,Maltman:1994uu}, except for 
$\sin{\Theta_{\pi\eta}} = 0.034 \pm 0.013$ reported in~\cite{Bagchi:1990ah}.

With the WASA facility at COSY, in the tagging reaction $p p \rightarrow 
p p \eta^\prime$ a statistical accuracy of $1\,\%$ can be achieved for 
the lowest $ \sin{\Theta_{\pi\eta}}$ estimate within a running time of 12 
weeks, improving the precision of the existing experimental value by an 
order of magnitude (see Table~\ref{tab_etap_rates}).
Background influencing the precision of the extracted value is primarily 
expected from non--resonant $3\,\pi$ production.
Although the background will be distributed smoothly in the vicinity of an 
$\eta^\prime$ signal in the two proton missing mass, the 
final precision of the result will depend on the four--momentum resolution 
achieved for both protons in the forward detector and the decay system.
Modifications of the WASA detector setup to ensure a required performance 
at higher energies are discussed in section~\ref{subsubdetmodi}. 

With selective central detector trigger conditions on either six neutral 
particles from the decays $\eta(\pi^0) \rightarrow 2\gamma$ or two neutral 
and two charged particles, both ratios ${\cal R}_i$ can be determined 
simultaneously. 
Since the Dalitz plots of the hadronic decay channels are measured 
completely, the energy dependence of the decays is inherent to the data,
and can be used both to test predictions from ChPT, and as a constraint on
the the scalar meson sector as discussed in section~\ref{subsubetap}.

\begin{figure}[tb]
\parbox{0.49\textwidth}{\epsfig{file=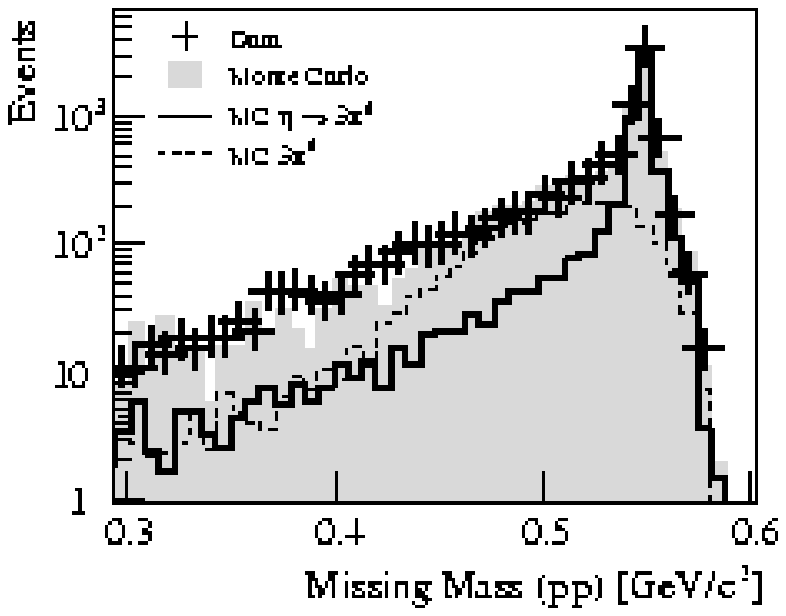,width=0.49\textwidth}}
\hfill
\parbox{0.49\textwidth}{\vskip-7ex\caption{\label{fig_etap_multipi}
  \small a) Measured $pp$ missing mass (crosses) and Monte Carlo event
  mixture (shaded area) of prompt $3\,\pi^0$ (dashed line) and resonant
  $\eta \rightarrow 3\,\pi^0$ production (solid line).
  b) Symmetrized experimental Dalitz plot for the decay $\eta \rightarrow
  3\,\pi^0$.
  c) Efficiency corrected, normalized radial density distribution of the
  Dalitz plot. Crosses denote data, the solid line corresponds to the fit
  of the slope parameter $\alpha$ (figures from~\cite{Pauly:2004me}).}}
\parbox{0.04\textwidth}{\raisebox{1ex}[0ex][0ex]{\mbox{}}}\hfill
\parbox{0.95\textwidth}{\raisebox{1ex}[0ex][0ex]{a)}}\hfill
\parbox{0.49\textwidth}
  {\epsfig{file=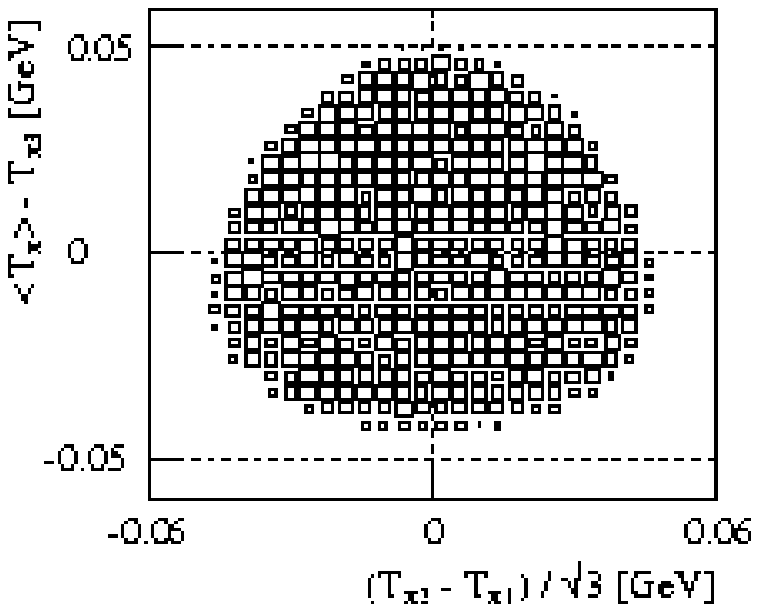,width=0.49\textwidth}}
\hfill
\parbox{0.49\textwidth}
 {\epsfig{file=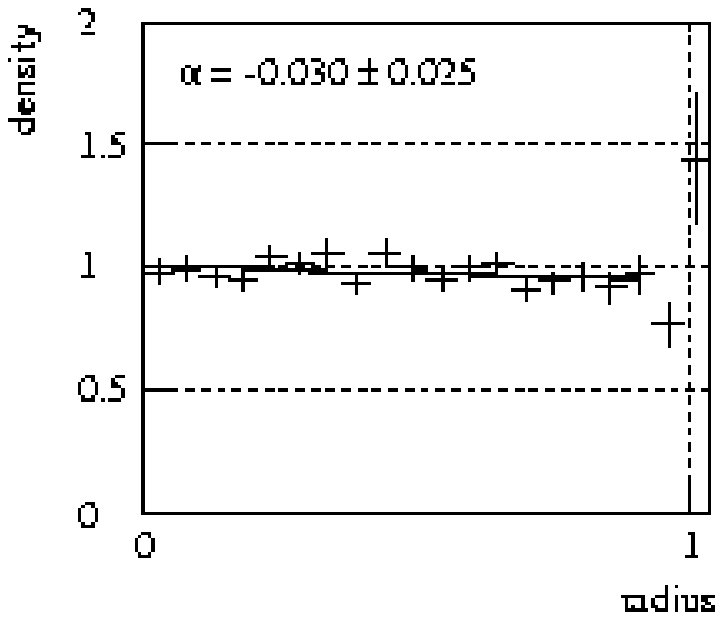,width=0.49\textwidth}}
\parbox{0.04\textwidth}{\raisebox{1ex}[0ex][0ex]{\mbox{}}}\hfill
\parbox{0.50\textwidth}{\raisebox{1ex}[0ex][0ex]{b)}}\hfill
\parbox{0.45\textwidth}{\raisebox{1ex}[0ex][0ex]{c)}}
\end{figure}

The present performance of the WASA detector for the detection of
multipionic final states is demonstrated in Fig.~\ref{fig_etap_multipi}
for a data sample obtained at CELSIUS at a kinetic beam energy of
$1360\,\mbox{MeV}$, i.e.\ at an excess energy of $\approx 40\,\mbox{MeV}$
with respect to the $\eta$ production threshold in proton--proton
scattering~\cite{Pauly:2004me}.

The experimental $pp$ missing mass spectrum is well reproduced by
contributions from $pp \rightarrow p p \eta \rightarrow pp 3\pi^0$ and
prompt $3\pi^0$ production (fig.~\ref{fig_etap_multipi}a).
The complete kinematical information is contained in the Dalitz plot for
the $\eta \rightarrow 3\,\pi^0$ decay, that is shown in its symmetrized
form ($<\mbox{T}_{\pi}> - \mbox{T}_{\pi\,3}$ vs.\ $\mbox{T}_{\pi\,2} -
\mbox{T}_{\pi\,1}/\sqrt{3}$ ) in Fig.~\ref{fig_etap_multipi}b.
The slope parameter $\alpha$ (fig.~\ref{fig_etap_multipi}c), which reflects
the strong and energy dependent $\pi \pi$ interaction, is extracted from
a linear fit of the normalized radial density
distribution~\cite{Nefkens:2002sa}.
The preliminary value $\alpha = -0.03 \pm 0.025$ is limited by the
statistics of 11700 events, but agrees in sign with the result
$\alpha = -0.031(4)$ of the Crystal Ball analysis based on
$10^6\,\mbox{events}$~\cite{Tippens:2001fm}.

\paragraph{Search for evidence of the box anomaly of QCD in $\eta^\prime 
 \rightarrow \pi^+ \pi^- \gamma$ decays.}

The experimental signature for the detection of the $\pi^+ \pi^- \gamma$ 
decay mode of the $\eta^\prime$ consists of two charged and one neutral 
particle in the central detector, with the $\eta^\prime$ being tagged from 
the missing mass with respect to the two protons identified in the forward 
detector.
This signature can be used for a selective trigger condition.
Moreover, with the requirement of one $\gamma$ in the central detector, 
the experiment trigger for a measurement of the hadronic decays 
$\eta^\prime \rightarrow \eta (\pi^0) \pi^+ \pi^-$ discussed above is a 
mere subset, i.e.\ both experiments could run simultaneously.

\begin{figure}[tb]
\begin{center}
\includegraphics[width=0.64\textwidth]{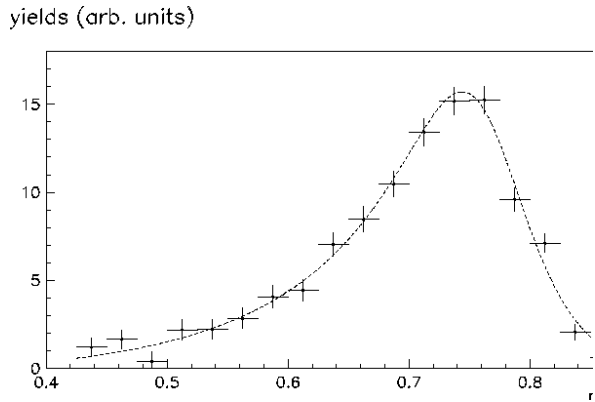}
\caption{\label{fig_cb_box} 
  Distribution of the $\pi^+ \pi^-$ invariant mass for $\eta^\prime
  \rightarrow \pi^+ \pi^- \gamma$ decays from~\cite{Abele:1997yi}.
  Crosses denote experimental data, the dashed line is the result of the
  fit used to extract box anomaly parameters with a statistical significance
  of $4\,\sigma$.}
\end{center}
\end{figure}

The experiment aims at measuring the $\pi^+ \pi^-$ invariant mass 
distribution in the decay $\eta^\prime \rightarrow \pi^+ \pi^- \gamma$ with 
high accuracy.
One (Two) order(s) of magnitude higher statistics compared to the presently 
available data can be achieved within a running time of two days (three 
weeks).
Following the experimental technique used in~\cite{Abele:1997yi} the number 
of $\eta^\prime \rightarrow \pi^+ \pi^- \gamma$ can be determined in each 
bin of the $\pi^+ \pi^-$ invariant mass separately.
Consequently, the $\pi^+ \pi^-$ spectrum will be background free 
(Fig.~\ref{fig_cb_box}).

The quantity $E_{\eta^\prime}$ associated with the matrix element of a 
non--resonant contribution as a consequence of the box anomaly is related 
at the chiral point to the parameters of the pseudoscalar nonet by one of 
the Chanowitz relations~\cite{Chanowitz:1975jm,Chanowitz:1980ma}:
\begin{equation}
\label{eq_chanowitz}
E_{\eta^\prime} \left(0\right) = 
  - \frac{e}{4\,\pi^2\,\sqrt{3}} \frac{1}{f_{\pi}^2} 
  \left[ 
    \frac{\sin{\Theta_{ps}}}{f_8} + \sqrt{2}\,\frac{\cos{\Theta_{ps}}}{f_0}
  \right] \, .
\end{equation}
$\Theta_{ps}$ denotes the pseudoscalar octet--singlet mixing angle, 
$f_{\pi}$ is the pion leptonic decay constant, and, in analogy, $f_0$ and 
$f_8$ are defined as couplings of the pseudoscalar singlet and octet 
states $\eta_0$ and $\eta_8$ to the divergences of the singlet and octet 
axial--vector currents, respectively.
$E_{\eta^\prime}$ can be derived from a fit to the spectrum of the 
$\pi^+ \pi^-$ mass $\mbox{m}_{\pi\pi}$ (Fig.~\ref{fig_cb_box} 
and~\cite{Benayoun:1993ty,Benayoun:1995mr,Abele:1997yi}) in the 
$\pi^+ \pi^- \gamma$ decay mode of the $\eta^\prime$ using
(for a more sophisticated analysis of this decay, see~\cite{Borasoy:2004qj},
which should be used in the final analysis)
\begin{equation}
\label{eq_pipi_box}
\frac{d\,\Gamma_{\eta^\prime}}{d\,\mbox{m}_{\pi\pi}} = 
  \frac{1}{48\,\pi^3} \left| 
    \frac{2\,\mbox{G}_{\rho}\left(\mbox{m}_{\pi\pi}\right)\,
          \mbox{g}_{\eta^\prime\rho\gamma}}
         {\mbox{D}_{\rho} \left(\mbox{m}_{\pi\pi}\right)} + 
  E_{\eta^\prime} \right| 
  \mbox{k}_{\gamma}^3\,\mbox{q}_{\pi}^3 \, ,
\end{equation}
where $\mbox{k}_{\gamma}$ and $\mbox{q}_{\pi}$ are the four--momenta of the 
outgoing $\gamma$ and $\pi^\pm$, $\mbox{g}_{\eta^\prime\rho\gamma}$ denotes 
the coupling constant for the $\rho$ contribution to the 
$\pi^+ \pi^- \gamma$ decay mode, and 
$\mbox{D}_{\rho} \left(\mbox{m}_{\pi\pi}\right)$ 
and 
$\mbox{G}_{\rho}\left(\mbox{m}_{\pi\pi}\right)$ 
are the $\rho$ propagator and its coupling to the $\pi^+ \pi^-$ channel.

On the one hand, the extraction of the box anomaly contribution is strongly 
dependent on the model employed in the description of the $\rho$.
On the other hand, from the two--photon widths of $\eta$ and $\eta^\prime$ 
as described by the Chanowitz 
equations~\cite{Chanowitz:1975jm,Chanowitz:1980ma} using additional 
information from radiative $J/\Psi \rightarrow \eta(\eta^\prime) \gamma$ 
decays (AFN relation~\cite{Novikov:1979uy,Akhoury:1987ed}), the parameters 
of the pseudoscalar nonet are completely determined, and can be used to 
predict the box anomaly constant $E_{\eta^\prime}$.
Consequently, the $\rho$ shape and the box anomaly parameters are closely
related to each other --- the approach might allow for an alternative approach
to extract the $\rho$ line shape in a clean way.

\begin{figure}[tb]
\begin{center}
\includegraphics[width=0.5\textwidth]{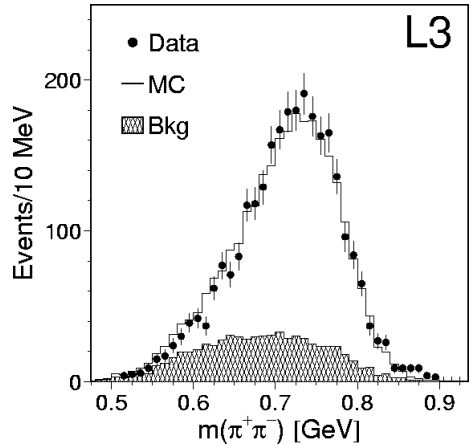}
\caption{\label{fig_l3_box}
  $\pi^+ \pi^-$ effective mass distribution in the decay
  $\eta^\prime \rightarrow \pi^+ \pi^- \gamma$ from
  L3~\cite{Acciarri:1998yx}. Points denote experimental data, the solid
  line is the best fit result with $\mbox{m}_{\rho} = 766\,\mbox{MeV}$ and
  $\Gamma_{\rho} = 150\,\mbox{MeV}$. The estimated background (shaded area)
  is mainly attributed to decays of the $a_2(1320)$.}
\end{center}
\end{figure}

So far, experimentally no consistent picture has been obtained:
The box anomaly contribution extracted from the Crystal Barrel
data~\cite{Abele:1997yi} is rather disfavoured (confidence level $3\,\%$)
by the L3 result (Fig.~\ref{fig_l3_box} and~\cite{Acciarri:1998yx}),
which is consistent with a pure $\rho$ line shape (C.L.\ $37\,\%$).
On the other hand, the Crystal Barrel data would require a mass
$\mbox{m}_\rho = 790\,\mbox{MeV}$ when neglecting the anomaly related
non--resonant contribution, i.e.\ $20\,\mbox{MeV}$ above the presently
accepted value~\cite{PDBook}.
In this respect, these two results with the up to now highest statistics
are inconsistent.

\subsection{\boldmath $a_0^+$(980) production in $pp{\to} d\pi^+\eta$}
\label{subsec:a0f0}

$a_0^+$-production has first been investigated at COSY by measuring
the reaction $pp \to dK^+ \bar{K}^0$ at a beam momentum of
$p=3.46$~GeV/c ($Q=46$ MeV) with ANKE~\cite{Kleber:2003kx}. The
measured total cross section for the $pp\to dK^+\bar{K}^0$ reaction is
$(38\pm 2\pm 14)$~nb and, thus, $\approx$1000 events could be collected
within five days of beam time using a cluster-jet target ($L= 2.7\cdot
10^{31}\, \mathrm{cm}^{-2} \mathrm{s}^{-1}$ during these
measurements).  According to the partial-wave analysis the $K^+ \bar
{K}^0$ pairs are mainly (about 83\%) produced in a relative $S$-wave,
i.e.\ via the $a_0^+$ channel~\cite{Kleber:2003kx}.  This observation
is in good agreement with a model prediction~\cite{Grishina:2004rd}
for the total $a_0^+$-production cross section in $pp$
collisions. This model has also been used to estimate the
cross-section ratio for resonant (via the
$a_0^+$)~\cite{Grishina:2001zj,Grishina:2000xp} and
non-resonant~\cite{Grishina:2001zj,pieta00} $\pi^+\eta$ production:
\begin{equation}
 R^{{\mathrm{res/nres}}}_{d} \left|_{p=3.46{\mathrm{GeV/c}}} 
                             \right. = 
     \frac{\sigma (pp \to d a_0^+)}{\sigma (pp \to d  \pi^+ \eta)} \approx 
                             0.3 \ldots 0.5.
 \label{ratio_d}
\end{equation}
This prediction for $R^{{\mathrm{res/nres}}}_{d}$ is in line with data
from ANKE for the reaction $pp\to d\pi^+X$~\cite{Fedorets}. Since most
of the non-resonant $\pi^+\eta$ pairs are expected at lower invariant
masses, we anticipate that the resonant signal can well be identified in
case of the $pp\to d\pi^+\eta$ reaction.

The model can also be used to estimate $R^{{\mathrm{res/nres}}}_{pn}$
for the $pp \to pn \pi^+ \eta $ reaction (see
Refs.~\cite{Bratkovskaya:2001ts,Kondratyuk:2002yf} for $a_0^+$- and
Ref.~\cite{pieta00} for non-resonant $\pi^+ \eta$ production):
\begin{equation}
 R^{{\mathrm{res/nres}}}_{pn} \left|_{p=3.46{\mathrm{GeV/c}}} 
       \right. = \frac{\sigma (pp \to pn a_0^+)}
                      {\sigma (pp \to pn \pi^+ \eta)} 
        \approx 0.015 \ldots 0.03.
 \label{ratio_pn}
\end{equation} 
Thus $R^{{\mathrm{res/nres}}}_{pn}$ is expected to be about one order
of magnitude smaller than the corresponding ratio
$R^{{\mathrm{res/nres}}}_{d}$ with deuteron formation in the final
state. Therefore, an experimental study of the $a_0^+$ in $pp$
reactions with WASA requires the identification of the $pp\to
d\pi^+\eta$ reaction, and the $pp \to pn
\pi^+ \eta$ reaction will be the most ``dangerous'' source of background. 
Consequently, the following simulation calculations focus on the
problem of deuteron-vs.-proton discrimination.

\begin{figure}[tb]
\begin{center}
 \resizebox{0.5\textwidth}{5.5cm}{\includegraphics{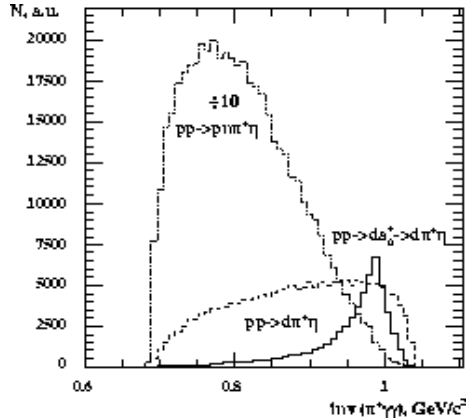}}
 \caption{Invariant mass $(\pi^+\eta)$ for $a_0^+$ production in
 $pp{\to}da_0^+{\to} d\pi^+\eta$ and for the two background processes,
 $pp \to d\pi^+\eta$ and $pp \to pn\pi^+\eta$ (downscaled by factor 10).}
 \label{fig:invmass}
\end{center}
\end{figure}

The reaction $pp{\to}da_0^+{\to} d\pi^+\eta$ with subsequent decay
$\eta{\to}\gamma\gamma$ can be measured with WASA by detecting
deuterons in the forward detector (FD) in coincidence with the $\pi^+$
and photons in the central detector (CD).  The reaction is identified
by reconstructing the masses $m(\eta)=m(\gamma\gamma)$ and
$m(d)=m.m.(\pi^+\gamma\gamma)$. In order to investigate the acceptance
and background suppression the simulations were performed for
$a_0^+$(980) production with a deuteron in the final state and both
background reactions using the cross section ratios
$\sigma(a_0^+){\colon}\sigma(d\pi^+\eta){\colon}\sigma(pn\pi^+\eta)=
1.1\,\mu\mathrm{b} {\colon} 3.5\,\mu\mathrm{b} {\colon}
96\,\mu\mathrm{b}$~\cite{Grishina:2000xp,Grishina:2001zj,pieta00}.
Fig.~\ref{fig:invmass} shows the initial $(\pi^+\eta)$ invariant
mass distributions for these reactions: a Flatte distribution for the
$a_0^+$(980)~\cite{Grishina:2000xp,Kondratyuk:2002yf}, a distribution
according to the model in
Refs.~\cite{pieta00,Kondratyuk:2002yf,Kudryavtsev:2003au} for the
non-resonant $d\pi^+\eta$, and a phase-space distribution for the
non-resonant $pn\pi^+\eta$ production.

The WASA acceptance for forward going particles is $\theta \approx
3^\circ \ldots 18^\circ$. This acceptance covers about $93\%$ of
deuterons from $pp{\to}da_0^+$, which are distributed in the range
$\theta = 0^\circ \ldots 17^\circ$.  The acceptance for pions and
photons is $\theta \approx 20^\circ \ldots 169^\circ$ (for the
simulations angles $\theta \approx 20^\circ \ldots 143^\circ$ ---
SEC and SEF, the central and forward part of the calorimeter --- were used). 
The result of the acceptance estimate is shown
in the Table~\ref{accep}. For the background reactions the acceptance
is smaller due to the wider angular distributions.  We have only
considered the decay channel $\eta{\to}\gamma\gamma$ since for the
other two main decay channels $\eta{\to}\pi^+\pi^-\pi^0$ and
$\eta{\to}3\pi^0$ the acceptance is two times smaller.

\begin{table}[h]
\begin{center}
\begin{tabular}{c|c} 
  Reaction &  Acceptance \\
           & ($\eta{\to}\gamma\gamma$ decay only)\\
\hline
$pp{\to}da_0^+{\to} d\pi^+\eta$ & 0.42 \\
$pp \to d\pi^+\eta$ & 0.34 \\
$pp \to pn\pi^+\eta$ & 0.32 \\
\end{tabular}
\caption{Simulated acceptance of WASA for $a_0^+$(980) production 
   and the two background processes. Deuterons/protons are detected in
   the FD ($\theta\approx 3^\circ \ldots 18^\circ$), pions and photons
   in the CD ($\theta \approx 20^\circ \ldots 143^\circ$).}
\label{accep}
\end{center}
\end{table}

In the simulations the $\pi^+$ detected in the CD have angular resolution
better than $\sigma(\theta)= 0.4^\circ$ (for angles larger than
30$^\circ$) and momentum resolution between $\sigma(p)/p=
3{\ldots}6\%$.  
For photons in the CD the resolution is
$\approx 5^\circ$(FWHM), due to the crystal sizes in the SEC.  The energy
resolution varies from 10~MeV (FWHM) for large angles and low initial
energies up to 40~MeV (FWHM) for angles less than 30$^\circ$ and
initial energies close to 1~GeV.  Such resolutions yield
$\gamma\gamma$ invariant mass distributions with a FWHM of
43~MeV/c$^2$ in the $\eta$ mass region, which is in agreement with 
WASA at CELSIUS data~\cite{Koch}.

\begin{figure}[tb]
\begin{center}
\resizebox{1.0\textwidth}{!}{\includegraphics{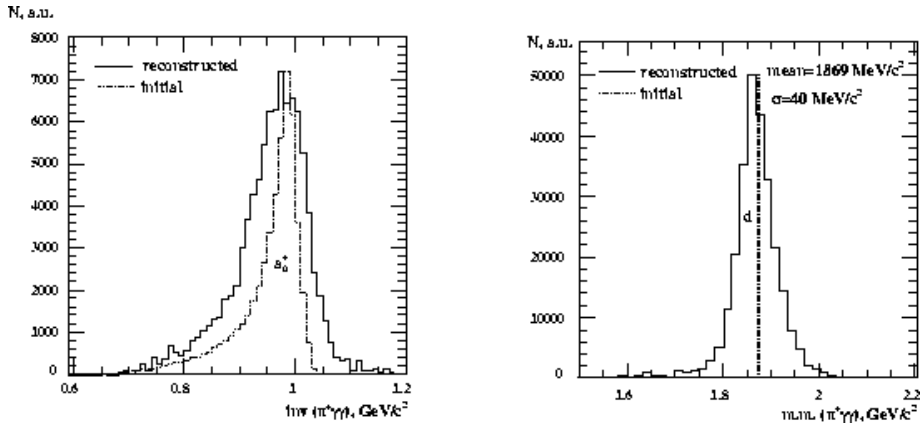}}
 \caption{Initial (dotted line) and reconstructed (solid) invariant
 mass $(\pi^+ \gamma\gamma)$ (left) and missing mass
 $(\pi^+\gamma\gamma)$ (right) for the reaction $pp{\to}da_0^+{\to}
 d\pi^+\eta$.}  
\label{fig:rec_init}
\end{center}
\end{figure}

\begin{figure}[tb]
\begin{center}
\includegraphics[width=\textwidth]{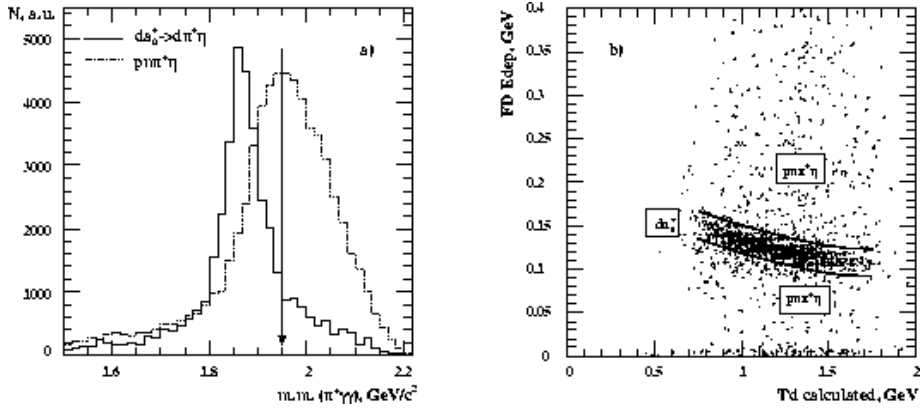}
 \caption{Reconstructed missing mass $(\pi^+\gamma\gamma)$ for the
 reactions $pp{\to}da_0^+{\to} d\pi^+\eta$ and $pp \to pn\pi^+\eta$
 (left).  The cut is indicated by the arrow.  Energy losses in the FD
 vs.\ kinetic energy of the deuterons/protons (calculated from the
 detected $\pi^+$ and the two photons assuming that the forward
 particle is a deuteron).  The cut is indicated by the two lines.}
 \label{fig:mx_td}
\end{center}
\end{figure}

Fig.~\ref{fig:rec_init} shows the initial and reconstructed
invariant mass $(\pi^+ \gamma\gamma)$ and missing mass
$(\pi^+\gamma\gamma)$  for the reaction $pp{\to}da_0^+{\to}
d\pi^+\eta{\to} d\pi^+\gamma\gamma$.  Pions and photons detected in
coincidence in CD were used for the reconstruction, which provide a
$(\pi^+\gamma\gamma)$ missing mass resolution of $\sigma=40$
MeV/c$^2$.

The expected background from the reaction $pp \to pn\pi^+\eta$ is two
order higher than the $a^+_0$ signal. Due to the large momenta of the
protons and deuterons they cannot be stopped in the FD and their
initial kinetic energy cannot be reconstructed.  Moreover, their energy
losses are close to minimum ionizing and the standard WASA at CELSIUS
$\Delta{E}/E$ method cannot be used for $p/d$ discrimination.

\begin{figure}[tb]
\begin{center}
\includegraphics[width=\textwidth]{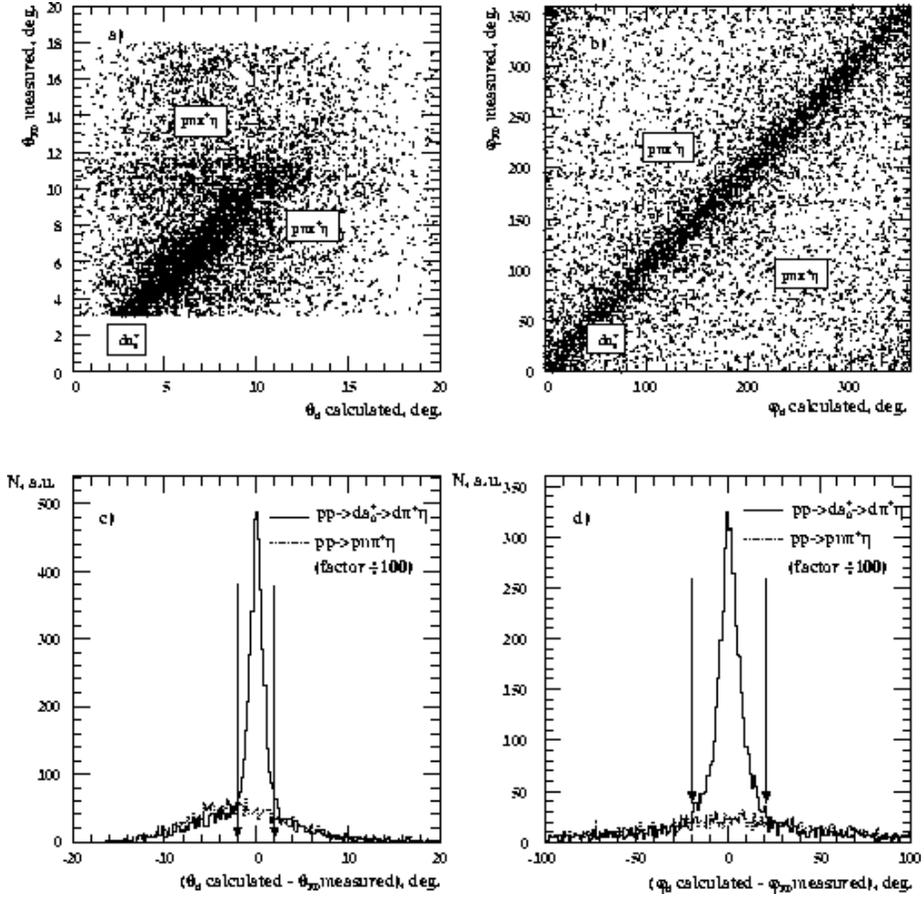}
  \caption{(a,b) Measured vs.\ calculated forward azimuthal and polar
  angles assuming that the forward particle is a deuteron.  (c,d)
  Difference between measured and calculated angles of the forward
  particles.  The background $pn\pi^+\eta$ events (dotted line) is
  downscaled by a factor 100. The cuts are indicated by the arrows.
\label{fig:angles}} 
\end{center}
\end{figure}

In order to suppress the proton background a set of criteria has been
applied.  The first cut was applied on the reconstructed
$(\pi^+\gamma\gamma)$ missing mass, see Fig.~\ref{fig:mx_td}. The cut
at 1.95~GeV suppresses protons by a factor $\approx 1.6$. The
two-dimensional distribution energy loss vs.\ kinetic energy of the
forward particle shows a correlation for $a_0^+$ events, see
Fig.~\ref{fig:mx_td}. Applying a gate with ${\pm}20$~MeV around these
events, background protons can be suppressed by a factor of $\approx
3.9$.

Another strong criterion for proton suppression is the difference
between the measured azimuthal and polar angles of the forward
particles and the expected deuteron angles calculated from the $\pi^+$
and two photons.  For the $a_0^+$ events these angles coincide within
the resolutions of the detector and the reconstruction procedure. For
protons from the $pn\pi^+\eta$ events the correlation is much
weaker. This is clearly seen in two dimensional plots of measured vs.\
calculated angles. If one assumes that all forward particles are
deuterons, then the real deuterons are seen as lines whereas the
protons are smeared out (Fig.~\ref{fig:angles} (a,b)).  Cuts were applied
to the difference between measured and calculated azimuthal and polar
angles of forward particles (Fig.~\ref{fig:angles} (c,d)).  An
azimuthal angle cut of ${\pm}20^\circ$ and a polar angle cut of
${\pm}2^\circ$ suppresses protons by a factor $\approx 21$.

Taking into account all mentioned cuts and the difference in
acceptances the proton suppression factor is
$1.3{\times}1.6{\times}3.9{\times}21{\approx}170$. We therefore expect
that the non-resonant $pn\pi^+\eta$ events can sufficiently be
suppressed without modifications of the existing FD for the higher
particle momenta at COSY.

Assuming a luminosity of $L=10^{31}\,\mathrm{cm}^{-2}\mathrm{s}^{-1}$, 
an overall efficiency due to detector acceptance, reconstruction algorithms 
and $d/p$ cuts of $\approx 0.07$, the effect of dead-time and detector 
efficiencies of $\approx 0.5$ and using the cross section estimate from
Refs.~\cite{Grishina:2001zj,Grishina:2000xp},
$\sigma(pp{\to}da_0^+{\to}d\pi^+\eta){\times}BR(\eta\to2\gamma)=
1.1\,\mu\mathrm{b} {\times} 0.393$, the final count rate is
$\approx 0.15\,\mathrm{s}^{-1} \approx 90000\,\mathrm{week}^{-1}$.

\subsection{\boldmath Pentaquarks}
\label{sec:day2_pentaquark}

At present, several experimental groups, including TOF at COSY, are
performing high statistics experiments to confirm the existence of the 
$\Theta^{+}$.

Provided the existence of the $\Theta^+$ is confirmed, we
will be interested in basic properties of this hadron and
details of the production mechanism. 

In the following we concentrate on the $pd\rightarrow p\Lambda \Theta ^{+}$
reaction as one example to illustrate the feasibility of high statistics
experiments involving both neutral and charged decay products.

Out of the two decay modes of the $\Theta^{+}$, the $K^{0}p$ channel
gives the best experimental conditions for being measured in the WASA
detector, so we shall here concentrate on this decay branch.

For the $\Lambda$ there are two important decay modes, $p\pi^{-}$
and $n\pi^{0}$ with branching ratios of 63.9\% and 35.8\%,
respectively. The choice of using the charged decay mode to identify
the $\Lambda$ is advantageous due to the following reasons:

\begin{itemize}
\item the energy resolution of WASA for charged particles is 
   generally better than for photons,
\item the combination of the forward tracker and the MDC provides a 
   clean identification of the $\Lambda$ by reconstruction of its
   decay vertex,
\item both decay products are measured,
\item higher branching ratios as compared to neutral decay modes.
\end{itemize}

Regarding the kaon decay, half of the events are lost due to
$K_{L}^{0}$ having $c\tau=15.5$~m and thus escaping detection. For the
$K_{S}^{0}$ two modes are important, $\pi^{+}\pi^{-}$ ($BR = 68.6$\%)
and $\pi^{0}\pi^{0}$ ($BR = 31.4$\%). Selecting the charged mode has the
advantage of higher rate. However, it leads to a final state with six
charged particles. It might be difficult to distinguish between the
$\pi^{-}$ produced in the $\Lambda$ decay and the $K_{S}^{0}$ decay,
which could impair the reconstruction of the $\Lambda$.  Therefore,
the neutral decay mode will be used.

Taking all branching ratios together gives the overall branching
ratio, or efficiency of detecting $\Theta^{+}$ using this particular
final state, of 0.5$\times$0.639$\times$0.314 = 0.1.
This final state with four charged particles, $\pi^{-}$ and three
protons, and the four photons from $\pi^{0}\pi^{0}$ provides a well
defined trigger as well as favorable analysis conditions.

A rough estimate of the WASA acceptance for the $pd\rightarrow
p+\Lambda+\Theta^{+}\rightarrow p+p\pi^{-}+p\gamma\gamma\gamma\gamma$
reaction has been obtained using phase space Monte Carlo
simulation. The first things needed are the energies and angles of all
charged final state particles. They are shown in
Fig.~\ref{fig:pentaquark_f2} below. The proton produced directly is
denoted $p$ while those produced in the decays of the $\Lambda$ and
$\Theta^{+}$ are denoted $p_{\Lambda}$ and $p_{\Theta}$,
respectively. It is seen that the maximum energies for $p$ and
$p_{\Lambda}$ are below 300~MeV, which is the highest energy deposited
by protons in the FD. For such protons one may expect a very good
energy resolution.

\begin{figure}[tb]
\begin{center}
  \resizebox{10cm}{!}{\includegraphics[scale=1]{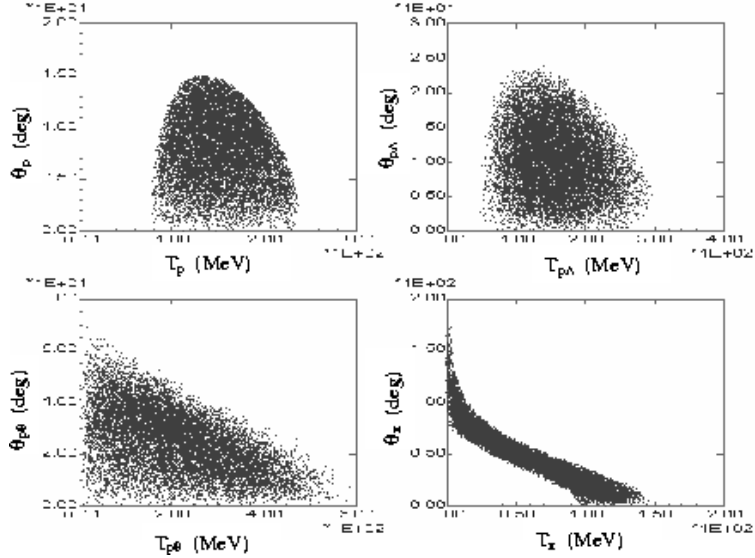}}
  \caption{Energy and angular ranges of the
    four final state charged particles in the reaction $pd\rightarrow
    p+\Lambda+\Theta^{+}\rightarrow p+p\pi^{-}+p\gamma\gamma\gamma\gamma$ at
    1360 MeV proton beam energy.}
  \label{fig:pentaquark_f2}
\end{center}
\end{figure}

Assuming that reconstruction of the $\Lambda$ requires the
$p_{\Lambda}$ proton to be detected in the FD
($2.5^{\circ}<\theta_{p\Lambda}<18^{\circ}$) and the $\pi^{-}$ either
in the FD or MDC ($22^{\circ}<\theta_{\pi}<158^{\circ}$) one finds
that this requirement is fulfilled in $\approx 85$\% of cases. This is
roughly the efficiency of the $\Lambda$ reconstruction,
$\epsilon_{\Lambda}$. It should be noted that the resolution of the
MDC is limited at the forward and backward angles due to a reduced
number of wire planes being crossed by the particles. All 17 layers
are crossed only for angles between 44$^{0}$ and 134$^{0}$. For the
$\pi^{-}$ going into the FD or into this limited angular range of the
MDC, the $\epsilon_{\Lambda}$ is reduced to 55\%.

The requirement that the proton $p_{\Theta}$ from $\Theta^{+}$ enters
the FD or the plastic barrel is fulfilled in $\epsilon_{p\Theta}=87$\%
of cases.

Finally, the requirement of the four photons (from the $K_{S}^{0}$
decay into $\pi^{0}\pi^{0}$) to fall into the angular range of the
electromagnetic calorimeter is fulfilled with an efficiency
$\epsilon_{4\gamma}=62$\%. This gives the overall acceptance of $\approx
45$\% for $\epsilon_{\Lambda}=85$\% and 29\% for
$\epsilon_{\Lambda}=55$\%.

Remembering that the resolution of the WASA detector is far better for
charged particles than for photons one would like the $\Theta^{+}$
mass to be calculated as the missing mass to the reconstructed $p$,
$p_{\Lambda}$ and $\pi^{-}$ (for $\pi^{-}$ only angles will be used,
see below). In such a case one has to take into account losses of
protons due to nuclear interactions with the detector material, which
is expected to reduce the acceptance by roughly a factor of two.

As already mentioned, the $\Theta^{+}$ mass will be found as the
missing mass calculated for the $p$ and $p_{\Lambda}$ reconstructed in
the FD and $\pi^{-}$ either in the FD or MDC. Measurement of the
energy and angle of the $p$ proton is straightforward since the
reaction vertex is at the known target position.  This is not the case
for the $p_{\Lambda}$ because of the $\Lambda$'s $c\tau $=7.89 cm. In
order to reconstruct the $\Lambda$ momentum one has to find the
$\Lambda$ decay vertex using either the forward tracker, when both
$p_{\Lambda}$ and $\pi^{-}$ enter the FD, or the forward tracker and
the MDC.  Reconstruction of the vertex requires finding two tracks
which, together with the target point, are in the same plane. When two
tracks fulfilling the above criteria are found, one can calculate the
absolute value of the $\pi^{-}$ momentum, $k$, from the condition:

$(\sqrt{m_{\pi}^{2}+k^{2}}+E)^{2}-(k\hat{u}+\vec{p})^{2}=m_{\Lambda}^{2}$,

\noindent where $E$, $\vec{p}$ represent the proton total energy and
momentum,\ $\hat{u}$\ is the unit vector representing direction of the
presumed $\pi^{-}$ and $m_{\Lambda}$\ is the $\Lambda$ mass. One
should verify that the reconstructed $\Lambda$-momentum points to the
target area.

\begin{figure}[tb]
\begin{center}
  \resizebox{10cm}{!}{\includegraphics[scale=1]{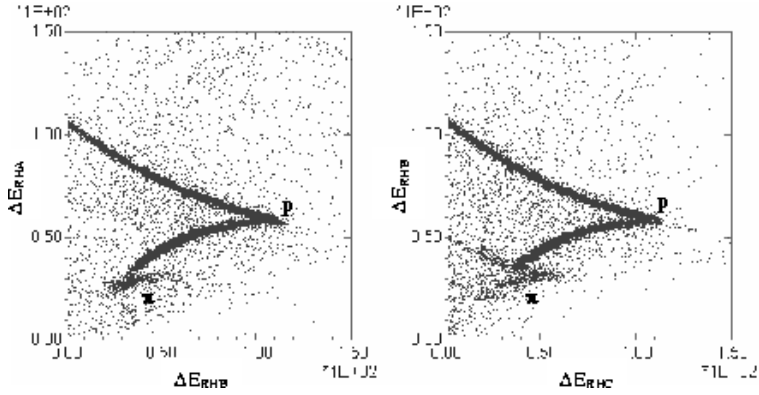}}
  \caption{$\Delta E$-$E$ plots for three consecutive layers of the
  range hodoscope in the FD as obtained in the M-C simulation of the
  reaction $pd\rightarrow p+\Lambda+\Theta^{+}\rightarrow
  p+p\pi^{-}+p\gamma\gamma\gamma\gamma$ at 1360 MeV proton beam
  energy.}
\label{fig:pentaquark_f3}
\end{center}
\end{figure}

A further check is provided by the $\Delta E$-$E$ method when the pion
enters the FD (see Fig.~\ref{fig:pentaquark_f3}) and extracting the
momentum from the measurement of the track curvature in the magnetic
field, when the pion enters the MDC.

\begin{figure}[tb]
\begin{center}
  \includegraphics[width=0.7\textwidth]{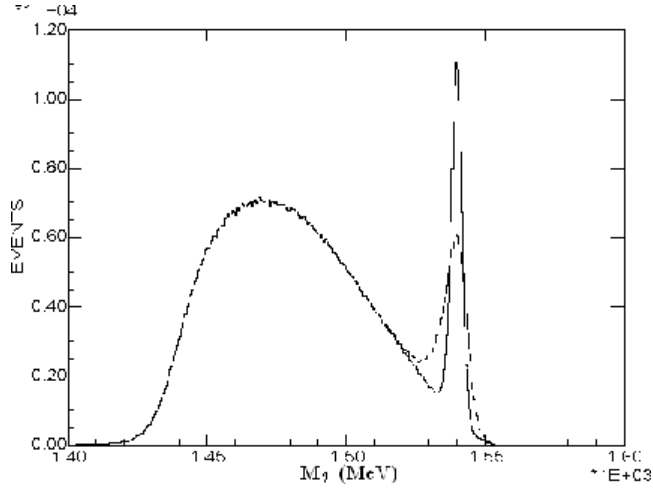}
  \caption{Phase space Monte Carlo simulation of the reaction
  $p+d\rightarrow p+\Lambda+\Theta^{+}$ at 1360 MeV beam energy.  The
  $\Theta^{+}$ peaks are seen on a wide background from non-resonant
  $p+d\rightarrow p+\Lambda +K^{0}+p$ reaction, which has been assumed
  to be 10 stronger than the $\Theta^{+}$ production. For both peaks
  the energy resolution for protons has been assumed to be equal to
  3\% and the $\pi^{-}$ angular resolution to amount to
  $3^{\circ}$. The narrower peak, FWHM=4.6 MeV/c$^2$, corresponds to the
  proton angular resolution of $1^{\circ}$ and for the wider one the
  resolution of $3^{\circ}$ has been assumed. The width obtained in
  this case amounts to 9.6~MeV/c$^2$.}
\label{fig:pentaquark_f4}
\end{center}
\end{figure}

At present the position resolution of the forward tracker amounts to
$\approx 2$~cm in the $x$-$y$ plane (perpendicular to the beam)
\cite{Dyring} so one can estimate the resolution in angle to be of the
order of $\arctan(2/140)\approx 1^{\circ}$, where 140~cm is the
distance between the target position and the FD tracker. For the
proton $p$, the resolution in angle is given by the size of the
beam-target overlap and accuracy of the FD tracker. For the $\pi^{-}$,
which in most cases enters the MDC, one can expect the resolution in
$\theta$ to be of the order of $\arctan(1/22) = 2.6^{\circ}$.

A rough estimate of the resolution in $M_{\Theta^{+}}$ can be obtained
by smearing the energies of the $p$, $p_{\Lambda}$ and angles of the
$p$, $p_{\Lambda}$ and $\pi^{-}$ with the experimental
resolutions. Taking for the energy resolution 3\%, and $1.0^{\circ}$,
and $3.0^{\circ}$ for the angular resolution for protons and $\pi^{-}$
respectively, leads to the results shown in the
Fig.~\ref{fig:pentaquark_f4} (solid line). The width of the peak at
half maximum amounts to 4.5~MeV/c$^2$, which is nearly five times better
than the resolution quoted in Ref.~\cite{Nakano:2003qx}.  This is a
valuable feature which could improve our knowledge of the true width
of the $\Theta^{+}$ resonance expected to be much smaller than the
20~MeV/c$^2$ measured in the experiments so far. The wide background seen in
the figure corresponds to the phase space simulation of the
non-resonant reaction $p+d\rightarrow p+\Lambda+K^{0}+p$.

For the $\Theta^{+}$ cross section of 30 nb, the beam-target
luminosity of 10$^{31}\,$/cm$^{-2}$s$^{-1}$ and the acceptance of 1.5\% (with
the overall branching ratio of 10\% included) one would expect the
$\Theta^{+}$ event rate of $30 \times 10^{-33}\times 10 ^{31}\times
0.015 \mathrm{s}^{-1} = 0.0045\mathrm{s}^{-1}$ at a beam energy of
1360 MeV. Simulation shows that the detector acceptance depends rather
weakly on the beam energy. Thus increasing the beam energy to 1450 MeV
would increase the event rate by a factor of $\approx$7 due to the change
of the cross section~\cite{Goeran}. With a duty factor of 50\%
obtaining 10$^{4}$ events would require less than 200 hours.

\subsection{\boldmath Isospin violation in $\vec{d}d\,\to\,\alpha\pi^0$}
\label{sec:mediumterm_dd2alphapi0}

Due to the small cross section (a few pb) and the high background from
the reaction $dd \to \alpha\mathrm{X}$ the detection and
unambiguous identification of the $\alpha\pi^0$ channel is rather
difficult. Hence, first attempts to measure this reaction have
produced only upper limits (see Ref.~\cite{Banaigs:1987mx}). The only
positive report at a deuteron energy of 1.1~GeV~\cite{Goldzahl:1991ge}
has been questioned, because the background from the double radiative
capture process $dd \to \alpha \gamma\gamma$ could have been
misinterpreted~\cite{Dobrokhotov:1999cs}. Consequently, in the recent
experiment performed at IUCF~\cite{Stephenson:2003dv} the $\alpha$ and
the two $\gamma$'s have been measured in coincidence to provide a
clean signal of $dd \to \alpha\pi^0$ channel.

At COSY, studies of this reaction were initially proposed for 
BIG KARL~\cite{MAG97}. However, a sufficient background 
suppression without photon detection would have been hard to
achieve. The idea was resubmitted within a proposal for an
electromagnetic calorimeter at ANKE~\cite{PD00}. This combination
--- like WASA --- would have provided a clean identification of the 
forward-going 
$\alpha$ particle and the $\pi^0$ decaying into two 
photons.

\begin{figure}[tb]
\begin{center}
\resizebox{1.0\textwidth}{!}{\includegraphics{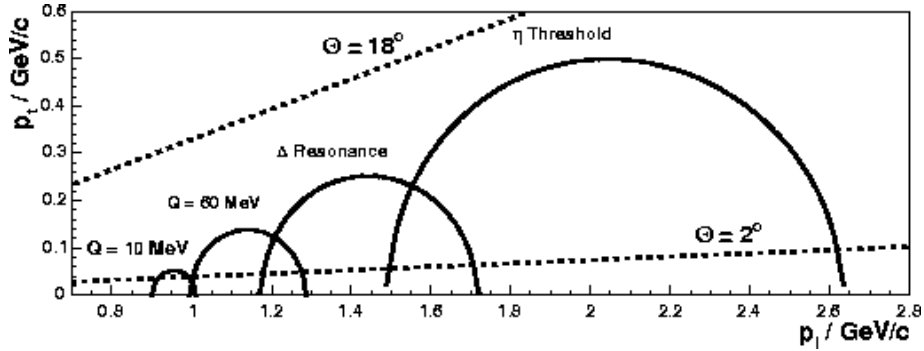}}
\caption{Kinematics of the reaction $\mathrm{dd}\to\alpha\pi^0$ for
    various $Q$ values. The transverse momentum of the $\alpha$ versus
    its longitudinal momentum in laboratory is plotted.  The lines
    indicate the azimuthal angle covered by WASA.
\label{fig:alphaacc}}
\end{center}
\end{figure}

Compared with the latter proposal, the WASA detector provides a
similar acceptance for photons ($\Theta = 20^\circ \ldots 169^\circ$)
but a significantly wider angular acceptance for $\alpha$ particles
($\Theta \approx 3^\circ \ldots 18^\circ$). This will allow the
extraction of angular distributions starting from $Q\approx60$~MeV up
to and beyond the $\eta$ threshold (see Fig.~\ref{fig:alphaacc}). In
addition, the fully symmetric WASA detector covers the entire azimuthal 
angular range of 2$\pi$ and, thus, is well suited to measure 
polarization observables.
The simulations are done for the measurement at $Q=60$~MeV (studying
the development of $p$-waves).

As indicated previously, the reaction $dd \to\alpha\pi^0$ will
be identified by detecting the $\alpha$ and the decay
$\pi^0\to\gamma\gamma$ in coincidence. Here, the high energy loss of
the $\alpha$ in combination with two neutral hits in the calorimeter
provides a very efficient first level trigger. For $\alpha$ momenta
above 1~GeV/c all particles pass the (currently used) Forward Window
Counter (FWC) of WASA and are stopped in one of the layers of the
Forward Trigger Hodoscope (FTH). Subsequent layers (e.g. of the
Forward Range Hodoscope, FRH) can be used as veto layers for
particles, which are not stopped. Angles are reconstructed by means of
the Forward Proportional Chambers (FPC).

\begin{figure}[tb]
\begin{center}
\resizebox{1.0\textwidth}{5.5cm}{\includegraphics{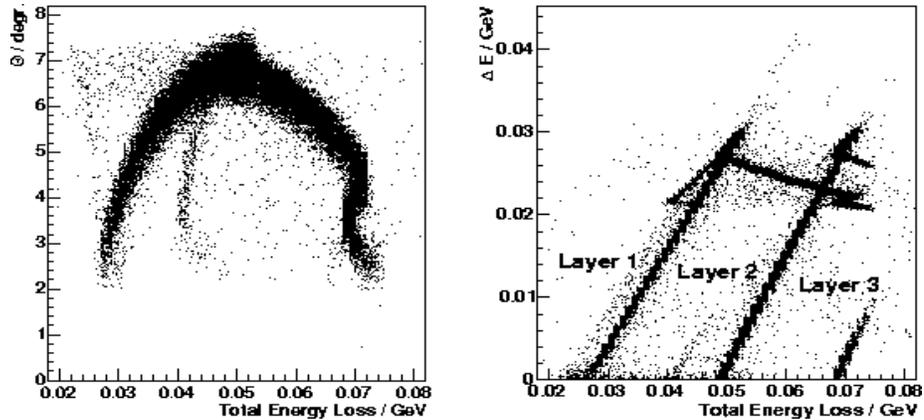}}
\caption{Energy losses of $\alpha$ particles from the reaction 
     $dd\to\alpha\pi^0$.  Left: Reconstructed angle versus
     total energy loss. Right: Energy loss in the three single layers
     of the Forward Tracking Hodoscope versus total energy loss.
\label{fig:e_loss}} 
\end{center}
\end{figure}

\begin{figure}[tb]
\begin{center}
\resizebox{1.0\textwidth}{5.5cm}{\includegraphics{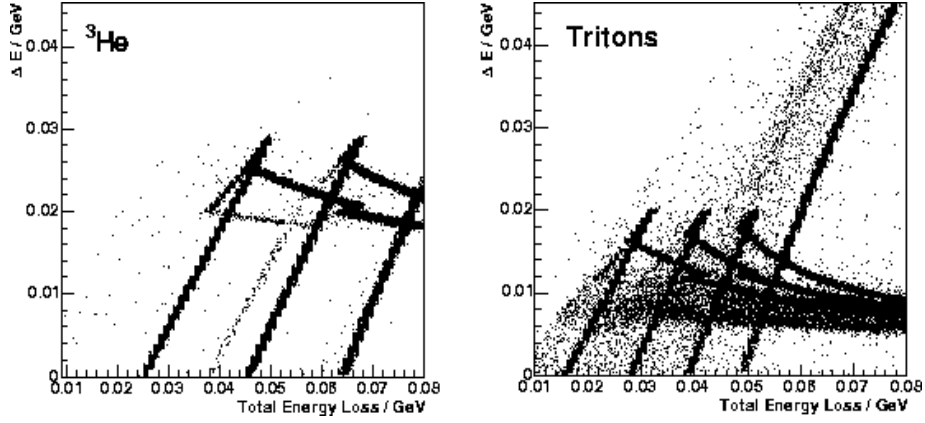}}
\caption{Energy loss in the individual layers versus total energy loss.
    On the left $^3$He, on the right tritons. While the energy loss
    pattern for $^3$He is rather similar to $^4$He, the one for
    tritons can be used efficiently for particle discrimination.}
\label{fig:e_loss_other}
\end{center}
\end{figure}

Fig.~\ref{fig:e_loss} (right side) and Fig.~\ref{fig:e_loss_other}
show the energy losses in the individual layers of the tracking
hodoscope for $\alpha$ particles as well as for $^3$He and
tritons. While tritons can already be discriminated by these energy
loss patterns, $^3$He is quite similar to $\alpha$
particles\footnote{Protons and deuterons are not shown. Their energy
loss is --- compared with tritons --- again smaller and the bands for
stopped particles are shifted further to lower energy
losses.}. However, a further reduction of the $^3$He content can
already be done by requiring the correct kinematics as shown in
Fig.~\ref{fig:e_loss} (left side).

\begin{figure}[tb]
\begin{center}
\resizebox{1.0\textwidth}{5.5cm}{\includegraphics{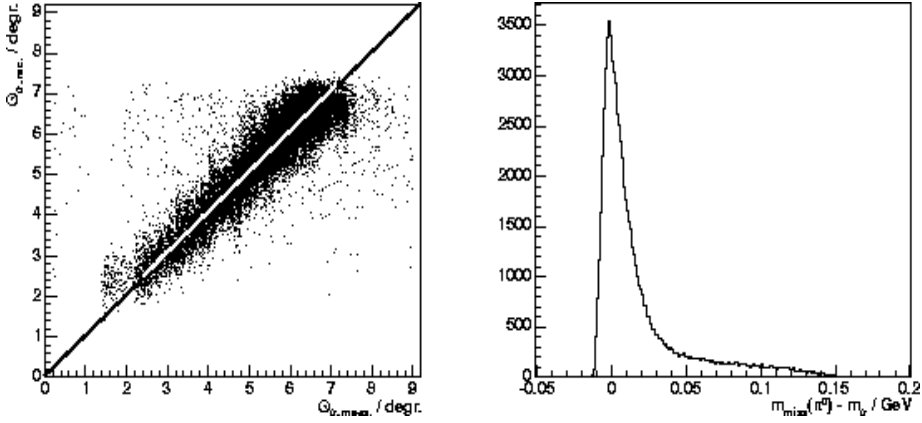}}
\caption{$\alpha$ kinematics reconstructed from measured $\pi^0$.
    Left: Reconstructed polar angle versus measured polar
    angle. Right: Reconstructed $\alpha$ missing mass.
\label{fig:pions}} 
\end{center}
\end{figure}

Together with the measured $\pi^0$, the
reaction is kinematically over-constrained. Tests on the energy and
momentum conservation prevents further particles from being missing.
While the standard procedure would be an overall kinematic
fit, Fig.~\ref{fig:pions} shows --- as an example --- the correlation
between the measured polar angle of the $\alpha$ and the one
reconstructed from the pion kinematics as well as the missing mass
with only the pion detected.

Taking into account the current geometry of WASA, the combined
acceptance for $\alpha$ and $\pi^0$ does not vary much and will be
$\epsilon_\mathrm{acc}\approx0.5$ for all beam momenta from 1.2~GeV/c
up to $\eta$ threshold. For the following estimates an additional
factor of 0.5 is considered representing dead-time corrections,
detector- and analysis efficiencies.

Assuming a luminosity of $L = 10^{32}\,\mathrm{cm}^{-2}\mathrm{s}^{-1}
= 10^{-4}\,\mathrm{s}^{-1}\mathrm{pb}^{-1}$ the final count rate will
be 15 events per week and pb. Since the goal of this experimental program is
to extract differential cross sections, the total yield has to be at
least a few hundred events for each beam energy. Using a total cross
section of $\sigma\approx75\mathrm{pb}$ at $Q=60$~MeV (i.e.\ scaling
the IUCF result by $s$-wave phase space) leads to beam times of about
1 or 2 weeks per measurement. Consequently, although a lower
luminosity would still allow to measure one data point within a longer
(but still reasonable) beam time, it would hardly be possible to carry
out the full experimental program.  This program
would cover measurements in two different energy ranges, namely at
$Q\approx60$~MeV and in the $\Delta$ region ($Q\approx160$~MeV).
However, initially one data
point at the lowest beam energy should be measured. It is aimed to run
with polarized beam to disentangle the contribution of different
partial waves (especially $p$-waves) as discussed in 
section~\ref{sec:dd2alphapi0}. Further runs will then depend
on the analysis of these data.

As shown above the reaction $dd\to\alpha\pi^0$ can be
measured with the existing WASA setup without any modifications on the
detector system itself. The electronics has to be upgraded: the
trigger logic has to be adopted in order to use an efficient energy loss
trigger and the data acquisition systems has to be replaced to be able
to run at high luminosities (see section~\ref{subsubdaq}).

\subsection{\boldmath $a_0^0$-$f_0$ mixing in $pn{\to} d\pi^0\eta$ and
               $dd{\to}\alpha \pi^0\eta$}
\label{subsuba0f0mixing}

At a later stage of the experimental program with WASA at COSY the
reactions $pn\to d\pi^0\eta$ and $dd\to \alpha\pi^0\eta$ will be
measured.  The main challenges for the identification of the isospin
violating effects in these reactions will be:
\begin{description}
\item[\boldmath $pn\to d\pi^0\eta$:] The measurement of this reaction 
  is similar to the $pp$ experiment described in Section~\ref{subsec:a0f0}. However, an
  additional complication comes from the fact that deuterium has to
  be used as an effective neutron target. The
   signal for isospin violation in this reaction would be provided from the 
   measurement of
   the angular forward-backward asymmetry of the $(\pi^0\eta)_{l=0}$
   system.
\item[\boldmath $dd\to \alpha\pi^0\eta$:] This reaction is only 
  possible due to isospin violation and the yield will be rather
  small. First estimates show that the cross section should be of the order of
  100~pb~\cite{dd_proposal}. A better prediction can be obtained from
  data on the reaction $dd{\to}\alpha K^+K^-$ which are expected from
  ANKE in winter 2004/05.  It is
  thus suggested to perform these measurements (at lower beam
  energies, see Sect.~\ref{sec:mediumterm_dd2alphapi0}) before the
  $dd\to \alpha\pi^0\eta$ reaction will be studied at maximum COSY
  momentum ($\approx 3.7$ GeV/c).
\end{description}

\subsection{\boldmath Study of hyperon resonances}
\label{sec:mediumterm_lambda1405}
In order to understand the nature of the $\Lambda$(1405) hyperon
resonance it is proposed to study $\Lambda$(1405) production in
proton-proton collisions, and in a second step also in proton-nucleus
collisions.  The primary goal of this study is to measure the
$\Lambda$(1405) production cross section and, in particular, its
spectral shape.  As a by-product also information on the
$\Sigma$(1385) resonance is obtained.

Due to its large acceptance for both charged particles and photons,
the WASA detector allows to investigate all isospin combinations of
$\Sigma\pi$ and $\Lambda\pi$ that are populated in hyperon resonance
decay subsequent to the production reactions
$pp\rightarrow\Lambda(1405)K^+p$ and
$pp\rightarrow{}[\Sigma(1385)K]^+p$.  With the use of a deuterium
target the implementation of a proton spectator detector close to the
interaction point would be required, in order to allow the full
measurement of all particles in the final state (equivalent to $pp$
collisions) also in $pn\rightarrow\Lambda(1405)K^0p$ and
$pn\rightarrow{}[\Sigma(1385)K]^0p$ reactions.  Note that the latter
reaction also allows to excite the $\Sigma^-$(1385) state.

The photon detection capability of WASA matches particularly well to
the study of the $pp\rightarrow\Lambda(1405)K^+p$ reaction with the
decay channel $\Lambda(1405)\rightarrow\Sigma^0\pi^0\rightarrow{}
(\Lambda\gamma{})(\gamma\gamma{})$ which is not populated in
$\Sigma^0$(1385) decay, and thus should reflect the $\Lambda$(1405)
spectral distribution in an undisturbed way.  Therefore special
emphasis should be given to this channel, but as discussed in
Sect.~\ref{sec:hyperons} also the $\Sigma^+\pi^-$ and $\Sigma^-\pi^+$
channels deserve investigation in order to extract a possible
isovector contribution.  For the observed final state in the decay
channel $\Lambda(1405)\rightarrow{}\Sigma^0\pi^0\rightarrow{}
(\Lambda\gamma{})(\gamma\gamma{})\rightarrow{}(p\pi^-\gamma{})
(\gamma\gamma{})$ the expected branching ratio is rather large, and
amounts to 21\%.

\begin{figure}[tb]
\begin{center}
\resizebox{10cm}{!}{\includegraphics{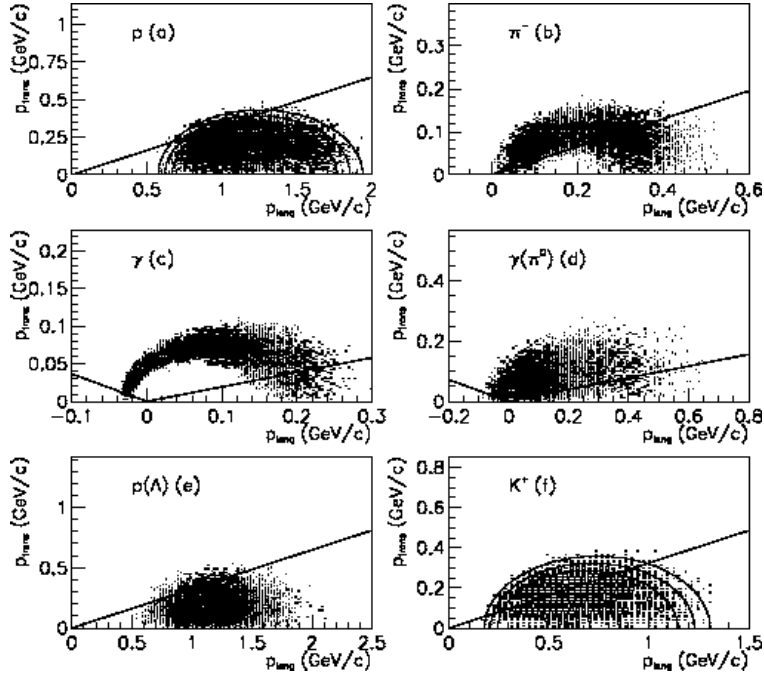}}
\caption{Simulated transverse vs.\ longitudinal momentum distribution
   for the final state particles in the reaction
   $pp\rightarrow\Lambda(1405)pK^+\rightarrow\Sigma^0\pi^0pK^+
   \rightarrow\Lambda\gamma\pi^0pK^+\rightarrow{}
   p^{(e)}\pi^{-(b)}\gamma^{(c)}\gamma\gamma^{(d)}p^{(a)}K^{+(f)}$.
   The superscripts denote the corresponding panels.  The ellipses in
   panels (a) and (f) show the kinematic limits with $\Lambda$(1405)
   production at its pole mass as well as 25~MeV/c$^2$ below and above.  The
   straight lines separate the acceptance of the central and the
   forward spectrometer of WASA.}
\label{fig:l1405-sim}
\end{center}
\end{figure}

Fig.~\ref{fig:l1405-sim} shows the simulated transverse versus
longitudinal momentum distribution for the particles in the final
state in this reaction induced by 3.6~GeV/c protons, assuming a pure
phase space distribution of the events, and taking into account a
$\Lambda$(1405) Lorentz mass distribution of a full width
$\Gamma{}=50$~MeV/c$^2$.  It demonstrates that both protons in the final
state, from the primary vertex and from the $\Lambda$ decay, have a
high probability to be emitted into the acceptance of the forward
detector, whereas the dominant fraction of the pions will be detected
by the central detector of WASA.  The $K^+$ mesons populate both
regions of the phase space seen by the central and by the forward
detector.  Only a small fraction of photons is not covered by the CsI
calorimeter acceptance.

In order to add new information to the experimental knowledge of the
spectral shapes of the $\Lambda$(1405) and $\Sigma$(1385) resonances,
the mass resolution has to be clearly better than the widths of these
resonances of $\approx{}50$~MeV/c$^2$ and $~\approx{}40$~MeV/c$^2$, respectively.  The
resolution of better than $\approx{}10$~MeV/c$^2$ allowed by the WASA detector
is considered to be sufficient for this study.  Further improvement of
the resolution is possible if new tracking detectors, such as
$\mu$-strip silicon arrays, are implemented close to the target. This
would allow, in addition to the reduction of non-strange background,
to make use of additional kinematic constraints by a precise
measurement of tracks including displaced vertex information.

The optimum proton beam energy is given by the requirement that the
kinematic limit be at sufficient distance from the pole of the
$\Lambda$(1405) resonance in order to guarantee the undisturbed
measurement of its spectral distribution.  At energies too close to
threshold it may be difficult to disentangle the $\Lambda$(1405)
spectral function and the effect of an energy dependent final state
interaction.  A reasonable beam momentum is $p_p=3.6$~GeV/c, well
within the COSY momentum range for internal beams, at which the
kinematic limit in the $\Lambda^{\star}$ or $\Sigma^{\star}$ mass is
$M_{max}=1.53$~GeV, more than 100~MeV/c$^2$ above the $\Lambda$(1405) pole.

The expected hyperon resonance production cross sections are of the
order of $1\,\mu$b~\cite{AnkeWS}.  With a luminosity
$L=10^{31}\,{\rm{}cm}^{-2}{\rm{}s}^{-1}$, assuming 10\% overall
efficiency, one thus expects approximately 18000 counts per day for
the $pp\rightarrow\Lambda(1405)K^+p$ reaction with a
$(p\pi^-\gamma{})_{\Sigma^0}(\gamma\gamma{})_{\pi^0}K^+p$ final state
discussed above.  In case the cross section and/or efficiencies should
be smaller, the maximum design luminosity
$L=2\cdot{}10^{32}\,{\rm{}cm}^{-2}{\rm{}s}^{-1}$ of WASA gives room for
an increase of the luminosity.

The second part of the program related to comparative studies of the
in-medium properties of the $\Lambda$(1405) and $\Sigma$(1385) hyperon
resonances requires the usage of nuclear targets, and thus an
extension of the existing pellet target to an operation with gases
like nitrogen, argon and xenon which still needs to be developed.
Furthermore, for these studies additional tracking information close
to the target allowing hyperon and $K_s$ identification will be
necessary, since a full reconstruction of the final state is not
possible with a nuclear target, which results in a loss of kinematic
constraints.  For this purpose the WASA detector would have to be
extended by the implementation of $\mu$-strip silicon detectors in
close geometry to the interaction point.

\subsection{\boldmath Rare and very rare $\eta$ and $\eta^\prime$ decays}
\label{sec:mediumterm_eta_etaprime}
\label{subsubetapmedium}

\paragraph{High precision branching ratio for $\eta^\prime \rightarrow 
\gamma \gamma$.}
In spite of the large branching ratio this experiment will be
performed at a later time since it requires technical modifications to
achieve high precision as discussed below. 
The radiative decay widths of $\eta (\eta^\prime) \rightarrow 
\gamma \gamma$, together with additional information from radiative 
$J/\Psi$ decays to $\eta (\eta^\prime) \gamma$ or from radiative 
transitions of vector ($V$) and pseudoscalar ($P$) mesons in $V 
\rightarrow P \gamma$ and $P \rightarrow V \gamma$ processes, allow to 
extract the fundamental constants of the pseudoscalar nonet (see 
e.g.~\cite{Ball:1996zv}). 
The unprecedented statistical accuracy, that is expected to be feasible 
for radiative decays using the WASA facility at COSY (see 
table~\ref{tab_etap_rates}) suggests to increase the precision on branching 
ratios of radiative decays presently available.

For the decay $\eta^\prime \rightarrow \gamma \gamma$, the statistical 
accuracy of the branching ratio might be increased by an order of 
magnitude in a dedicated run of two to three weeks, yielding a statistical 
error below the percent level.
However, to obtain precision values for the branching ratio requires 
to control the systematic error with comparable accuracy.
A suitable experimental approach is described for the decay $\eta 
\rightarrow \gamma \gamma$ in~\cite{Abegg:1996wz}, with a direct 
measurement of the branching ratio.
Thus, all systematic uncertainties related to the ''normalization'', 
i.e.\ the simultaneous measurement of a known branching ratio, are 
eliminated, and, analogously to~\cite{Abegg:1996wz} the branching ratio 
is given by
\begin{equation}
\label{eq_etap_2g_direct}
\frac{\Gamma\left(\eta^\prime \rightarrow \gamma \gamma\right)}
     {\Gamma_{tot}} = 
 \frac{\mbox{N}\left(\eta^\prime \rightarrow \gamma \gamma\right)}
      {\mbox{N}\left(p p \rightarrow p p \eta^\prime \right)} \times
 \left( A_{\eta^\prime \rightarrow \gamma \gamma}\,
        \epsilon_{\eta^\prime \rightarrow \gamma \gamma}^{analysis}\,
        \epsilon_{\eta^\prime \rightarrow \gamma \gamma}^{electronics}
 \right) \, ,
\end{equation}
where $A_{\eta^\prime \rightarrow \gamma \gamma}$ denotes the detector 
acceptance and $\epsilon_{\eta^\prime \rightarrow \gamma \gamma}$ the 
efficiencies in the analysis and in the electronics to reconstruct and 
reliably digitize an $\eta^\prime \rightarrow \gamma \gamma$ decay.
The precise number of events, background subtraction and (different) 
efficiency corrections only have to be considered for the $\gamma \gamma$ 
decay mode.
Experimentally, the approach requires to trigger on the $pp$ system 
associated with $\eta^\prime$ production, i.e.\ on two protons in the 
forward detector.
With an estimate of a $50\,\mbox{kHz}$ rate from a trigger on two charged 
particles in the forward detector at luminosities of 
$10^{32}\,\mbox{cm}^{-2}\mbox{s}^{-1}$, and in view of the attainable event 
rates for the planned data acquisition system (see section~\ref{subsubdaq}) 
in the order of $10\,\mbox{kHz}$ a high precision measurement of the 
$\eta^\prime \rightarrow \gamma \gamma$ branching ratio seems feasible 
after all, but will require further improvements of the selectivity of  
the forward detector trigger. 
Developments have already started for WASA at CELSIUS to implement a 
missing mass trigger for the $\Delta \mbox{E}/\mbox{E}$ technique by 
implementing energy reconstruction and angular correction of the energy 
loss information on the trigger level, and direct branching ratio 
measurements should become feasible after some experience and development 
with WASA at COSY.

\paragraph{Double vector meson dominance in $\eta \rightarrow 
e^+ e^- e^+ e^-$.}
Experimentally, the $\eta$ will be tagged by using the missing mass of the 
two protons detected in the forward detector.
Tracks of electrons and positrons are measured in the Mini Drift Chamber 
(MDC) inside the super-conducting solenoid, and their energy is determined 
by means of the CsI calorimeter.

Background from $2\,\pi^0$ production via $\pi^0 \pi^0 \rightarrow 
e^+ e^- e^+ e^- \gamma \gamma$ with two undetected $\gamma$s can be 
effectively removed by kinematic cuts.
The $\eta$ decay modes to $e^+ e^- \gamma$ ($\Gamma(\eta \rightarrow 
e^+ e^- \gamma)/\Gamma_{tot} = 6.0 \pm 0.8 \cdot 10^{-3}$~\cite{PDBook}) 
and to $2\,\gamma$ ($\Gamma(\eta \rightarrow \gamma \gamma)/\Gamma_{tot} = 
39.43 \pm 0.26\,\%$~\cite{PDBook}) contribute via photon conversion in the 
beam tube.
Both channels are effectively reduced to a level below $1\,\%$ by 
reconstructing the vertices of both $e^+ e^-$ pairs~\cite{Bosch:1996wr}.

Based on the QED prediction for the branching ratio of 
$2.52 \cdot 10^{-5}$~\cite{Jarlskog:1967np} and the acceptance of the WASA 
setup, at a luminosity of $10^{32}\,\mbox{cm}^{-2}\mbox{s}^{-1}$ we expect a 
rate of about $3 \cdot 10^3$ events per week.
In a 3-months run one will collect a data sample of 30000 events which
will allow for a high statistics measurement of the form factor dependence
on the virtual photon masses. In a second phase the data for the decay 
$\eta\rightarrow e^+ e^- e^+ e^-$ can be collected together with the
$\eta\rightarrow e^+ e^-$ experiments.

\paragraph{Search for CP violation in the rare decay $\eta \rightarrow 
\pi^+ \pi^- e^+ e^-$.}
$\eta$ tagging as well as electron and positron detection will be performed 
as described in Section~\ref{subsubetapday1}.
Electrons are separated from pions using the momentum over energy ratio 
with an accuracy of $5 \cdot 10^{-3}$.
The fraction of misidentified charged leptons has been estimated to be 
less than $2.5 \cdot 10^{-5}$.
From the complete four--momentum information for all four charged decay 
products, the angle between the $\pi^+ \pi^-$ and $e^+ e^-$ production 
planes can be reconstructed.

The major source of background arises from the conversion of photons from 
the $\eta \rightarrow \pi^+ \pi^- \gamma$ decay in the beam tube.
With an overall conversion probability of $0.28\,\%$ the background is 
important only in the region of small $e^+ e^-$ invariant masses below 
$40\,\mbox{MeV/c$^2$}$.
In comparison, background from other sources like direct production is 
negligible.

The experiment can be carried out together with studies of the
$\eta\rightarrow e^+ e^- e^+ e^-$ decay channel (see paragraph
above) and statistics of nearly $4\cdot 10^5$ 
$\eta\rightarrow \pi^+ \pi^- e^+ e^-$ events could be obtained,
corresponding to an accuracy of $2\cdot 10^{-3}$ for the
asymmetry. 

\paragraph{Search for physics beyond the Standard Model in the very rare $\eta 
\rightarrow e^+ e^-$ decay.}

The experiment aims at reaching a sensitivity that allows to test whether 
the $\eta$ decay to the $e^+ e^-$ mode occurs at a rate compatible with the 
Standard Model prediction of 
$5 \cdot 10^{-9}$~\cite{Savage:1992ac,Ametller:1993we}, which makes the 
decay susceptible to contributions from physics beyond the Standard Model, with a 
present experimental upper limit at $7.7 \cdot 10^{-5}$~\cite{PDBook}.

\begin{figure}[tb]
\begin{center}
\includegraphics[width=0.6\textwidth]{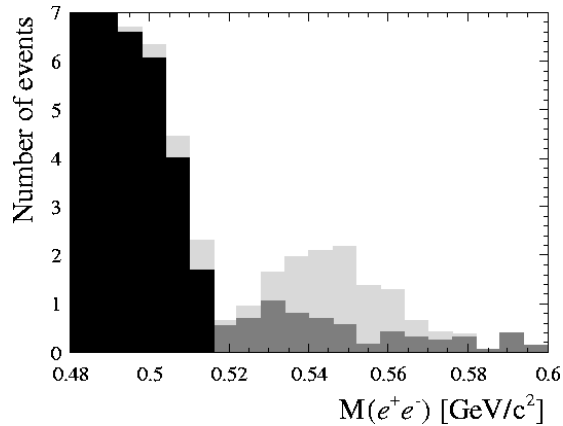}
\caption{\label{fig_eta_ee_bg} 
  Background contribution from $\eta \rightarrow \gamma e^+ e^-$ 
  (black), $\eta \rightarrow \gamma \gamma$ (dark grey) and sum of the 
  $\eta \rightarrow e^+ e^-$ signal and the two background sources (light 
  grey).  The simulation assumes the Standard Model prediction for the 
  branching ratio of $5 \cdot 10^{-9}$ for $\eta \rightarrow e^+ e^-$ and 
  the PDG values for the background channels, scaled to a total of 
  $10^{10}$ $\eta$ events produced.}
\end{center}
\end{figure}

The experimental signature will consist of four charged tracks, with two 
protons in the reaction $p p \rightarrow p p \eta \rightarrow p p e^+ e^-$ 
tagged in the forward detector and the $e^+ e^-$, with an isotropic 
distribution in the $\eta$ rest frame measured in the central detector with 
magnetic field.
In view of the low branching ratio the experiment requires high acceptance
and high luminosity, charged particle tracking, and the possibility to 
reject events with photons by means of an electromagnetic calorimeter,
which makes the WASA facility at COSY a unique device for this rare decay 
channel.
The most important physics processes that can contribute to a possible 
background are the $\eta \rightarrow \gamma e^+ e^-$ Dalitz decay, the 
radiative decay mode $\eta \rightarrow \gamma \gamma$, and prompt 
$e^+ e^-$ production. 
The continuous spectrum of $e^+ e^-$ pairs up to the $\eta$ mass from the 
$\gamma e^+ e^-$ decay mode ($\Gamma(\eta \rightarrow \gamma e^+ e^-) / 
\Gamma_{tot} = 6.0 \pm 0.8 \cdot 10^{-3}$~\cite{PDBook}) is effectively 
reduced to a level of less than $10\,\%$ compared to the $\eta \rightarrow 
e^+ e^-$ signal by rejecting photons in the central detector 
(Fig.~\ref{fig_eta_ee_bg}).

$\gamma$ conversion in the material between the interaction point and the 
plastic barrel (PSB) leads to a background contribution from the $\eta 
\rightarrow \gamma \gamma$ mode ($\Gamma(\eta \rightarrow \gamma \gamma) / 
\Gamma_{tot} = 39.43 \pm 0.26\,\%$~\cite{PDBook}), if both $\gamma$s 
convert asymmetrically in energy.
Most of the relevant material is concentrated by the $1.2\,\mbox{mm}$ 
Be beam--pipe, giving rise to a signal--to--background ratio\footnote{The 
background is expected to be reduced further by vertex reconstruction, 
which is not implemented in the simulations.} of 1--2 to 1 in the $\eta$ 
mass range (Fig.~\ref{fig_eta_ee_bg}).
A statistically significant signal from the rare $\eta \rightarrow e^+ e^-$ 
decay on the level of the Standard Model prediction, as indicated in 
figure~\ref{fig_eta_ee_bg}, could be obtained in a running time of 2 months.
It should be noted, that an appropriate trigger setting will allow to 
run the experiment in parallel with other $\eta$ decay studies, and, that 
already after one hour of beamtime, the experiment can increase the 
level of sensitivity in comparison to the present upper limit by an order 
of magnitude.
Furthermore, it should be emphasized, that these investigations are only 
feasible due to the optimization of the WASA setup with respect to the 
amount of material between the interaction point and the central detector 
(see section~\ref{subwasa}), to fully exploit the advantages of a 
windowless internal target.

However, the theoretical and experimental uncertainties for the 
background from prompt $e^+ e^-$ production in the $\eta$ mass range are large.
Calculations~\cite{Stepaniak:1998wm} based on the approach described 
in~\cite{Titov:1994vg} lead to a signal--to--background ratio for the decay 
$\eta \rightarrow e^+ e^-$ of 1 to 30 with respect to prompt production.
Thus, in order to have a reliable estimate for the prompt production 
background, it is suggested to prepare the $\eta \rightarrow e^+ e^-$ 
search by measuring the prompt contribution $pp \rightarrow p p e^+ e^-$ 
first.

At the same time, since WASA allows for exclusive measurements, this 
preparatory study can shed light on both the relative strength of 
different sources of $e^+ e^-$ pairs, and meson and baryon isobar form 
factors.
The first aspect is intimately motivated by the expectation that dileptons 
carry information from the early stage of heavy ion collisions.
Discrepancies between theoretical calculations  (see 
e.g.~\cite{Zetenyi:2001fu,Shekhter:2003xd}) and the presently available 
data for dilepton production in proton--proton and proton--deuteron 
collisions measured with rather small geometrical acceptance at the DLS 
spectrometer~\cite{Wilson:1997sr} --- especially in the mass range between 
0.2 and $0.5\,\mbox{GeV/c$^2$}$ --- could be verified experimentally using WASA 
at COSY.
Assuming a luminosity of $10^{32}\,\mbox{cm}^{-2}\mbox{s}^{-1}$ reasonable 
statistics of about $10^4$ events can be achieved in five days of data 
taking, i.e.\ a program of four kinetic beam energies ranging from 
1.04 to $2.5\,\mbox{GeV}$ to overlap both with the DLS data and theoretical 
calculations is feasible within three weeks, and will provide the 
necessary information for a reliable background estimate from prompt 
production concerning the search for the rare $\eta \rightarrow e^+ e^-$
decay.

\paragraph{C violation in $\eta \rightarrow \pi^0 e^+ e^-$.}
The experiment aims at a sensitivity down to the Standard Model prediction 
for the branching ratio of 
$0.2 - 1.3 \cdot 10^{-8}$~\cite{Cheng:1967pr,Ng:1993sc,Jarlskog:2002zz}, 
to search for new C--violating effects in the electromagnetic interaction.
With a value of $\Gamma(\eta \rightarrow \pi^0 e^+ e^-) / \Gamma_{tot} < 
4 \cdot 10^{-5}$~\cite{PDBook} the present upper limit is three orders of 
magnitude weaker than the theoretical estimate.

Monte Carlo simulations have been carried out to investigate 
the feasibility of the experiment and background conditions.
In conclusion, a search sensitivity of $10^{-9}$ is possible considering 
the full background.
A sample of ten events from C--conserving mechanisms for the rare decay 
can be expected after a period of three weeks to five months, depending on 
the actual value of the branching ratio.

A C--violating admixture at the level of $10^{-3}$ would already double 
the $\eta \rightarrow \pi^0 e^+ e^-$ decay rate.
However, at present this effect is small compared to the uncertainty in 
theoretical estimates for the C conserving decay mechanism, which is due to 
insufficient knowledge of value and structure of the $\eta \rightarrow 
\pi^0 \gamma \gamma$ amplitude.
Consequently, an improved precision in the $\eta \rightarrow 
\pi^0 \gamma \gamma$ decay rate is important for the interpretation of the 
experimental result of the proposed study of the $\pi^0 e^+ e^-$ decay
mode, and a high statistics study of $\eta \rightarrow \pi^0 \gamma \gamma$ 
should be carried out simultaneously.

\paragraph{Measurement of the $\eta^{\prime}$ transition form factor.}
With $\eta^\prime$ tagging by means of two detected protons in the forward 
detector and the reconstruction of the decay system of the 
Dalitz conversion decay $\eta^\prime \rightarrow \gamma \gamma^* 
\rightarrow \gamma e^+ e^-$ in the central detector, the WASA setup is well 
suited for measuring the $\eta^\prime$ transition form factor.
With the tracking information from the Mini Drift Chamber, the momenta 
of electron and positron are determined.
Thus, the differential cross section 
\begin{equation}
\frac{d\,\sigma}{d\,\mbox{q}^2} = 
  \left[ \frac{d\,\sigma}{d\,\mbox{q}^2} \right]_{pointlike} \,
  \left[ \mbox{F}(\mbox{q}^2)\right]^2
\end{equation}
with respect to the four--momentum transfer squared $\mbox{q}^2$ of the 
time--like virtual photon, which is equal to the invariant mass squared 
of the $e^+ e^-$ pair, can be measured, and the transition form factor 
$\mbox{F}(\mbox{q}^2)$ can be extracted.
The form factor is usually fitted with the simplest pole--type formula
\begin{equation}
\label{eq_dalitz_slope}
\mbox{F}(\mbox{q}^2) = 
  \frac{1}{1-\frac{\mbox{\footnotesize q}^2}{\Lambda^2}} \approx 
  1 + \frac{\mbox{q}^2}{\Lambda^2}
\end{equation}
with the slope parameter $\Lambda$.

At present, only an upper limit exists for the $\gamma e^+ e^-$ decay mode 
of the $\eta^\prime$ ($\Gamma(\eta^\prime \rightarrow \gamma e^+ e^-) / 
\Gamma_{tot} < 9 \cdot 10^{-4}$).
With an estimated branching ratio of $3 \cdot 10^{-4}$, as discussed 
in~\cite{Briere:1999bp}, we expect 45 fully reconstructed $\eta^\prime$ 
Dalitz conversion decays per day.
Following the discussion for the $\eta^\prime \rightarrow 
\gamma \pi^+ \pi^-$ decay in section~\ref{subsubetapday1} we only 
consider the radiative decay $\eta^\prime \rightarrow \gamma \gamma$ with 
subsequent conversion of one $\gamma$ as a possibly important source of 
background.
With fits to the $\eta^\prime$ signal in each bin of the $e^+ e^-$ 
invariant mass scale, the differential cross section can be extracted 
background free, unless there is a background process for the $eta^\prime 
\rightarrow \gamma e^+ e^-$ transition.
However, since for $\eta$ decays the $\gamma \gamma$ mode was found to 
be negligible above very low invariant lepton pair 
masses~\cite{Stepaniak:2002ad}, and since the ratios 
$\Gamma(\eta(\eta^\prime) \rightarrow \gamma e^+ e^-) / 
\Gamma(\eta(\eta^\prime) \rightarrow \gamma \gamma)$ are approximately the 
same for $\eta$ and $\eta^\prime$, and since the photon conversion 
probability is not changing drastically, it can safely be neglected as a 
background contribution.

The beam time estimate depends on the desired accuracy for the slope 
parameter $\Lambda$ in eq.~\ref{eq_dalitz_slope}.
From the analogous discussion of the $\eta$ transition form factor 
in~\cite{Stepaniak:2002ad}, a running time of two months would allow a 
determination of the slope parameter with a $10\,\%$ error.
A further improvement by a factor of two in accuracy would require a nine 
months period of data taking.
It should be noted, that due to the trigger condition of two charged and 
one neutral particle in the central detector data taking could be done 
in parallel with other $\eta^\prime$ decay studies.
For example, the specific trigger requirement for the $\eta^\prime$ 
Dalitz conversion decay exactly matches the trigger setup for the 
$\pi^+ \pi^- \gamma$ mode discussed in section~\ref{subsubetapday1} in 
terms of particle multiplicities.


\clearpage

\newcommand{\emparag}[1]{\paragraph{#1}~\\}

\section{Experimental facility}

\subsection{Cooler Synchrotron COSY }
\label{subcosy}

\begin{figure}[b]
\begin{center}
\includegraphics[clip,width=0.7\textwidth]{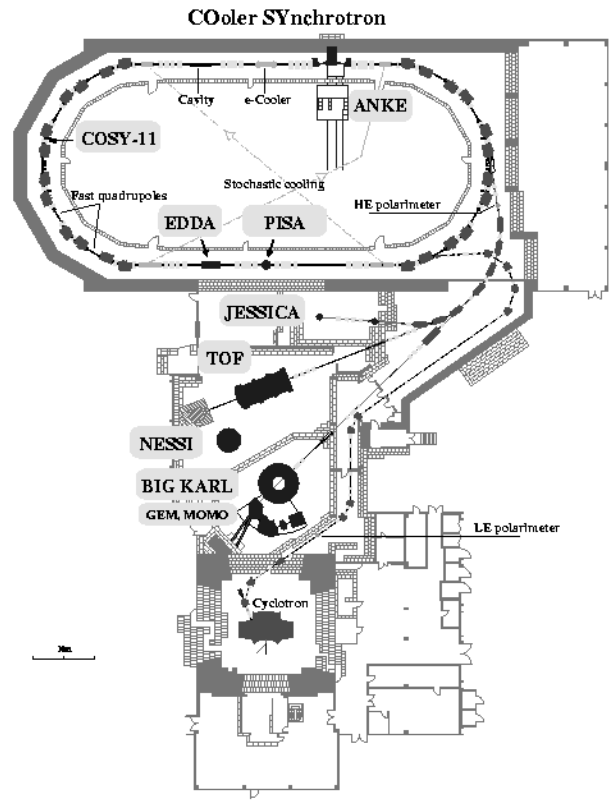}
  \caption{Floorplan of the accelerator complex. The WASA detector
           will be installed at the place of the old cavity in
           front of the electron cooler.}
\label{fig:cosy}
\end{center}
\end{figure}

COSY is a cooler synchrotron and storage ring operated at the IKP of FZJ. 
The accelerator complex (see Fig.~\ref{fig:cosy}) comprises an isochronous 
cyclotron (JULIC), used as an injector, a race track shaped cooler 
synchrotron with a circumference of 184~m, and
internal and external target stations~\cite{RMaier97}. COSY delivers
beams of polarized and unpolarized protons and deuterons in the momentum
range between 0.3~GeV/c and 3.7~GeV/c. The ring can be filled with up
to $10^{11}$~particles leading to typical luminosities of 
$10^{31}\,\mathrm{cm^{-2}s^{-1}}$ when using an internal cluster target.
Beams can be phase-space cooled by
means of electron cooling at injection energy as well as stochastic cooling
at high energies. Typical beam preparation times, including injection, 
accumulation and acceleration, are of the order of a few 
seconds, while the beam lifetime with a cluster target is between several 
minutes and an hour.

Currently, four internal experiments (ANKE, COSY-11, EDDA, PISA) and three 
external detector systems (BIG KARL, JESSICA, TOF) are operated by large
international collaborations.
On the average COSY is running for more than 7000~hours per year. 
Typically, it delivers 
beams for experiments with a reliability of 94\%.  

\emparag{Implementation of WASA at COSY}
The Cooler Synchrotron has two 40~m long straight sections joining
the arcs. After removal of the old RF cavity at the beginning of
2004 space for the implementation of WASA at an internal target
position in front of the electron cooler is available (see 
Fig.~\ref{fig:cosy}). This location has a number of advantages: most
importantly it minimizes both the interference with the existing COSY detectors
and the modifications needed to the WASA detector and the accelerator.

Operating WASA with the pellet target with an effective thickness of
$2.5\cdot 10^{15}\,\mathrm{cm^{-2}}$ average luminosities of 
$10^{32}\,\mathrm{cm^{-2}s^{-1}}$ and beam lifetimes of a couple of
minutes are expected~\cite{lehrach03}. Compared to the present
operation at CELSIUS the experimental conditions will be improved
due to the different accelerator characteristics: fast magnet ramping,
dispersion-free target position, stochastic cooling and 
a smooth microscopic time structure of the beam.

\subsection{The WASA detector}
\label{sec:setup}
\label{subwasa}

The 4$\pi$ detector facility WASA \cite{CWwww,Zabi02} was designed for 
studies of
production and decays of light mesons at CELSIUS.
Pions and eta mesons are produced in proton-proton and proton-deuteron
interactions. The highest beam-proton kinetic energy reachable at CELSIUS is 1.5 GeV. 

The pellet-target system is integrated in the setup and it provides small spheres of
frozen hydrogen and deuterium as internal targets. This allows high luminosity and
high detection coverage for meson decay products like photons, electrons
and charged pions.

\begin{figure}[!htb]
\includegraphics[width=\textwidth,clip]{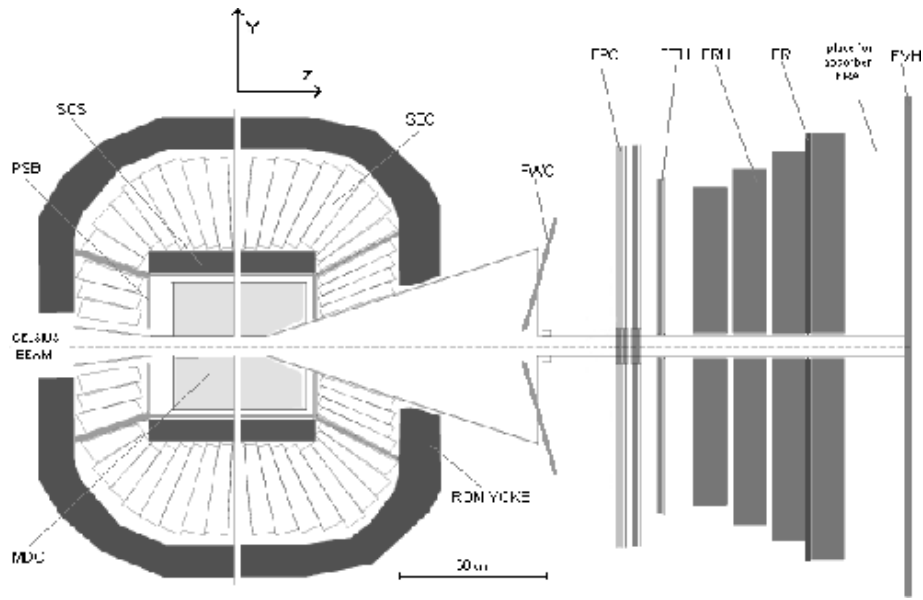}
\caption{\label{fig:wasa} Cross section of the WASA detector. The
central detector built around the interaction point (at the left) is
surrounded by an iron yoke. The layers of the forward detector are
visible on the right-hand side. The
individual components are described in the text.}
\end{figure}

WASA consists of a forward part for measurements of charged target-recoil particles
and scattered projectiles, and a central part designed for measurements of
the meson decay products. The forward part consists of eleven
planes of plastic scintillators and of proportional counter drift tubes.
The central part consists of an electromagnetic calorimeter of CsI(Na) crystals 
surrounding a superconducting solenoid. 
Inside of the solenoid a cylindrical chamber of drift tubes and a barrel 
of plastic scintillators are placed. A vertical cross section of the WASA detector is shown in 
Fig.~\ref{fig:wasa}. 

\subsubsection{Pellet target}
\label{sec:pellet}

\begin{figure}[!htbp]
\centering
\includegraphics[width=\textwidth]{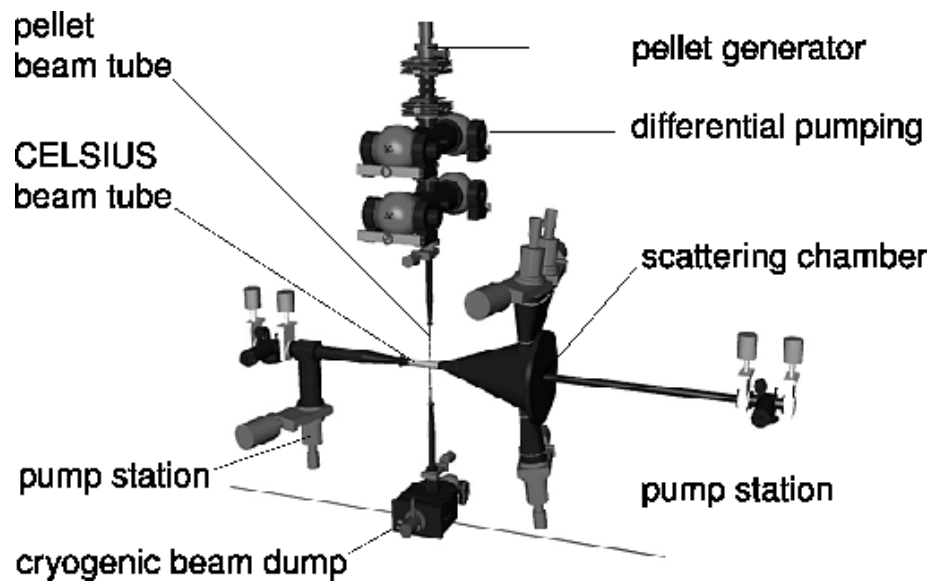}
\caption{\label{fig:pellet1}Layout of the pellet target system.}
\end{figure}

The pellet target system was a unique development for the 
CELSIUS/WASA experiment \cite{Tro95,Eks96}.
The main components of the system are shown in Fig.~\ref{fig:pellet1}.

The heart of the setup is the pellet generator where a jet of liquid hydrogen 
is broken up into droplets with a diameter around 35 $\mu$m 
by a vibrating nozzle. The droplets freeze by evaporation in a droplet
chamber and form a beam of pellets that enter a 7\ cm long
vacuum-injection capillary. After
collimation, the pellets are directed through a thin 2~m long pipe into
the scattering chamber and further down to a pellet beam dump. 
The inner diameter of the pipe is 5\ mm at its entrance
to the scattering chamber. This arrangement provides the necessary space
to put the 4$\pi$ detection system around the interaction region.

Pellet target thicknesses of up to $3\cdot 10^{15}$ atoms/cm$^2$ give
acceptable half-lives of the circulating ion beam as well as
acceptable vacuum conditions. Some of the parameters of the pellet 
target at the present stage of operation are listed in table \ref{tab:pellet}.
  The pellet target system operates regularly with pellets of normal hydrogen 
and
 deuterium.

\begin{table}[!htbp]
\centering
\small
\begin{tabular}{|l|c|}
\hline
{\bf Pellet target parameters} & {\bf  Present performance}\\
\hline
Pellet diameter            &  25 - 35 $\mu$m      \\
Pellet frequency           &  5-12 kHz   \\
Pellet-pellet distance      &  9-20 mm   \\
Effective target thickness &  $> 10^{15}$ atoms/cm$^2$\\
Beam diameter              & $2-4$ mm   \\
 \hline
\end{tabular}
\normalsize
\caption{\label{tab:pellet} Present performance of the
pellet target system.}
\end{table}

\subsubsection{Forward detector}

The forward detector  (FD) is designed mainly 
for detection and identification of scattered projectiles and charged recoil particles 
like protons, deuterons and He nuclei in $\pi$ and $\eta$ production reactions.  
Also neutrons and charged pions can be measured.
All FD plastic scintillators may supply information for the first level trigger logic. 
This part of the setup has already been used in a previous experiment at CELSIUS,
WASA/PROMICE, which is described more in detail  in \cite{Calen96}.
A summary of the most important features of the forward detector (FD) is given in
table \ref{tab:FD}. The individual components are described in some
detail in the following.

\begin{table}[!h]
\centering
\begin{tabular}{|l|c|}
\hline
\multicolumn{2}{|l|}{\large {\bf Forward detector}} \\
\hline
Total number of scintillator elements  & 280\\
Scattering angle coverage & $3^{\circ}$ - $17^{\circ}$ \\
Scattering angle resolution & $0.2^{\circ}$\\
Amount of sensitive material [g/$cm^{2}$] & 50\\ 
~~~~[radiation lengths] & $\approx$ 1 \\ 
~~~~[nuclear interaction lengths] & $\approx$ 0.6 \\
Thickness of vacuum window (st. steel) [mm] & $\approx$ 0.4\\ 
Maximum kinetic energy ($T_{stop}$) for stopping: & \\
~~~~$\pi^{\pm}$/proton/deuteron/alpha\qquad [MeV] & 170/300/400/900\\
Time resolution                     & $<$ 3 ns \\
Energy resolution for: & \\
~~~~ stopped particles & $\approx$ $3\%$\\
~~~~ particles with $T_{stop}$ $<T<$ 2$T_{stop}$ & 4 - $8\%$\\ 
Particle identification & $\Delta$E-E\\ 
\hline
\end{tabular}
\normalsize
\caption{\label{tab:FD} Some features of the Forward Detector.}

\end{table}

\begin{figure}[ht]
\hspace{3cm}\includegraphics[width=0.5\textwidth,clip]{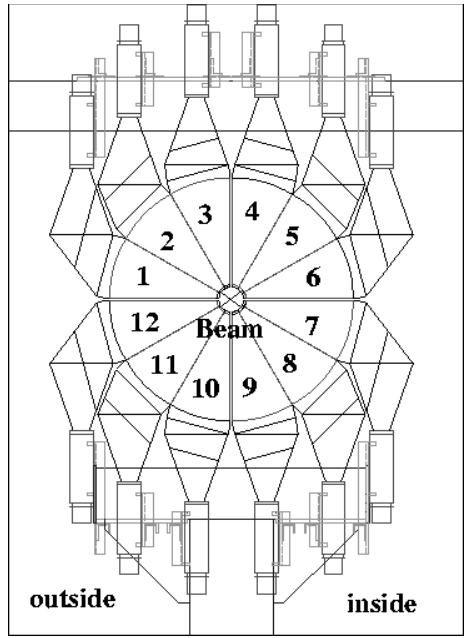}
\caption{\label{fig:fwc} Schematic view of the FWC.}
\end{figure}

\emparag{The Forward Window Counters (FWC)}
\noindent The FWC is the first detector layer in the FD
(along the beam direction) and consists of 12, 5 mm thick plastic scintillators
(Fig.~\ref{fig:fwc}).
It is mounted tightly on the paraboloidal stainless steel vacuum window. 
Therefore, the elements
are inclined by approximately
10$^{\circ}$ with respect to the plane perpendicular to the beam direction. 
The FWC signals are used in the first level trigger logic to reduce the
background caused by particles scattered in the downstream beam pipe and in the flange 
at the entrance to the FD. The signals are also used to select He ejectiles
on the trigger level.

\begin{figure}[ht]
\hspace{3cm}\includegraphics[width=0.5\textwidth,clip]{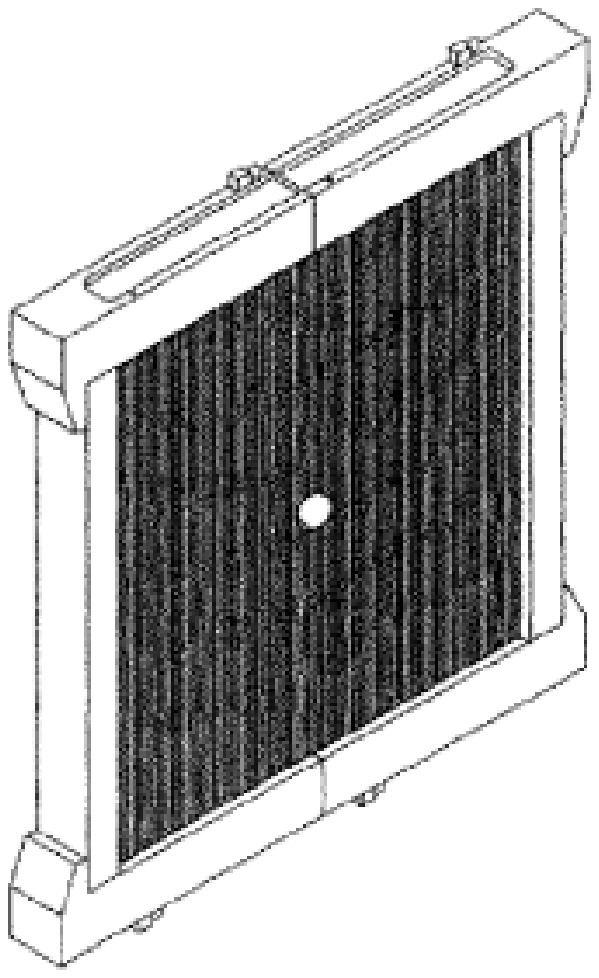}
\caption{\label{fig:fpc} Schematic view of one module of the FPC.}
\end{figure}

\emparag{The Forward Proportional Chambers (FPC)}
\noindent Immediately downstream of the FWC, there is a tracking device. 
It is composed of
4 modules, each with 4 staggered layers of 122 proportional drift
tubes (so called straws) of 8 mm diameter
(Fig.~\ref{fig:fpc}).
The modules are rotated by 45$^o$ with
respect to each other (in the plane perpendicular
to the beam axis). They are used for accurate reconstruction
of track coordinates and provide precise
angular information of the  particles originating from the target region.

\emparag{The Forward Trigger Hodoscope (FTH)}
\noindent The FTH, consisting
of three layers of  5mm thick plastic scintillators, is placed
next to the FPC. There are 24 Archimedian spiral shaped elements in the
first two planes and 48 radial elements in the third. 

\begin{figure}[ht]
\vspace{-2cm}
\includegraphics[width=0.5\textwidth,origin=b,angle=55.5,clip]{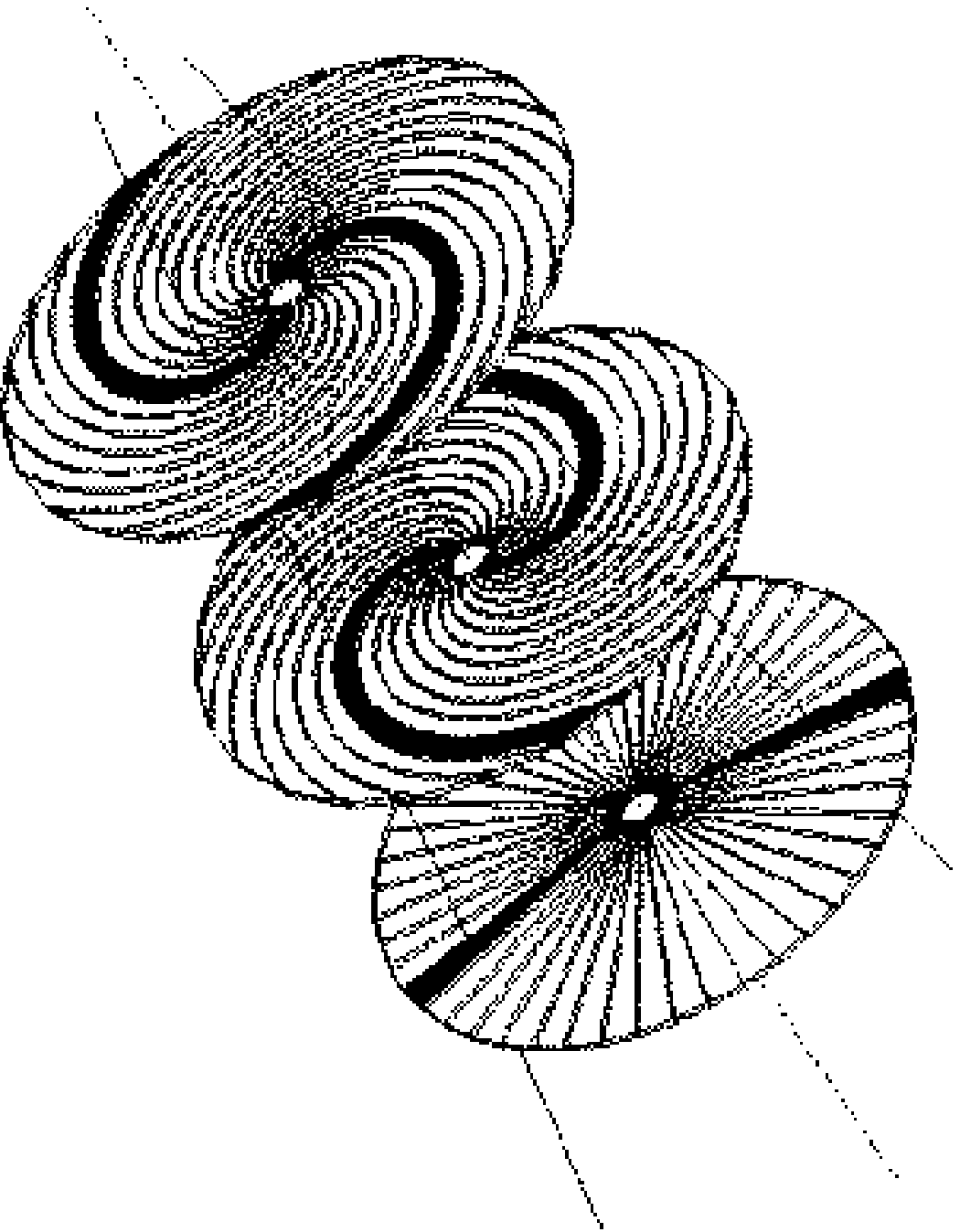}%
\includegraphics[width=0.3\textwidth,origin=t,clip]{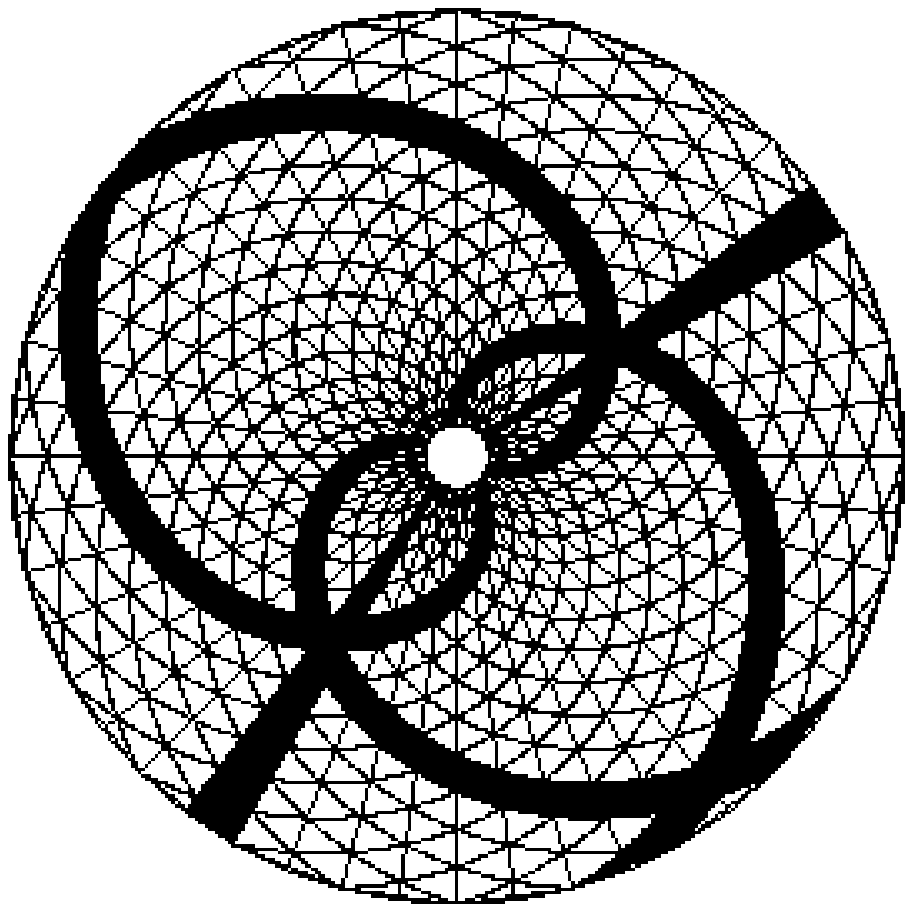}
\vspace{-2cm}
\caption{\label{fig:fth} Schematic view of the three forward trigger
hodoscope layers.}
\end{figure}

The FTH is mainly used  in the first level trigger logic. However, the
special geometry, combining all three layers, results in  a pixel
structure, which is useful for resolving multi-hit ambiguities. In
Fig.~\ref{fig:fth}
 the structure of the FTH is shown with hits from two passing
charged particles.

\begin{figure}[ht]
\hspace{3cm}\includegraphics[width=0.5\textwidth,clip]{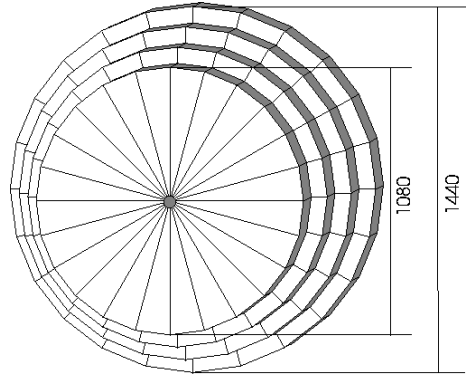}
\caption{\label{fig:frh} Schematic view of the four forward range
hodoscope layers. Dimensions are given in mm.}
\end{figure}

\begin{figure}[!hbt]
\centering
\includegraphics[width=0.7\textwidth,angle=-90,clip]{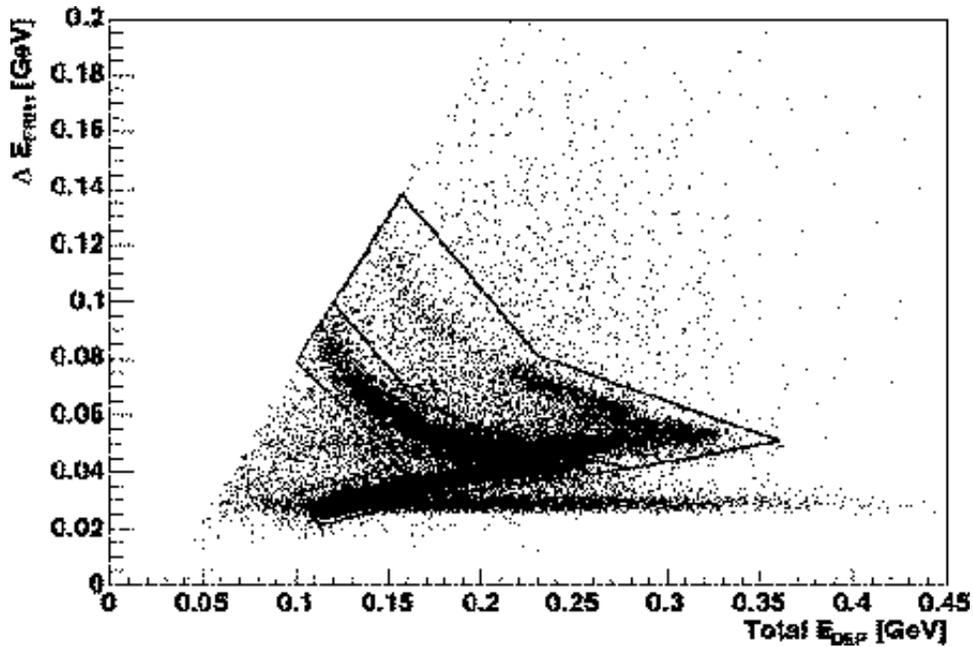}
\caption{\label{fig:FDdep} Example of particle identification in the
forward detector for data collected at 640~MeV. Energy deposited
in  the 1$^{st}$ layer of FRH is plotted versus total
energy deposited in FTH and FRH. The lines indicate cuts to be used for selecting 
protons (lower) and deuterons (upper).
}
\end{figure}

\newpage
\emparag{The Forward Range Hodoscope (FRH)}
\noindent Behind the FTH, the four layers of the FRH are positioned (Fig.~\ref{fig:frh}). Each plane is made
of 24, 11 cm thick, plastic scintillator modules. The FRH, together with FTH, 
is used for energy determination of charged particles and for particle
identification  by $\Delta$E-E technique. Fig.~\ref{fig:FDdep} shows how
protons and deuterons can be identified.

\begin{figure}[tb]
\begin{center}
\includegraphics[width=0.6\textwidth,clip]{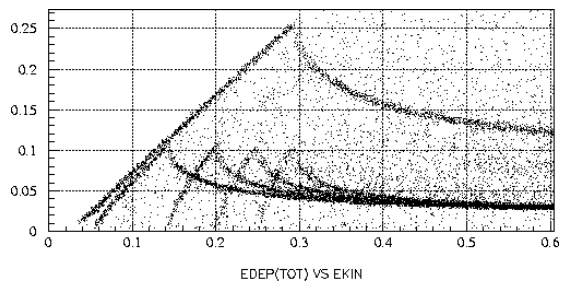}
\caption{\label{fig:fdedep} Energy absorbed  in  the  forward  detector
      scintillator material  for protons as a  function of their initial 
      kinetic energy.  The  upper curve  gives the  total energy
      deposited in the   plastic scintillator planes  and the lower curves show
      the energy distributed among the different FRH planes.}
\end{center}
\end{figure}

The identity and initial kinetic energy of a charged particle is reconstructed from the
      pattern   of   deposited  energy  in  the   different   detector  planes.
      Fig.~\ref{fig:fdedep} shows how the  initial kinetic energy of a proton is
      related to the  deposited energy in  the forward  detector scintillators.
      Even for  particles that are  stopped in a detector,
      the total deposited energy is  different from the initial kinetic energy.
      This is because some of  the energy is lost in  inactive material between
      the  detectors  elements, in the  scattering  chamber  windows, etc.
      Fig.~\ref{fig:fdedep}   also  shows that  the  variation of  the deposited
      energy is strong enough  to be useful for  energy reconstruction also for
      high energy particles not stopping in the detector material. For protons,
      this can be used in the  kinetic energy range  300~MeV to 800~MeV and for
      deuterons in a similar energy range starting from 400~MeV. Identification
      of punch-through particles can be done using either the veto hodoscope (FVH)
      or the $\Delta$E  information  from the last FRH  planes.

\begin{figure}[ht]
\hspace{3cm}\includegraphics[width=0.5\textwidth,clip]{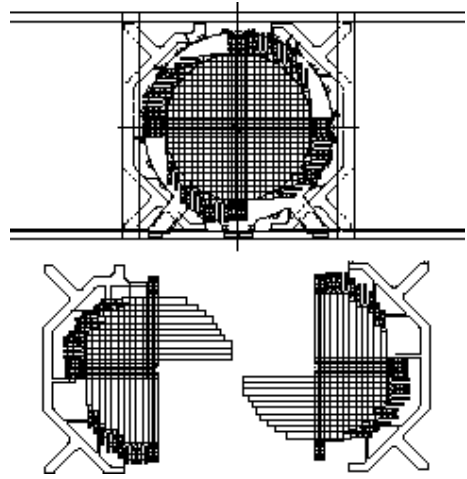}
\caption{\label{fig:fri} Schematic view of the FRI detector (upper picture) and its
two planes with orthogonally oriented scintillator bars (lower
picture).}
\end{figure}

\emparag{The Forward Range Interleaving Hodoscope (FRI)}
\noindent Between the third and fourth layers of the FRH there are two interleaving 
layers of 5.2 mm thick plastic scintillator bars (Fig.~\ref{fig:fri}).
Each layer has 32 bars, oriented horizontally in one and vertically in the other.

The main purpose of this addition to the FRH is to provide a two-dimensional position 
sensitivity inside the FRH
necessary for measurement of scattering angles for neutrons. The efficiency
for detection of neutrons of a few hundred MeV kinetic energy by the FRH is around 35 \%. 
In addition, the FRI can help in vertex reconstruction and to discriminate against background tracks
due to secondary interactions in the beam pipe and other structural material.

The FRI was recently commissioned; more information
about the design and performance of the FRI can be found in \cite{Pauly04}.

\begin{figure}[ht]
\hspace{3cm}\includegraphics[width=0.5\textwidth,clip]{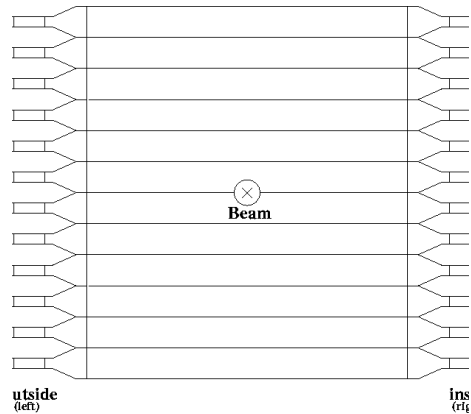}
\caption{\label{fig:fvh} Schematic view of the Forward Veto Hodoscope.}
\end{figure}

\emparag{The Forward Veto Hodoscope (FVH)}
\noindent The last detector layer of the FD is a wall of plastic 
scintillators (Fig.~\ref{fig:fvh}).
 It consists of 12 horizontal plastic scintillator bars,
equipped with  photomultipliers on  both sides. The hit position along a bar may 
be reconstructed from signal time information.
 In the first level trigger the signals are used for rejection 
(or selection) of particles punching through the FRH.

\emparag{The Forward Absorber (FRA)}
\noindent 
Optionally, a passive absorber layer made of iron can be introduced between the last layer of the FRH
and the FVH. The thickness of the absorber can be chosen from 5~mm up to  
100~mm.
 The absorber has been used for stopping the protons from the
$pp\rightarrow pp\eta$ reaction at a beam proton energy around 1360 MeV.
In this case the faster protons from elastic scattering and from pion production penetrate 
the FRA and induce signals in the FVH which can be used for veto purposes in the trigger.

\subsubsection{Central detector}

The central detector  (CD) is built around the interaction point and 
is designed mainly 
for detection and identification of the decay products of $\pi^\circ$ and 
$\eta$ mesons: photons, 
electrons and charged pions. It consists
of an inner drift chamber (MDC), a solenoid (SCS) providing magnetic field for
momentum measurements, thin plastic scintillators in a cylinder
geometry (PSB) and a CsI calorimeter (SEC).
The amount of structural material is kept minimal to reduce the
disturbances on the particles.
The beam pipe is made of 1.2~mm thick
beryllium and the total thickness  of the solenoid corresponds to 0.18 radiation
lengths only.
 
For the design the main requirements were the
following: 
\begin{itemize}
\item it has to handle high particle fluxes at luminosities around ${\rm
10^{32}\,cm^{-2}s^{-1}}$.
\item it should be able to measure photons with energies from a few MeV 
up to 800~MeV.
\item it should be able to measure, in a magnetic field of about 1 T,
the momenta of electrons and positrons in the range
${\rm p \approx 20-600~MeV/c}$ with an accuracy ${\rm \sigma_p/p < 2\%}$.
\end{itemize}

The momenta of heavier charged particles can also be measured in a 
similar momentum 
ranges, but with lower accuracy:\\ 
for pions and muons with ${\rm p \approx 100-600~MeV/c}$, ${\rm \sigma_p/p < 4\%}$ \\
for protons with ${\rm p \approx 200-800~MeV/c}$, ${\rm \sigma_p/p < 6\%}$ \\

The main components of the Central Detector, shown in Fig.~\ref{fig:wasa}, 
are presented below in some detail.

\emparag{The Superconducting Solenoid - (SCS)}
\noindent The SCS provides an axial magnetic field for the momentum 
measurements
for the tracks measured by the MDC. It also protects the CD from low-energy
delta electrons produced in the interactions of beam particles with the pellets.
The wall thickness of the SCS is minimized in order to allow high
accuracy of the energy measurements in the calorimeter.
The return path for the magnetic flux is provided by a  yoke 
made out of 5 tons of 
soft iron with low carbon content. The yoke shields the  
readout electronics from the magnetic field and serves also as support for the calorimeter 
crystals. The main parameters of the SCS are given in table
\ref{tab:scs}.

\begin{figure}[!h]
\includegraphics[width=\textwidth,clip]{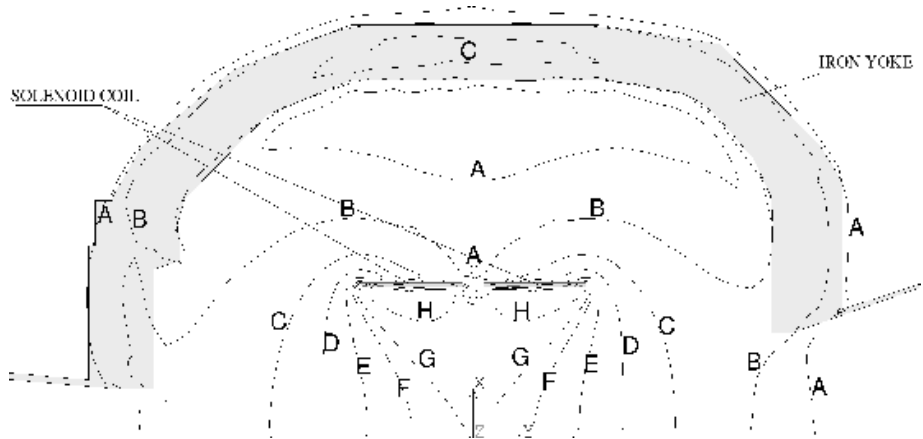}
\caption{\label{fig:sol}Calculated distribution of the magnetic flux
density for a coil current of 667~A \cite{Ruber01}. Contour maxima are indicated by lines
marked A-H, where: A=0.10T, B=0.25T, C=0.50T,
D=0.75T, E=1.00T, F=1.20T, G=1.30T, H=1.50T. }
\end{figure}

In order to map the magnetic field inside the volume enclosed by the
SCS, the magnetic field strength inside the MDC was measured with Hall
probes and, in addition, the field distribution was calculated with simulation
programs. The calculated  values were fitted to the measured
ones with an error of $\pm$1\% of B$_{total}$ \cite{Ruber01}.
The magnetic flux density distribution inside of the iron yoke
calculated for a current of 667~A is given in Fig.~\ref{fig:sol}.
The SCS is described in detail in the Ph.D. thesis of 
Roger Ruber \cite{Ruber99}.

\begin{table}[!h]
\centering
\begin{tabular}{|l|c|}
\hline
\multicolumn{2}{|l|}{\large {\bf Superconducting coil}} \\
\hline
 Inner/outer radius [mm] & 267.8 / 288.8\\ 
 Superconductor (stabilizer) & NbTi/Cu (pure Al)\\ 
 Total winding length & 465 mm\\
 Maximum central magnetic flux density, {\bf B$_{c}$} & 1.3 T\\
 Field uniformity in the MDC & 1.22 T $\pm$$20\%$\\
 Cooling & Liquid He, $4.5^{\circ}$K\\ 
\hline
\multicolumn{2}{|l|}{\large  {\bf Cryostat}}\\
\hline
 Material & Aluminium\\
 Inner / outer radius [mm] & 245 / 325\\ 
 Overall length [mm] & 555\\ 
\hline
 {\bf SCS wall thickness} (coil+cryostat) [radl] & {\bf 0.18}\\ 
\hline
\end{tabular}
\caption{\label{tab:scs} Main parameters of the superconducting
coil and its cryostat.}
\end{table}

\emparag{The Mini Drift Chamber - (MDC)}
\noindent The MDC is placed around the beam pipe and is used for 
determination of particle
momenta and reaction vertex. It is a cylindrical chamber covering 
scattering angles from 24$^o$ to
159$^o$. For large angle scattered protons from elastic proton-proton 
scattering, a vertex 
resolution ($\sigma$) of 0.2~mm perpendicular and 3~mm along the beam 
axis can be reached. 
A detailed description of the MDC 
can be found in the Ph.D. thesis of Marek Jacewicz~\cite{Jacewicz04}. 

The MDC consists of 1738 drift tubes, so called straws, arranged in  17 cylindrical layers.
 The diameter of the straws in the 5 inner most layers is 4~mm, 6~mm in the 
next 6 layers and 8~mm in the outer 6 layers.
 The straws are made of a thin (25~$\mu$m) mylar  foil coated with
 0.1~$\mu$m aluminum on the inner side only.
In the center of each straw there is a 20~$\mu m$ diameter sensing wire
made of gold plated tungsten (W(Re)), stretched
with a tension of 40~g. 
The wires are aligned with a precision of $\pm 20\;\mu$m.

This design  was chosen in order to cope with the expected high
particle flux allowing a maximum deposited energy of approximately 70
$\frac{MeV}{mm\cdot s}$ for the straws exposed most at the inner part of the chamber.

 The  layers are located between radii of 41 and
203~mm. The straws in nine
layers are parallel to the beam axis (z-axis) and the other eight layers have small
skew angles (6$^{o}$-9$^{o}$) with respect to the z-axis. These ``stereo'' layers form a
hyperboloidal shape. 

The straws in the five inner layers are divided unequally by
the center of the pellet pipe, while the other layers are
symmetrical. The straws in each layer are inter-spaced by small gaps in order to
prevent the mechanical deformation by  neighboring tubes.

\begin{figure}[!h]
\includegraphics[width=0.425\textwidth,]{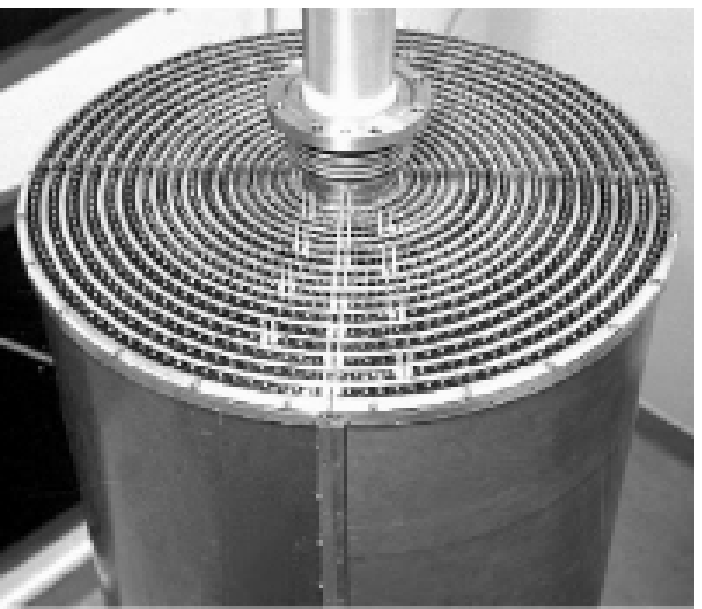}%
\includegraphics[width=0.575\textwidth,]{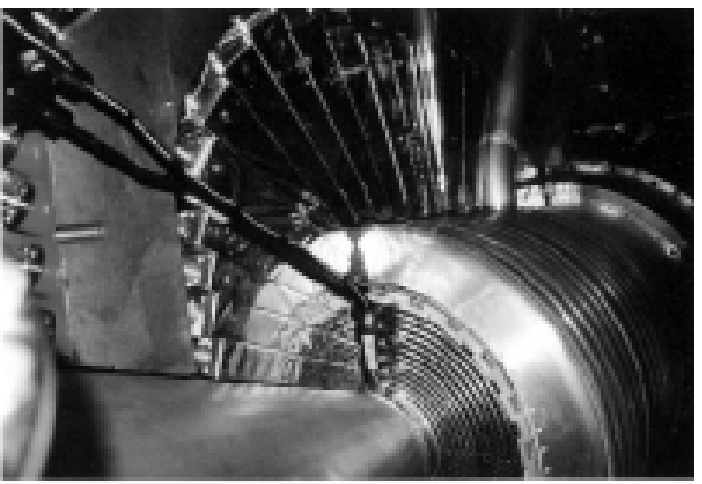}
\caption{\label{fig:mdccyl} (Left) The fully assembled MDC inside 
the Al-Be cylinder. (Right) The MDC surrounded by PSB elements.}
\end{figure}

The MDC is fitted inside a cylindrical cover made of 1~mm Al-Be and
is placed inside
the solenoid (Fig.~\ref{fig:mdccyl}).

\begin{figure}[!hbt]
\centering
\includegraphics[width=0.6\textwidth]{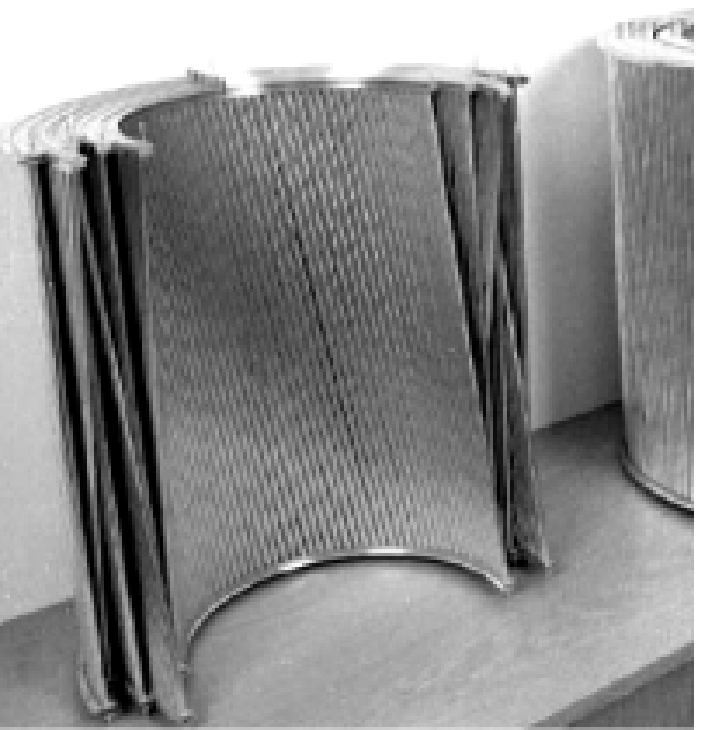}%
\includegraphics[width=0.4\textwidth]{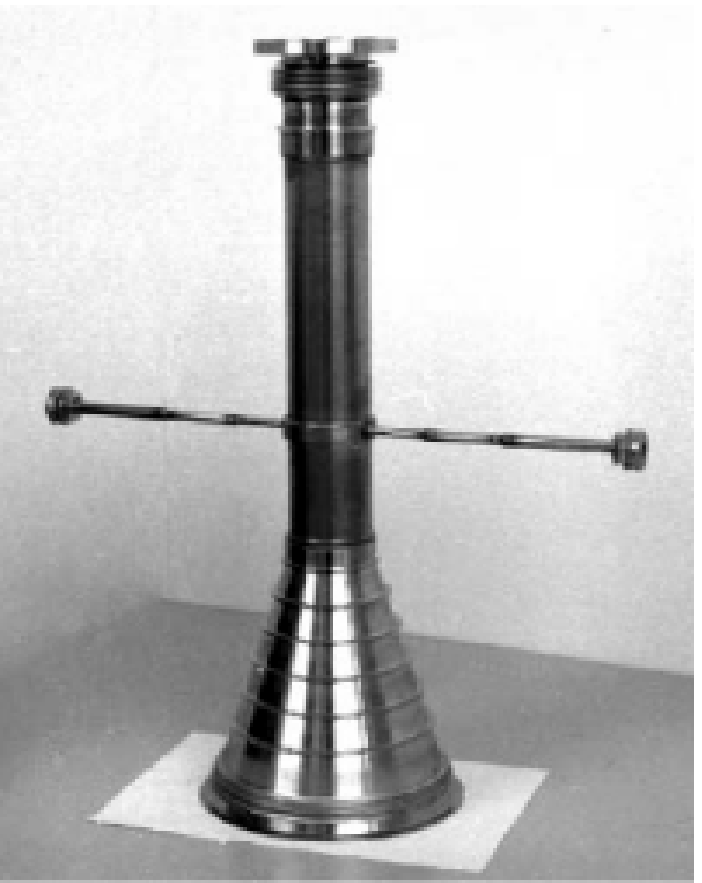}
\caption{\label{fig:mdcpipe} (Left) Drift tubes secured in the
end-plates. Note the ``stereo'' layers interleaved with parallel
layers. (Right) Be beam-pipe.}
\end{figure}

\begin{figure}
\includegraphics[width=\textwidth]{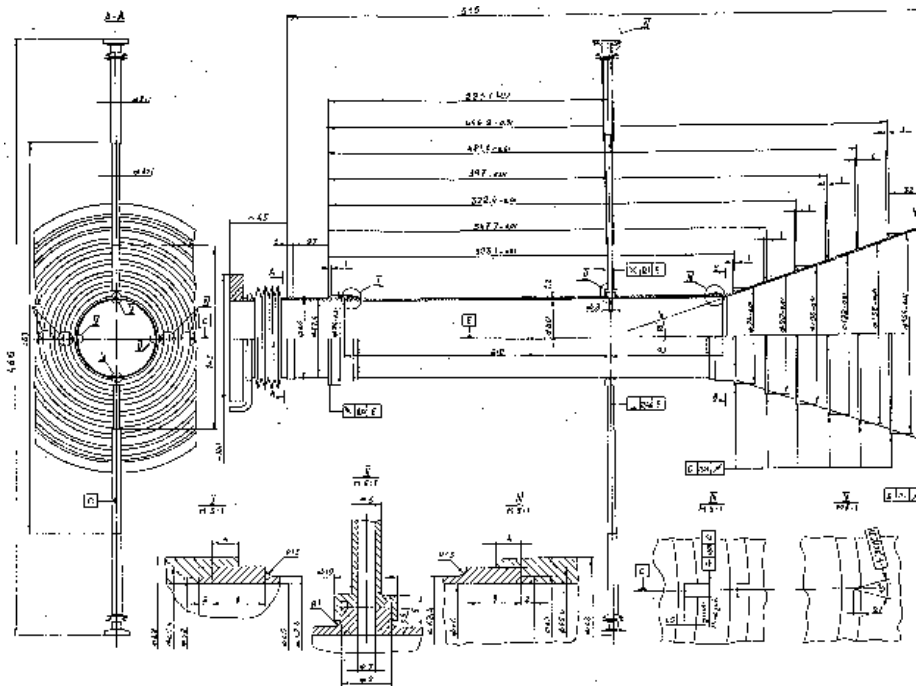}
\caption{\label{fig:mdcschem1} Beryllium beam-pipe including the crossing
 pellet target tube.}
\end{figure}

The straws in each (half) layer are mounted between $\approx$5~mm thick 
Al-Be end-plates. The layers are assembled around the Be beam-pipe 
(Fig.~\ref{fig:mdcpipe}) and the attached pipe for the pellets. 
The beam-pipe has a diameter of 60~mm and a wall thickness of 1.2~mm.
The design drawing of the Be beam pipe is shown in Fig.~\ref{fig:mdcschem1}.

\newpage
\emparag{The Plastic Scintillator Barrel - (PSB)}
\noindent The PSB is located inside the SCS coil and surrounds the MDC.
It provides fast signals for the first level trigger logic and, 
together with the mini drift chamber and the CsI calorimeter, it is
employed for charged particle identification by the $\Delta$E-p
and $\Delta$E-E methods and serves as a veto for $\gamma$ identification.

\begin{figure}[!hbt]
\centering
\framebox{
\includegraphics[width=0.25\textwidth,angle=-90,clip]{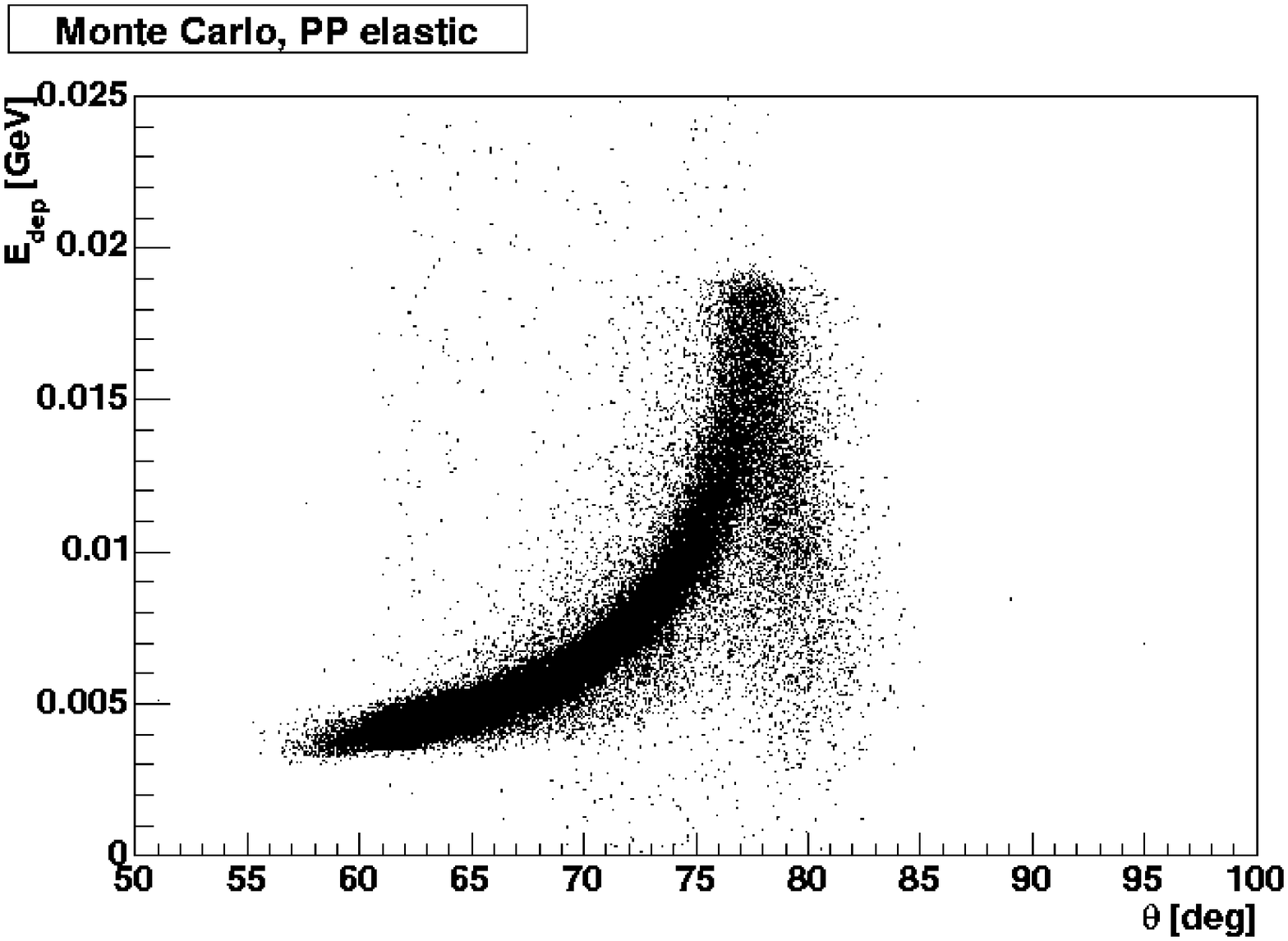}
}
\framebox{
\includegraphics[width=0.4\textwidth,angle=-90,clip]{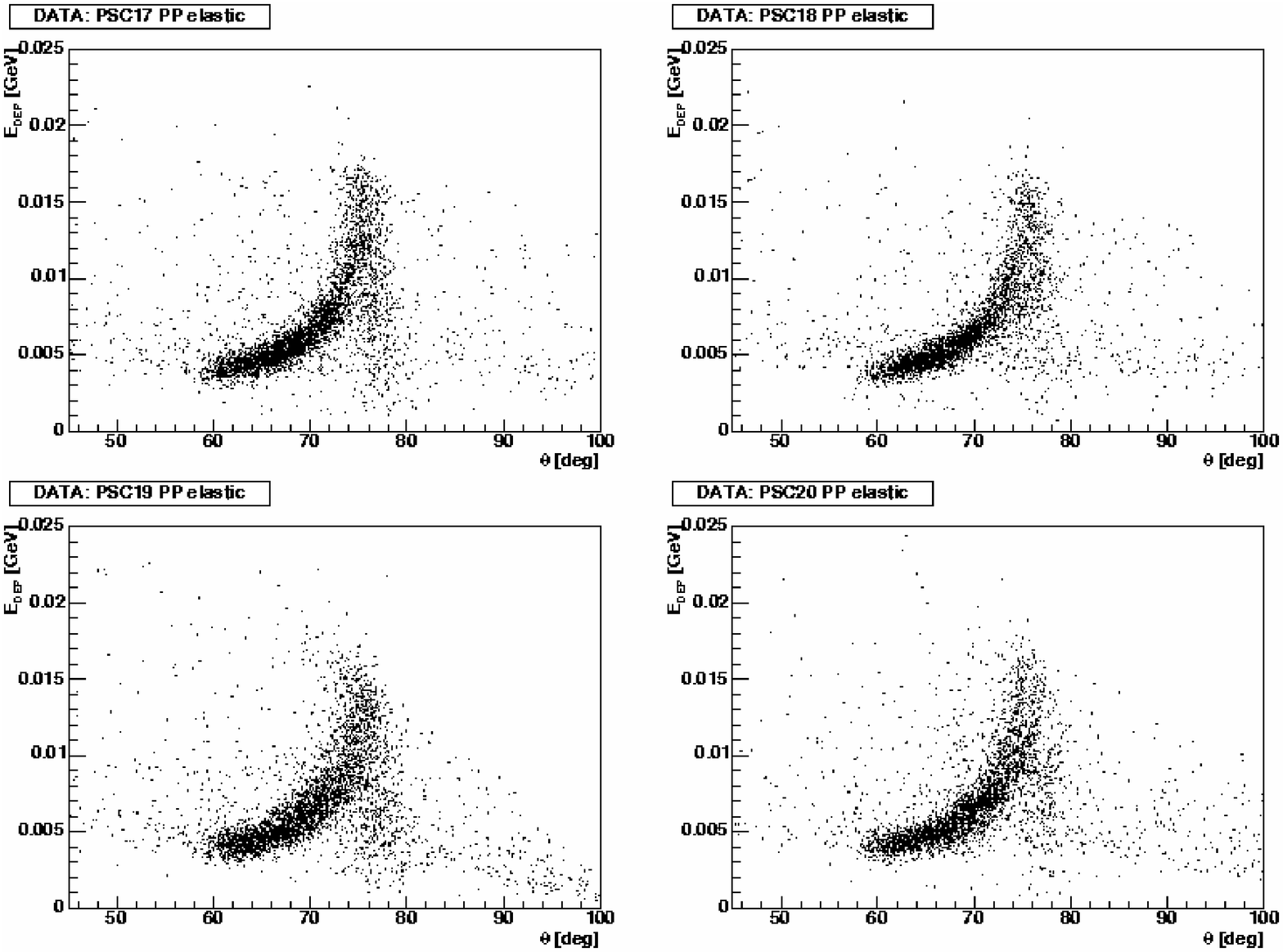}
}
\caption{\label{fig:pscal} (Left) 
Angular dependence of the deposited energy in
the PSB for simulated elastically scattered protons. 
The energy deposition increases with
increasing polar angle (corresponding to a decrease of the kinetic
energy of the proton), until particles begin to  stop in the plastic
scintillator material (at around $\theta$=77$^o$).
(Right) Experimental spectra corrected for light attenuation
for four of the PSB central elements.}
\end{figure}

The performance of the PSB has been studied using proton-proton elastic scattering events.
Fig.~\ref{fig:pscal} (left plot) shows the result of a Monte Carlo simulation 
of the angular dependence on the energy deposited in
the PSB. Fig.~\ref{fig:pscal} (right plots) shows typical experimental spectra
after a correction for nonuniform signal response has been applied.

\begin{figure}[!hbt]
\centering
\includegraphics[width=0.4\textwidth,angle=-90,clip]{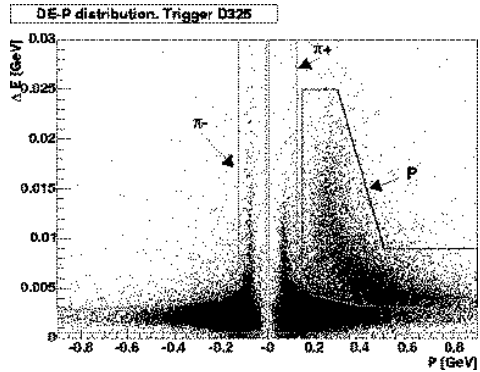}
\caption{\label{fig:CDdep} Example of particle identification in
the central detector for a raw data sample collected at 1360
MeV. Energy deposited in the PSB is given as a
function of signed momentum from the MDC. The regions for
protons and pions are marked. 
}
\end{figure}

In the initial experiments the momentum and energy resolution allowed
reasonable discrimination
between pions and protons, which is illustrated in Fig.~\ref{fig:CDdep}.
For high energy charged particles also SEC information is available and 
can be used for the identification.

\begin{figure}[!hbt]
\includegraphics[width=0.32\textwidth,clip]{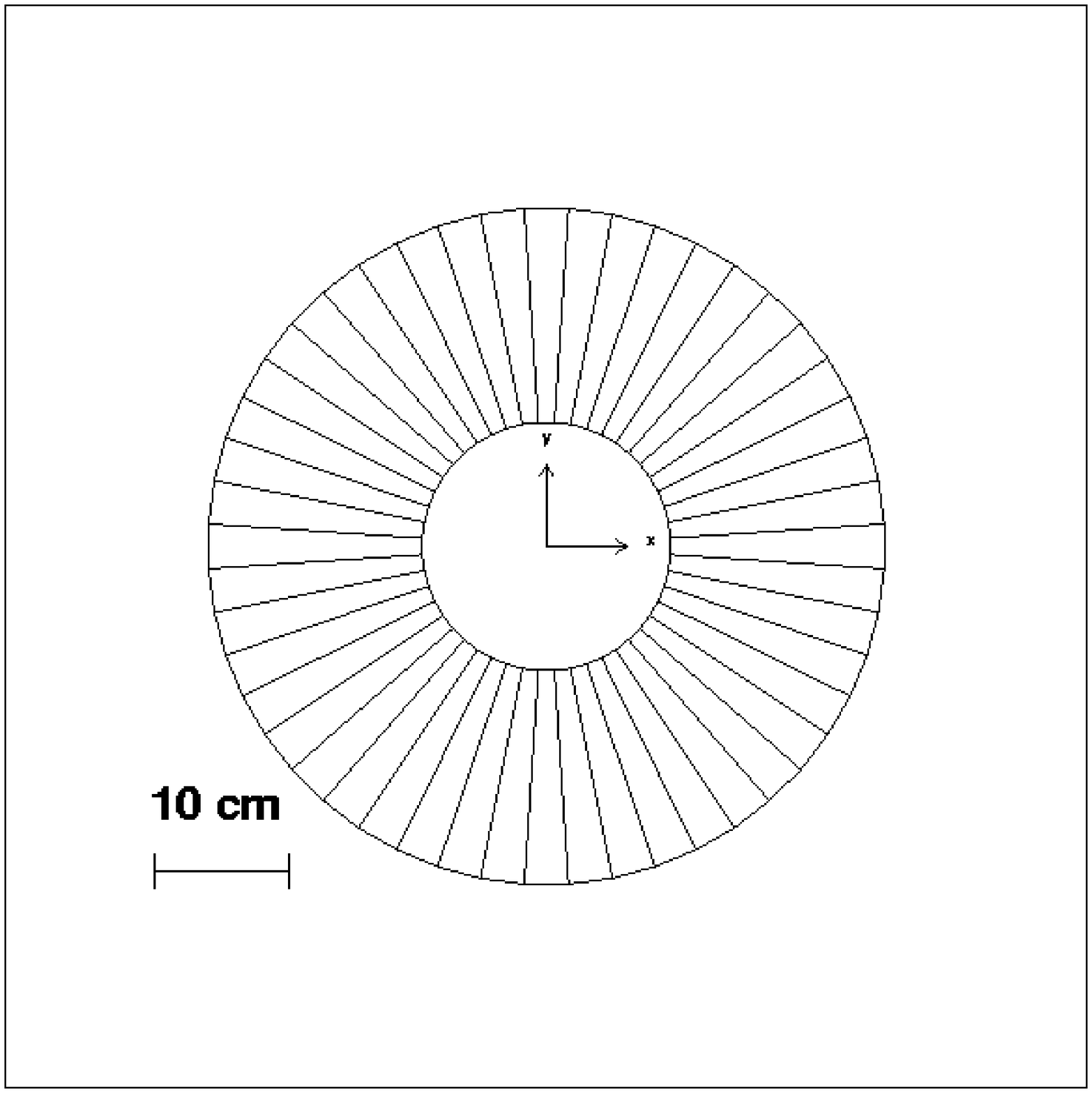}
\includegraphics[width=0.32\textwidth,clip]{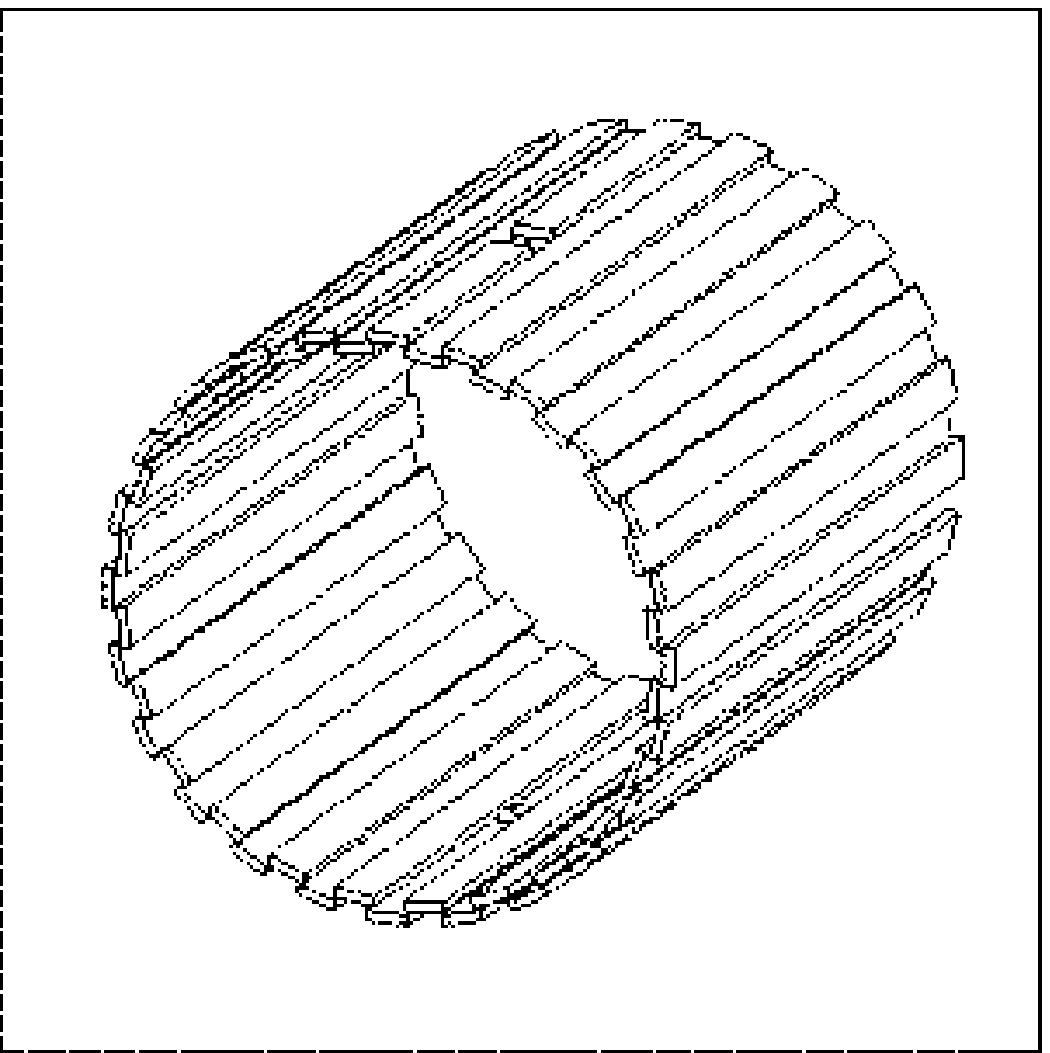}
\includegraphics[width=0.32\textwidth,clip]{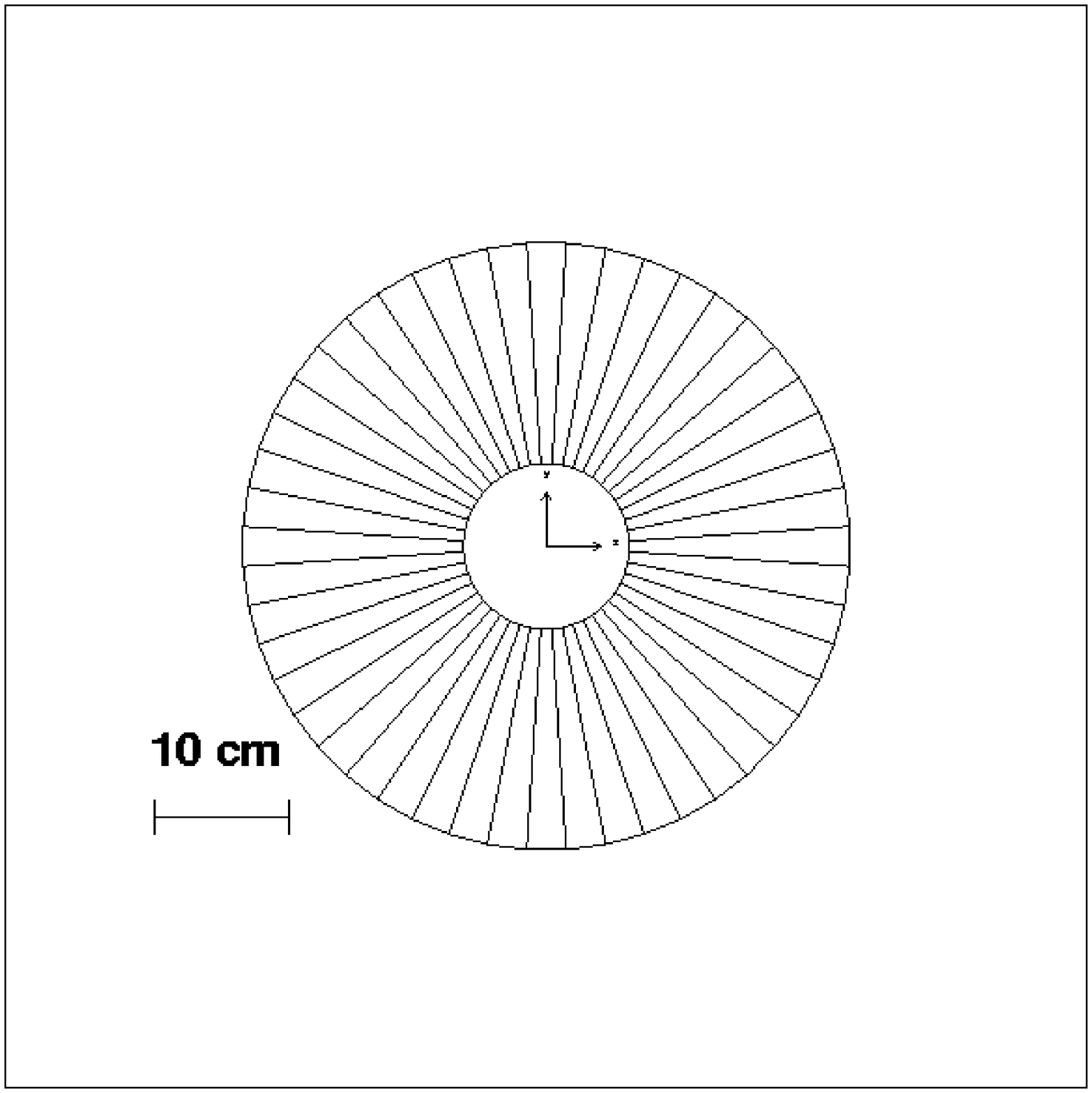}
\caption{\label{fig:ps} Forward, central and backward parts of the
PSB. In the central part, the gaps for the pellet target pipe are  visible.}
\end{figure}

\begin{figure}[!hbt]
\begin{minipage}[t]{0.35\linewidth}
\vspace{0.5cm}
\includegraphics[width=\textwidth,clip]{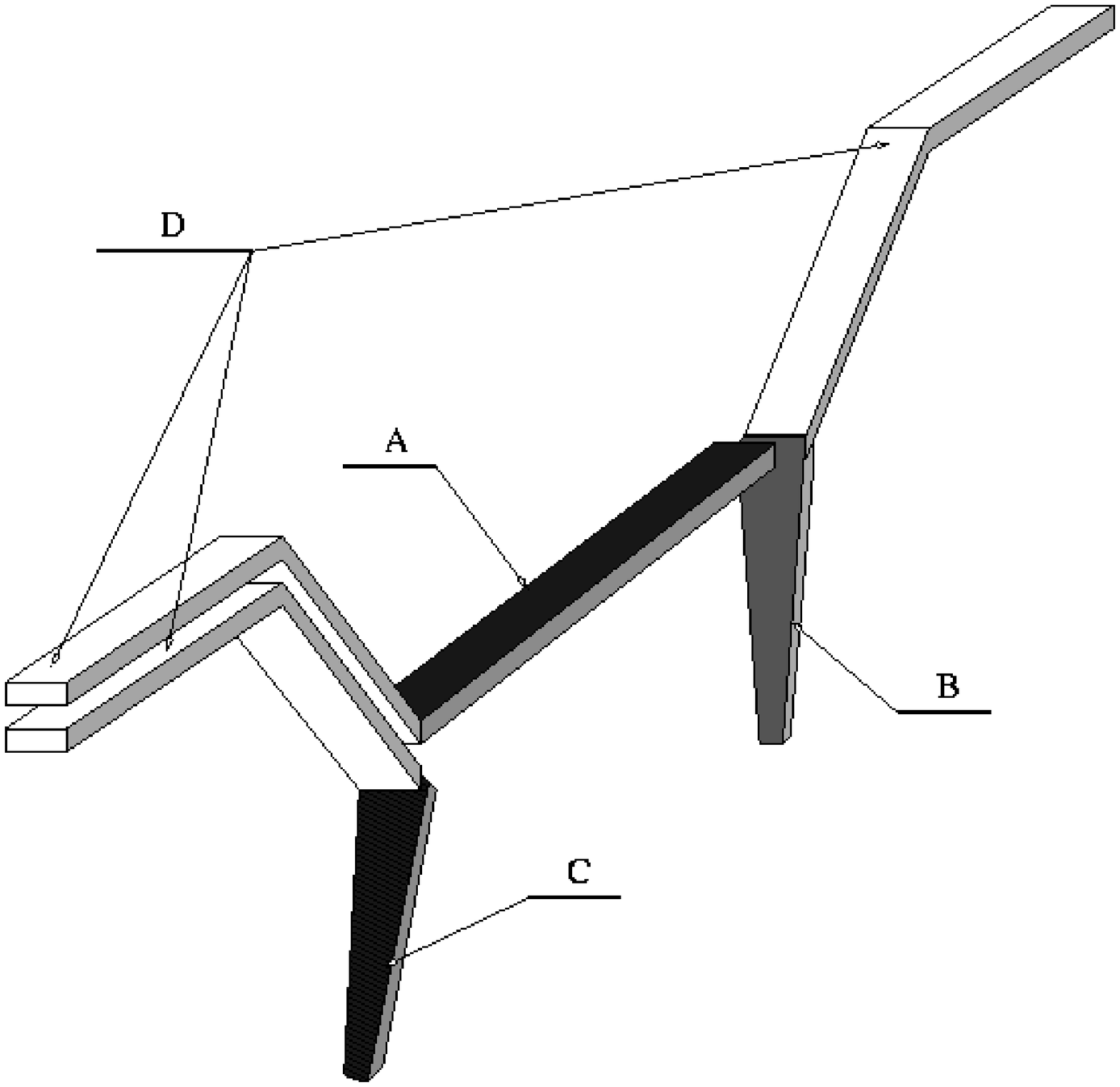}
\end{minipage}
\begin{minipage}[t]{0.65\linewidth}
\includegraphics[width=0.45\textwidth,angle=-90,clip]{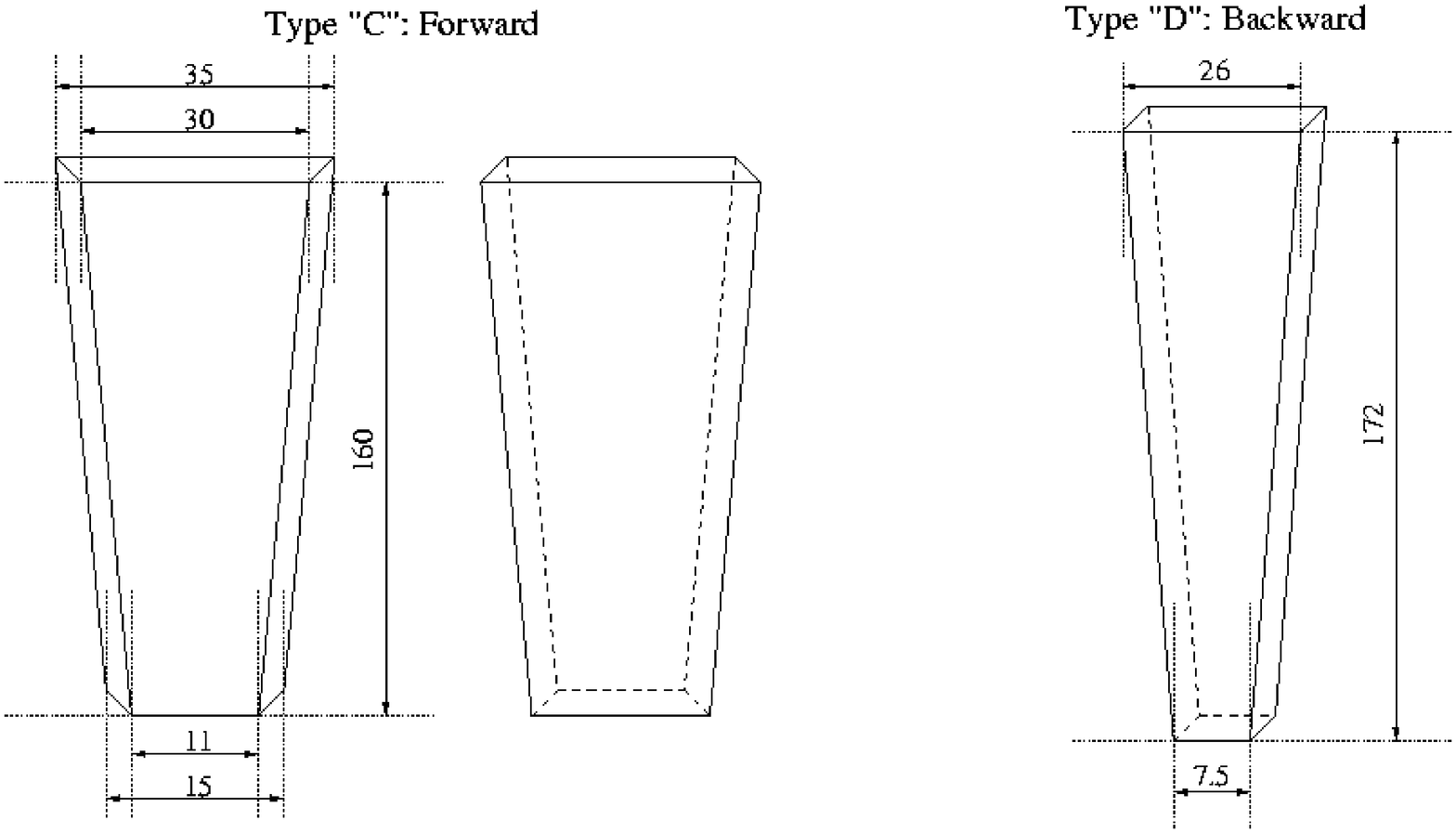}
\end{minipage}
\caption{\label{fig:psbsection} (Left) Layout of one section of the PSB
detector. {\bf A} denotes the rectangular counters of the barrel wall and
{\bf B } and {\bf C} are trapezoidal elements in the forward and in the
backward caps respectively. {\bf D} are bended light guides. (Middle)
Two shapes of the trapezoidal forward elements with dimensions marked
in mm. (Right) Shape and dimensions in mm of the trapezoidal backward
element.
}
\end{figure}

\begin{figure}[!hbt]
\centering
\includegraphics[width=\textwidth,clip]{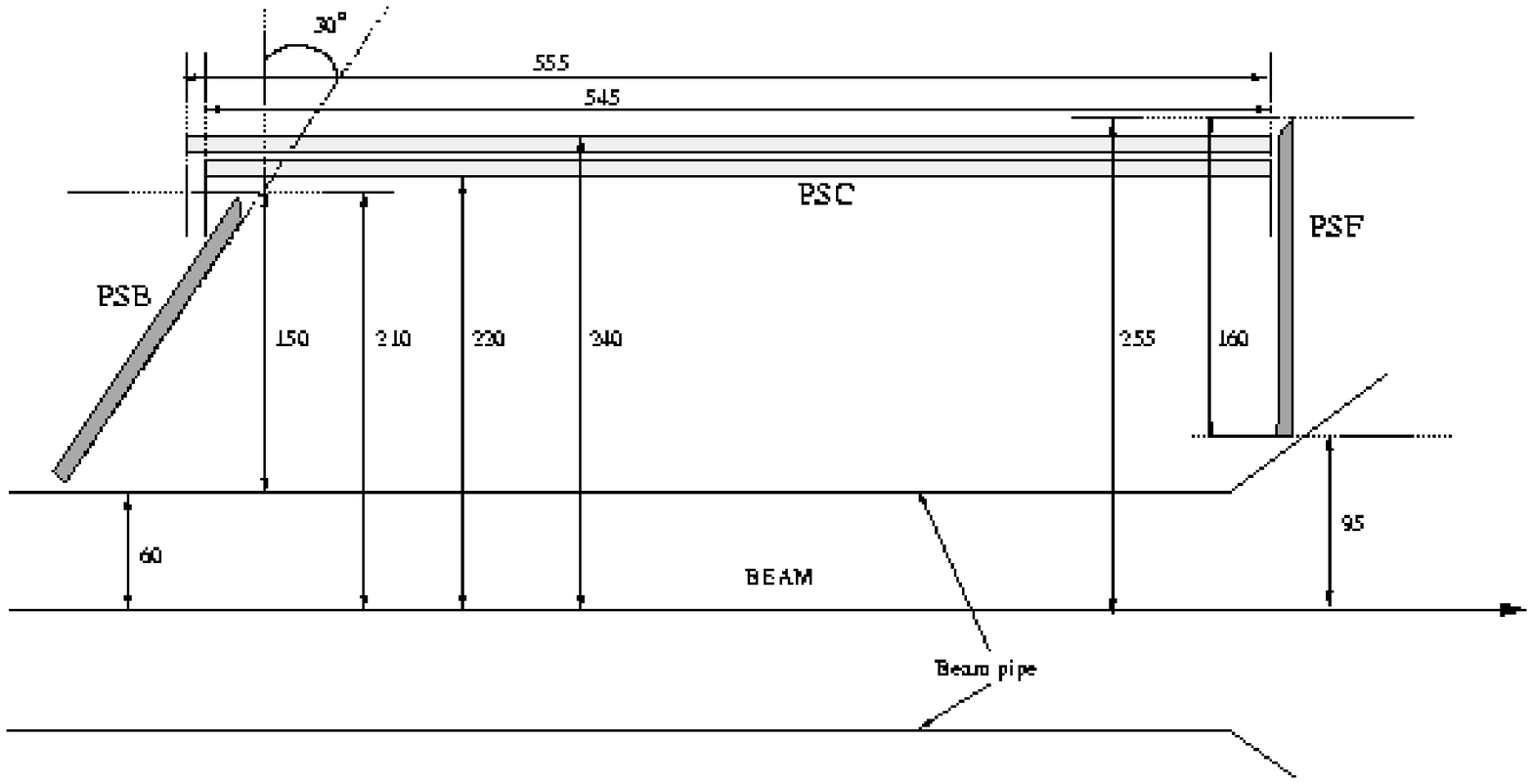}
\caption{\label{fig:psbmeas} Cross section of the PSB scintillators with 
dimensions marked in mm. No structural material is shown.
}
\end{figure}

The PSB consists of a cylindrical part and two end caps and contains in
total 146 elements shaped as strips of 8mm thickness. In the cylindrical part there are
48(+2) elements of 550~mm length and 38~mm width, forming 2 layers
with a small (on average
6~mm) overlap between  neighboring elements to avoid that particles
pass without registration. 
The end caps with an outer diameter of approximately 42~cm in the backward and
51~cm in the forward part contain 48
``cake-piece'' shaped elements each. The front end cap is flat while
the rear cap  forms a conical surface. Both end caps  have a central hole for
the beam pipe (19~cm diameter in the forward and 12~cm diameter in the
backward part). 
The PSB as modeled in the detector
simulation program is shown in Fig.~\ref{fig:ps}.
One sector of the PSB is shown  in Fig.~\ref{fig:psbsection}
and  \ref{fig:psbmeas}.

Each scintillator is glued to an acrylic light guide coupled to the photomultiplier tube (PMT).
The PMTs are placed outside of the iron yoke to shield them from the
magnetic field. For this purpose, approximately 50~cm long light guides are
used.

\emparag{The Scintillator Electromagnetic Calorimeter - (SEC)}
The CD calorimeter SEC is able to measure photons, electrons and positrons 
with energies up to 800~MeV. The energy threshold for detection of photons is
about 2~MeV. SEC
consists of 1012 sodium-doped CsI scintillating crystals placed
between the super-conducting solenoid and the iron yoke. The scattering
angles covered by the SEC are between 20$^{\circ}$ and
169$^{\circ}$. 

\begin{figure}
\begin{center}
\epsfysize7cm
\epsfclipon
\epsffile{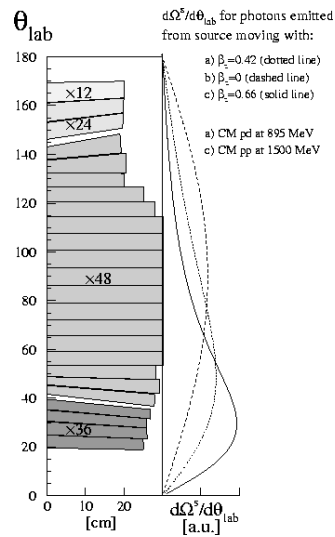}
\epsfclipoff
\end{center}
\caption[The angular coverage of the SEC]
{\label{secross}The angular coverage of the SEC. The CM system
solid angle vs.\,the LAB scattering angle is shown for pp and pd
interactions at 1500~MeV and 895~MeV.}
\end{figure}

\begin{figure}[htbp]
\begin{center}
\epsfxsize10cm
\epsfclipon
\epsffile[33 155 555 683]{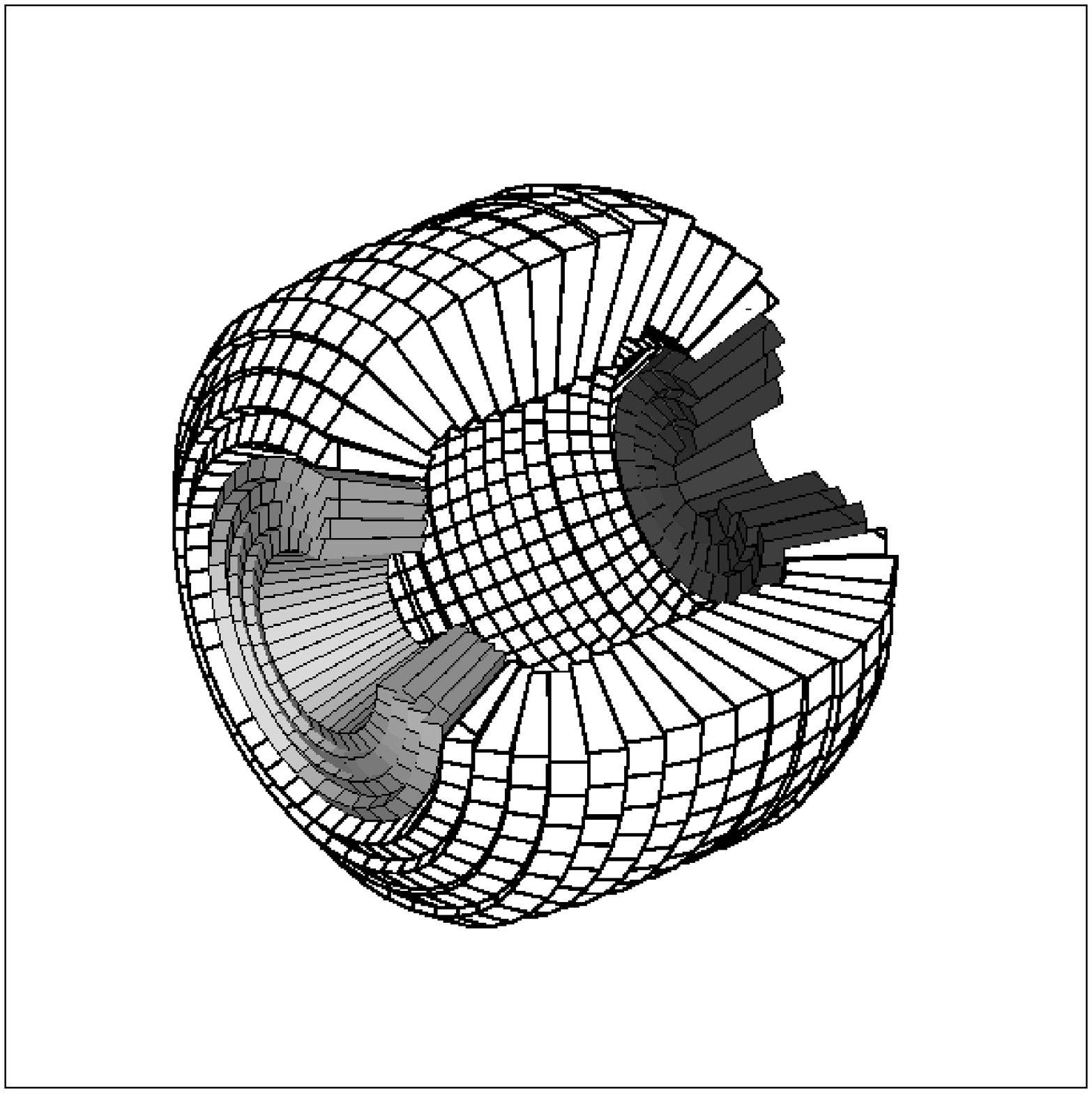}
\epsfclipoff
\end{center}
\vspace*{-10mm}
\caption[A schematic view of the SEC]
{\label{se}Schematic view of the SEC. It consists of the forward
part (shadowed
area to the left), the central part (not shadowed area in the middle)
and the backward part (shadowed area to the right). The beam is
coming from the right side.}
\vspace*{5mm}
\begin{center}
\epsfysize6cm
\epsfclipon
\epsffile{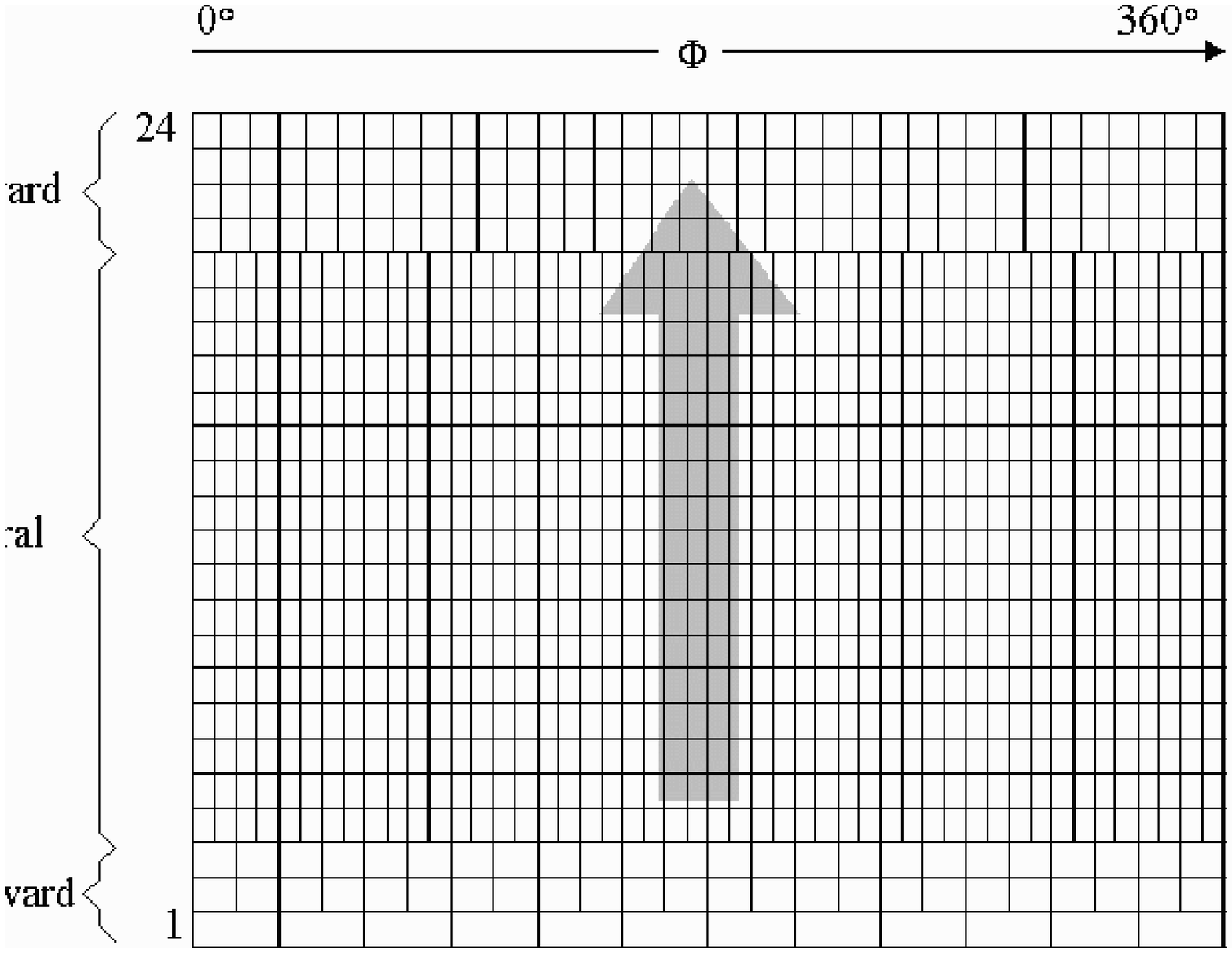}
\epsfclipoff
\end{center}
\caption[Planar map of the SEC]
{\label{planarmap}Planar map of the SEC.}
\end{figure}

The crystals are shaped as
truncated pyramids and are placed in 24 layers along the beam
(Fig.~\ref{fig:wasa} and \ref{se}). The
lengths of the crystals vary from 30~cm (16.2
radiation lengths) in the central part to 25~cm in
the forward and 20~cm in the backward part. 
Fig.~\ref{secross} shows the angular coverage together with the 
thickness of SEC. As a measure of the anticipated photon fluxes, the  
center of mass (CM) system solid angle vs.\,the laboratory (LAB) 
scattering angle is shown for some experimental conditions at WASA.

A planar map of the crystals is given in Fig.~\ref{planarmap}. One can
clearly distinguish the three different main parts of the calorimeter:
the forward, central and backward parts. The forward part consists of
4 layers with 36 elements each. It covers scattering angles from
nearly 20$^{\circ}$ to 36$^{\circ}$. The central part consists of 17
layers each having 48 elements, and covers scattering angles from
36$^{\circ}$ to 150$^{\circ}$. The backward part consists of three
layers. Two layers have 24 elements and one layer closest to the
CELSIUS beam pipe has only 12 elements. 
The small spaces between the forward-central and central-backward parts are
occupied by PSB light guides and mechanical support for the solenoid 
(back end only). The calorimeter covers nearly
360$^{\circ}$ in ${\rm \Phi}$ angle. Holes for the pellet pipe (2+2 crystals) 
and the solenoid chimney (4 crystals) are not shown in the figure. Some
design parameters of the calorimeter are given in table~\ref{sec2}.

\begin{table}[bht]
\begin{center}
\vspace{\baselineskip}
\begin{tabular}[b]{|l|l|} \hline
\multicolumn{2}{|c|}{\bf Scintillator Electromagnetic Calorimeter} \\ \hline 
Amount of sensitive material & 135~g/cm$^2$ \\ \hline
\hspace*{3mm} [radiation lengths] & ${\rm \approx 16}$ \\ \hline
\hspace*{3mm} [nuclear interaction length] & ${\rm \approx 0.8}$ \\ \hline
Geometric acceptance: & 96\% \\ \hline
\hspace*{3mm} polar angle &  ${\rm \approx 20^{\circ}-169^{\circ}}$\\ \hline
\hspace*{3mm} azimuth angle &  ${\rm \approx 0^{\circ}-360^{\circ}}$\\ \hline
Max kinetic energy for stopping &  \\ \hline
\hspace*{3mm} ${\rm \pi^{\pm}}$/proton/deuteron & 190/400/500 \\ \hline
Scattering angle resolution & ${\rm \approx 5^{\circ}}$(FWHM) \\ \hline
Time resolution  &  \\ \hline
\hspace*{3mm}charged particles & 5~ns(FWHM) \\ \hline
\hspace*{3mm}photons & ${\rm \approx}$40~ns(FWHM)\\ \hline
Energy resolution &  \\ \hline
\hspace*{3mm}charged particles  & ${\rm \approx 3\%}$(FWHM) \\ \hline
\hspace*{3mm}photons  & ${\rm \approx 8\%}$(FWHM) \\ \hline
\end{tabular}
\end{center}
\caption[SEC design parameters]
{\label{sec2}SEC design parameters.}
\end{table}

The SEC is composed of sodium-doped CsI scintillating crystals. This
type of scintillator material provides a large light yield, has short
radiation length and good mechanical properties.
CsI(Na) was chosen instead of the more commonly used CsI(Tl)
scintillators for the following reasons \cite{rub:90,sch:95}:

\begin{itemize}
{\item Its emission peak at 420~nm matches well the bi-alkali S11
photocatode of the selected  PM tubes, giving good photon statistics and
sufficiently fast response.
\item Its shorter scintillation decay time is preferable in high-rate
applications.
\item CsI(Na) gives much less afterglow than CsI(Tl).
\item CsI(Na) seems more resistant against radiation damage. When irradiated
by a proton
beam corresponding to 10 years of operation a test crystal did not
show any visible change in its structure. The CsI(Tl) test crystal, on
the contrary, lost its transparency.}
\end{itemize}

\begin{figure}[htbp]
\begin{center}
\epsfysize5cm
\epsfclipon
\epsffile{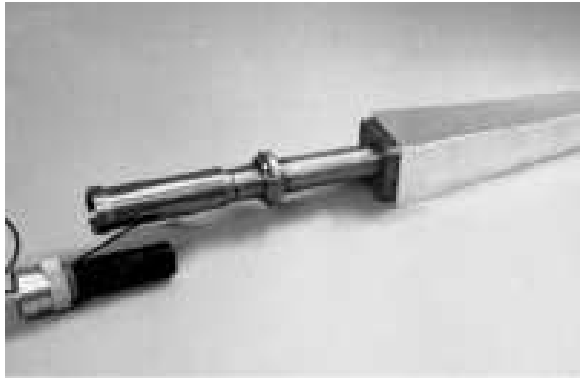}
\epsfclipoff
\end{center}
\caption[A fully equipped CsI module]
{\label{csimodule}A fully equipped module with CsI crystal,
light guide, PM tube and housing.}
\end{figure}

The crystals are connected
by plastic light guides, 120~mm to 180~mm long, with the photomultipliers 
placed the outside of the iron yoke.
In Fig.~\ref{csimodule}, a fully equipped single calorimeter module
consisting of a CsI crystal, a light guide, a PM tube and a high
voltage unit, enclosed inside a special housing, is shown. 

The SEC and its performance is described in more detail in the 
Ph.D. thesis of Inken Koch  \cite{Koch}.

\emparag{The light pulser system (LPS)}
\noindent The LPS delivers  reference light pulses via light fibers to all 
scintillation counters in order to monitor their gain during the experiment. 
Since both organic and inorganic
scintillators are used, two types of light sources were designed. A
xenon flash tube from Hamamatsu is used for the CsI elements of the calorimeter and three
LED-based light sources for all plastic scintillators. From those four sources the light signals are
transmitted to  individual elements via a  network of light 
fibers~\cite{Zabi94}.

\subsection{Modifications}
\label{subsubdetmodi}
\label{sec:modifications}

\subsubsection{Forward detector (FD)}

\begin{figure}[ht]
\begin{center}
\includegraphics[clip=1,width=0.5\textwidth]{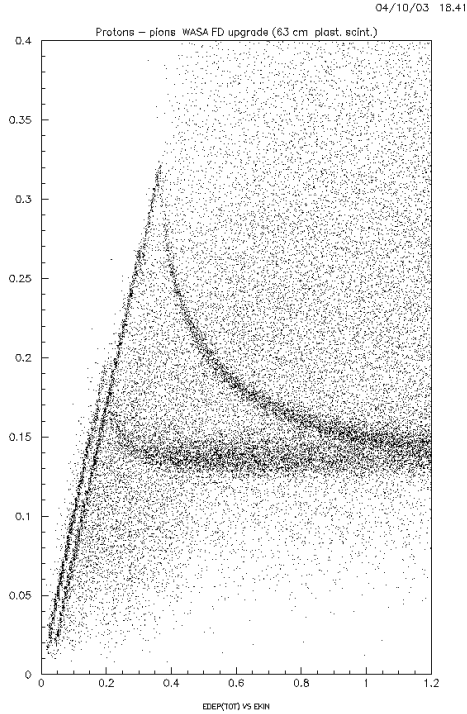}
\caption{\label{fig:fdetot} Energy absorbed in the (upgraded) FRH planes
      for protons (upper curve) and pions (lower curve)
      as a function of their kinetic energy at  the target.} 
\end{center}
\end{figure}

\begin{figure}[ht]
\hspace{1.3cm}
\rotatebox{-90}{
\includegraphics[width=0.5\textwidth,clip]{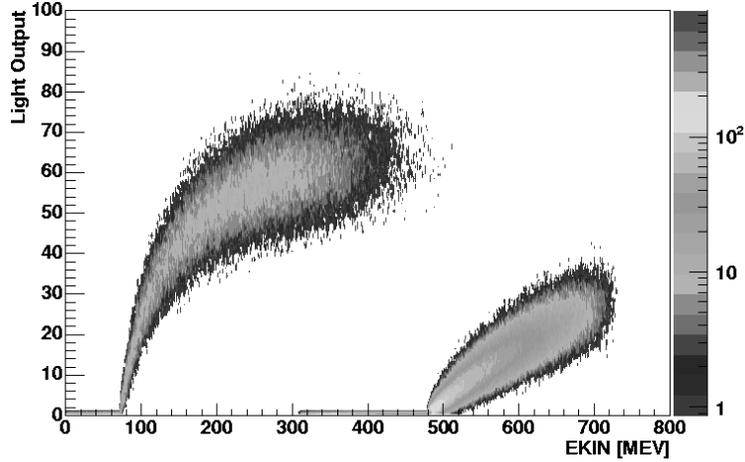}
}
\caption{\label{fig:fdcerenkov} Cherenkov light output for charged particles 
from the 
reaction $pp\to pp\eta'$ at 3.3 GeV/c as a function of kinetic energy.} 
\end{figure}

The FD was designed mainly for measurements of protons of kinetic energies
up to 500 MeV. An energy resolution $\sigma_E/E$ around 3\% is obtained 
at the highest energy. 
The track coordinates are measured by the FPC straw chambers. At CELSIUS, three
modules with 4 layers of straws each, are installed. 
To improve track coordinate measurements, energy measurements and particle
identification for the higher energies anticipated at COSY, the following 
modifications are planned:

\begin{itemize}

\item The first FRH layer will be removed. It gives the few centimeters
of extra space, needed for the
installation of one more (existing) FPC module. By moving the remaining 
FRH layers a few centimeters upstream, one gets a better energy measurement 
performance at the largest scattering angles ($17^{\circ}$) covered by the FD. 

\item Repair of FTH. The FTH and FRH have served at CELSIUS for twelve 
years and signs of severe aging effects have appeared. Most detectors
still work well, but in one of the spiral planes of FTH some elements 
do not give acceptable signals and the whole plane should be replaced.

\item Two new FRH planes should be installed to improve particle energy
reconstruction. This would mean that protons of about 350 MeV kinetic
energy, instead of 300 MeV at present, would be stopped in the FD
(figure \ref{fig:fdetot}). The energy resolution at high energy would
consequently also be improved. A $\sigma_E/E$ around 10\% could be
reached for protons at 900 MeV.

\item Downstream of the FRH, we consider the installation of a 
10~cm thick water Cherenkov detector, that is 
composed of $\approx$2~m long and 10~cm wide bars.  For
protons it would have a range of sensitivity from 500 MeV
($\beta$=0.75) to 1200 MeV ($\beta$=0.9), which well covers 
the region where the energy
determination based on the FRH depositions deteriorates. The
combination of the FRH and the Cherenkov information
will improve both the particle identification power and the
kinetic energy resolution.
As an example, the Cherenkov light output is plotted as a function of
kinetic energy for protons and pions from the
reaction $pp\to pp{\eta }'$ where ${\eta }'\to \eta \pi^{+} \pi^{-}$
at a beam momentum of 3.3 GeV/c (figure \ref{fig:fdcerenkov}). The
proton kinetic energy energy resolution is about 5\% for protons above
500 MeV.  Studies are underway to obtain more quantitative information
on the performance of the proposed Cerenkov detector.

\end{itemize}

\subsubsection{Central detector (CD)}

The CD was originally designed for measurements of gammas with energy up 
to 600 MeV and of electrons and positrons with momenta up to 400 MeV/c.

The charged particle momenta are determined by the track curvature in
a magnetic field of about 1 Tesla. The maximum lever arm is 160~mm for
these measurements and the design accuracy in the position
measurements in each of the 17 MDC layers is 100 $\mu$m. The accuracy
of the field map should be better than 1\%.  The design momentum
resolution for electrons and positrons, scattered at large angles is
$<$~2\% (${\rm \sigma_p/p}$). With the present implementation of the
MDC in the CELSIUS/WASA experiment an optimal resolution, ${\rm
\sigma_p/p} \approx$3\% for $e^{\pm}$, $\approx$4\% for $\pi^{\pm}$
and $\approx$8\% for protons could be obtained. In real experiments,
the momentum resolution has been evaluated for $\pi^{\pm}$ and protons
and ${\rm \sigma_p/p}$ values of $\approx$4.5\% and $\approx$12\% were
obtained \cite{Jacewicz04}. To improve the momentum measurements for
work at higher energies the following actions are planned:
\begin{itemize}
\item An increase of the maximum magnetic field to 1.3 Tesla. This puts great 
  demands on some parts of the magnet system, since there will be a
  70\% increase in the stored energy and in the mechanical forces
  involved, but it would not require any major hardware upgrades
  \cite{Ruber99}. However, it needs careful tests and "training" of
  the magnet by expert personnel.
\item The MDC electronics have to be upgraded in order to improve the 
  efficiency for detection of pions and electrons and to reach the
  design time resolution of $\approx$0.5~ns. The extent of this task
  depends on the actual electrical noise environment at the WASA site
  at COSY.
\end{itemize}

The design energy resolution for photons measured in the SE
calorimeter is about 5 \% ($\sigma_E/E$). At present an effective
energy resolution of 7-10 \% is obtained \cite{Koch}, and the main
limitation is a non-linear signal response. This is the product of
some different effects and the situation will be improved by
modification of PM bases, by usage of new QDCs and, if neccessary, by
changing some 150 PM tubes.

One source of background in CELSIUS/WASA is interactions in the
restgas in the scattering chamber. At present there are three big
cryopumps at WASA, two at the forward cone of the scattering chamber
and one at the backward cone. The vacuum in the scattering chamber
will be improved by adding one more big cryopump in the backward
direction. This needs a careful redesign of the backward part of WASA.

Another source of background is due to scattering events originating
from the vacuum chamber walls. Most of these events are caused by beam
halo particles. It is suggested to put a ring-shaped detector inside
of the beampipe just at the entrance to WASA. Signals from this
detector should be used in the trigger logic to "veto" events caused
by beam halo.  Further prestudies will be done to design such a
detector and evaluate its anticipated performance in more detail.

\subsubsection{Cost estimates}
\label{sec:cost}

\begin{table}[!htbp]
\centering
\begin{tabular}{|l|c|}
\hline
{\bf Cost estimate} & {\bf kEuro} \\
\hline 
FRH upgrade with two additional layers & 200\\ 
FTH repair / replacement of one layer  &  50\\ 
FD Cerenkov detector       & 100\\ 
SEC new PM bases / change of PM tubes   &  50\\ 
MDC system refurbishments   &  130\\ 
\hline
Additional upstream vacuum pump   & 40 \\ 
Upstream internal veto detector   & 30 \\ 
\hline
\end{tabular}
\caption[modcost]
{\label{modcost}Cost estimates for some planned detector modifications.}
\end{table} 

\subsubsection{DAQ- and Trigger-System}
\label{subsubdaq}

The WASA DAQ-System was designed more than 10 years ago and
implemented with electronics from the same period. 
Most of the digitizing modules were
delivered by the company LeCroy, which is no longer active on the market 
of nuclear electronics. Furthermore, many parts of the detector
electronics have a very low level of integration, 
leading to a high cabling effort at the detector side,
especially for the straws in the MDC and the Forward Detector. 
In addition, spare parts of the used components cannot be acquired 
anymore. 

Due to the progress in technology the readout speed of the
existing system is much slower than possible using a 
modern, state-of-the-art system. Since the event rate 
at COSY must be significantly higher than the one currently 
reached at CELSIUS, this becomes a serious problem. Considering also 
the maintenance problems and the age of the digitizing modules, 
it is obvious that successful operation of WASA at COSY can only 
be achieved by a major upgrade of the existing DAQ system. 

Currently, all experiments running at COSY are using the same
DAQ system following a uniform approach developed, implemented
and maintained by permanent staff of FZ J\"ulich. In view of
the time scale for ''WASA at COSY'' the upgrade of the 
WASA DAQ should be based on the third generation of the DAQ at 
COSY. The main advantages are:
\begin{itemize}
\item it implements the desired functionality, including a modern
   synchronization technique, avoiding the -- inherently slow -- sequential 
   readout of the digitizing modules, 
\item it is almost completely developed, thus fitting into the limited 
   time scale,
\item it has a good on-site support by the original developers, which
   is essential for a smooth and successful operation,
\item and it is compatible to all existing DAQ systems at COSY.
\end{itemize}

\begin{figure}[hbt]
 \begin{center}
  \resizebox{0.7\textwidth}{!}{\includegraphics{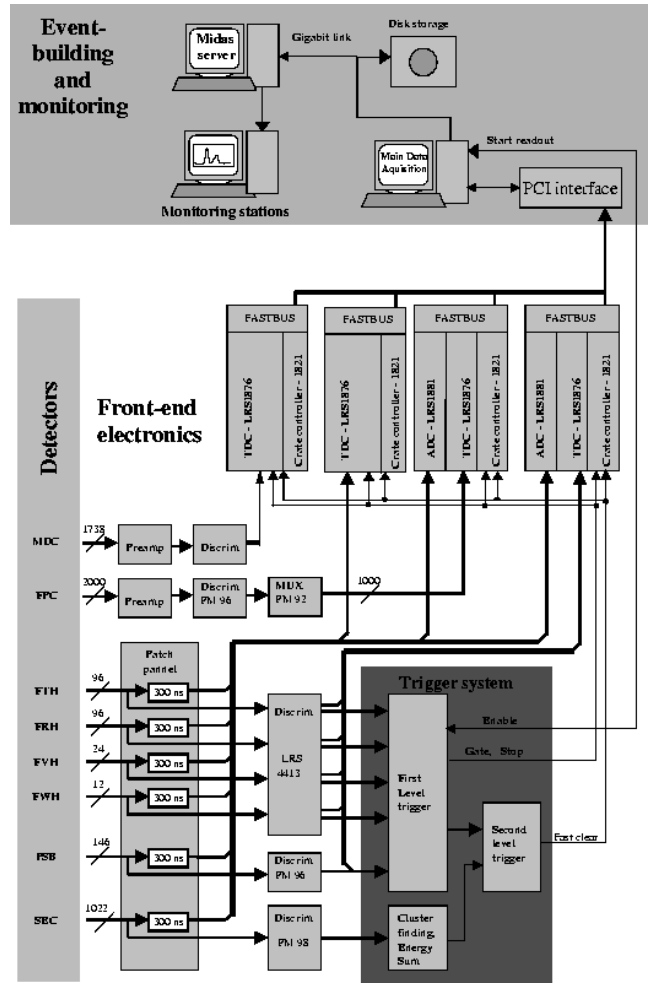}}
 \caption{Structure of the existing WASA DAQ and trigger 
          system~\cite{zel1,zel1a}.}
 \label{fig:zel1} 
 \end{center}
\end{figure}

\emparag{The current DAQ system}
The existing DAQ system of WASA at CELSIUS is mainly based on 
four FASTBUS crates containing the digitizing modules (see 
Fig.~\ref{fig:zel1}). The readout is done by PCs connected to
the FASTBUS crates via a proprietary parallel link based on
RS485. Since there is no synchronization system, all digitizing
modules have to be read out sequentially when a trigger
occurs. The readout of the (about) 1500 PMTs is done with 
LeCroy LRS1881 QDCs. The time information from the straws 
(2000 channels from the FPC and 1700 channels from the MDC) and 
the $\approx500$ fast plastic scintillators is digitized using 
LeCroy LRS1876 TDCs~\cite{zel1}.

The trigger system of WASA at CELSIUS is based on dedicated hardware 
that has been developed by TSL. Due to the high background
the trigger system is still evolving to reach higher performance. 
The first level trigger uses the discriminator outputs of the 
fast plastic scintillators. It is based on multiplicity and generates 
the gates for the QDCs and the stop for the TDCs. The second level 
trigger uses the discriminator outputs from the calorimeter. It is based 
on cluster multiplicity and energy deposition and generates the FastClear
for the digitizing modules.

\emparag{The new DAQ system}
Starting with the first experiments at COSY permanent staff from ZEL 
was responsible for a common DAQ system and its further development. 
In the meantime two evolutionary and elementary steps have been made 
to adapt it to the rapidly changing technologies~\cite{zel2}. 
Now, the implementation of a third generation of DAQ has been started
aiming for the highest possible event rates~\cite{zel3}. Therefore, 
state-of-the-art FPGA technologies are combined with fast communication 
paths to achieve system latencies as low as possible. Commonly used
in all three generations is a DAQ software called EMS~\cite{zel2}, which
guarantees the software compatibility of all implemented systems. EMS
follows a well proven client/server architecture and has been developed
during a period of more than 10 years (corresponding to a manpower effort
of nearly 30 man-years).

As in other DAQ systems (e.g.\ at CERN) this third generation of DAQ at
COSY is based on proprietary solutions in order to guarantee 
implementations optimized for speed. However, also CAMAC and VME can 
be integrated despite of the resulting loss of performance. The whole system 
is designed to run exclusively in a ``common stop mode'' to avoid many 
expensive and hard to handle delay units. Here, ``common stop mode'' means 
that the trigger is available some 100~ns to 1 $\mu$s after the signal 
has been digitized. Therefore, the acquisition boards must run in a 
self-triggering mode digitizing all interesting signals and storing them 
together with a time stamp. When the -- also time-stamped -- trigger arrives 
an FPGA selects the digitized signals inside a predefined time interval.

\begin{figure}[hbt]
 \begin{center}
  \resizebox{0.7\textwidth}{!}{\includegraphics{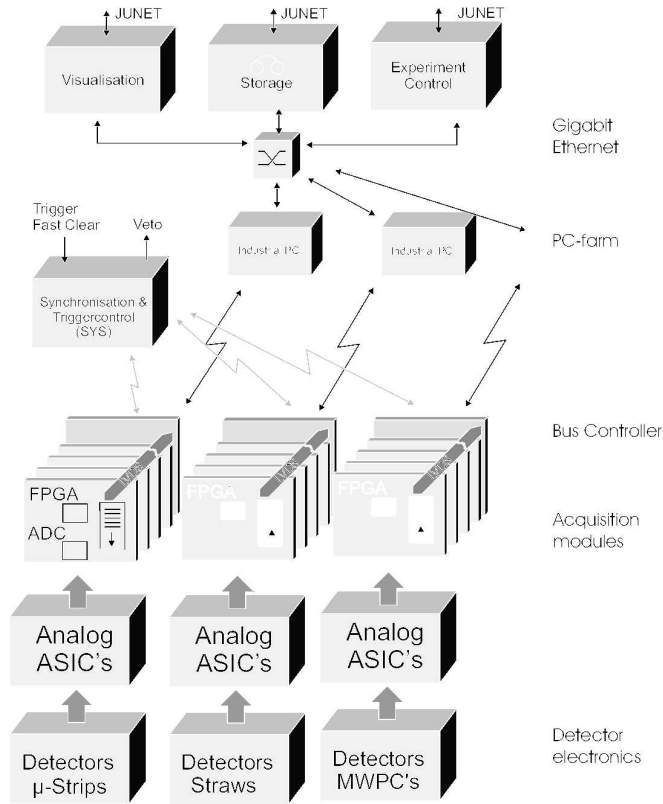}}
 \caption{Structure of the third generation DAQ system at COSY.}
 \label{fig:zel2} 
 \end{center}
\end{figure}

An essential key component required by this operation mode is a
Synchronization System (SYS) to reliably control and synchronize the
fast event flow~\cite{zel4}. The SYS delivers a global time base
relative to the trigger time. The control of the synchronization 
system as well as the readout of the digitization modules is done 
by a farm of embedded industrial PCs. On this farm, EMS-servers
controlling the front end modules are implemented. As a future extension 
also the cancellation of events (software trigger) is considered.  

The third generation of DAQ at COSY is based on a bus
controller (refer to Fig.~\ref{fig:zel2}) connected by a fiber
optical link to the PC farm and the synchronization system. The 
acquisition modules are connected by a high speed LVDS bus.
This bus can be implemented as a backplane or as a cable bus 
covering a distance of 15~meters and serving for control as well as 
for transmission of event data. The physical layer of the bus complies 
to the SCSI bus with a maximum transmission speed of 80 or 
160~Mbytes/s~\cite{zel5}. In general, acquisition, trigger processing, 
data reduction, buffer management, and sub-event building can be done by
means of FPGAs --- either on the acquisition boards or in the bus 
controller. All these functions are realized without the intervention 
of a general purpose processor and, thus, lead to the envisaged 
performance.

\begin{figure}[hbt]
 \begin{center}
  \resizebox{1.0\textwidth}{!}{\includegraphics{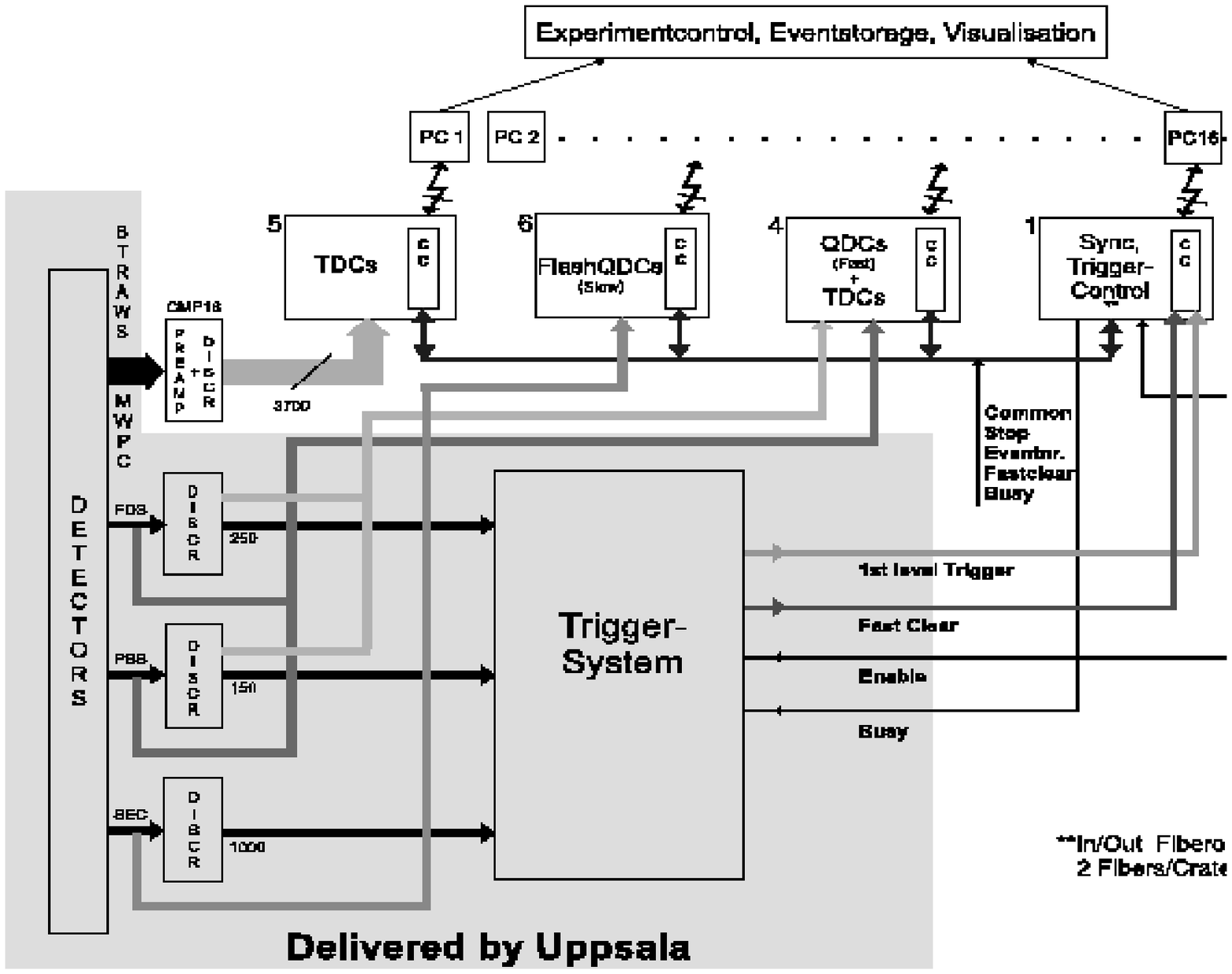}}
 \caption{Structure of the new DAQ and trigger system of WASA at COSY.}
 \label{fig:zel3} 
 \end{center}
\end{figure}

As shown in Fig.~\ref{fig:zel3} a new computing layer and a new layer
of digitization electronics will be implemented. The readout will be done 
by a farm of 15 PCs connected via an optical link (a common development 
of ZEL and the company SIS, with a physical layer identical to Gigabit 
Ethernet). The overall experiment control and storage is connected 
via Gigabit Ethernet. The whole system runs under control of the EMS 
software as used at all other experiments at COSY.

The digitization layer consists of crates with readout controllers, 
QDC and TDC modules and the synchronization system. The event buffering
capabilities of the digitization modules in combination with the
synchronization system allow the operation in common stop mode, thus 
avoiding delay lines. It is expected that with the new system
the event rate will be increased at least by a factor of 10. Because of 
performance requirements and costs the use of commercially available
digitizing modules is not possible.

The existing 64 channel TDC module (being used for the proportional
chambers of ANKE and the straws at TOF) can be used without modifications 
for the straws at WASA, provided the discriminator electronics will be
changed to a CMP16 based system. It is a 6U board
for the LVDS bus  based on the F1 chip system developed by ZEL
compatible with the third generation of DAQ at COSY~\cite{zel6}.

Optionally, also a VME board developed at TSL may be used, although 
some modifications and a further test phase are necessary~\cite{zel7}. 
This board has 64 channels, too, and is based on the CERN TDC32.  
The decision which TDC to use will have only minor impact on the overall 
cost estimate and project timing and will be done within the next year.

For the calorimeter (pulse duration about 2 $\mu$s) the QDC will
be implemented by sampling the analog signal with a 100 MHz FlashADC
and subsequent integration in an FPGA.  
Considering 
connector requirements and layout restrictions, it is planned to have 16
channels on one 6U base board conforming to the standards of the
LVDS bus system developed by ZEL.

The latter technique is not possible for the signals from the fast plastic 
scintillators, because the pulse duration is only $\approx 20$~ns. Here,
additional shaping or even the use of a QAC chip is required. The
technical issues are intensively discussed with people from the University 
of Giessen (TAPS collaboration), the KFKI Budapest and the TSL. 
Independent of the final decision, a mezzanine board will be manufactured 
fitting to the FlashADC board described above and performing the analog 
preprocessing for 16 channels.

The time stamps required by the free-running mode will be provided
by the FPGA on the motherboard for the ''slow'' QDC using the global
time information and by additional discriminators and TDCs for the
''fast'' QDC.

As indicated in Fig.~\ref{fig:zel3}, the discriminators and the
trigger electronics from the existing system shall remain.
Only with regard to the straws it is intended to replace the discrete 
preamplifier and discriminator electronics with new modules based 
on the CMP16 chip. The main advantage will be the increased sensitivity 
and the compact layout, which allows much shorter cable lengths between 
straws and preamplifier, thus reducing noise.

\begin{table}[t]
\begin{center}
\begin{tabular}[b]{|lr@{~channels\hspace*{4cm}}r@{~k\euro}r|}
\hline
  \multicolumn{4}{|l|}{Electronics for MDC and
              FPC based on the CMP16 ASIC\hspace*{1cm}}\\
    ~   &  3700 &95&\\
\hline
  \multicolumn{4}{|l|}{Digitizing modules}\\
    TDCs (MDC, FPC) &  3700 &  140&\\
    QDCs (slow) & 1000  &  100&\\
    QDCs (fast) &  400  &  80 &\\
    add. TDCs &  400  &  16 &\\
\hline
  \multicolumn{4}{|l|}{System components}\\
  \multicolumn{2}{|l}{17 Crates (incl.\ crate controllers, 
                           interfaces)} & 140 &\\
  \multicolumn{2}{|l}{Industrial PC farm (embedded)} & 30& \\
  \multicolumn{2}{|l}{Communication equipment} & 20& \\
\hline
  \multicolumn{2}{|l}{Synchronization system} & 70& \\
\hline
  \multicolumn{4}{|l|}{Miscellaneous}\\
  \multicolumn{2}{|l}{Racks, infrastructure test 
                       and measuring equipment} & 80& \\
  \multicolumn{2}{|l}{Experiment control, visualization 
                                 and storage} & 20& \\
  \multicolumn{2}{|l}{Additional cables and connectors} & 70& \\
  \multicolumn{2}{|l}{External developments} & 50& \\
  \multicolumn{2}{|l}{6\% risk reserve} & 60& \\
\hline
  \multicolumn{2}{|l}{{\bf Total}} & {\bf 1000}& \\  
\hline
\end{tabular}
\caption{Cost estimate for the development of a new data acquisition
         system for WASA at COSY.\label{tab:daq}}
\end{center}
\end{table}

A further improvement of the trigger system can be achieved by
implementing a third level trigger in the context of the new
DAQ system. While this is not possible in the limited time scale
of about one year, it should be considered as a future extension.
At the moment two basic hardware concepts are discussed:
\begin{itemize}
\item Implementation as a processor- or 
   FPGA-farm that operates directly on the DAQ data stream.
\item Splitting the data from the digitizing modules into two 
     independent branches, one for the trigger and one going to 
     the DAQ system~\cite{zel8}.
\end{itemize}

In a further future-oriented development project 
 it is planned to integrate the embedded PC
functionality in the LVDS bus controller itself
using ``system on chip'' (SOC) technology with a general purpose FPGA
and processor on the same chip. However, it is quite uncertain, whether
this project will be finished next year. Therefore it has not been
considered for the DAQ system of WASA at COSY.

\emparag{Cost estimate}
The cost estimate shown in table~\ref{tab:daq} does not include manpower, 
since all developments 
concerning the new DAQ system and improvements of the existing
WASA electronics can be made by existing permanent staff.
One exception could be the development of the analog part of the fast
QDCs. Most of the existing cables should be reused. However, additional
shielded cables are required in some cases and all connectors
will be replaced.

\clearpage
\section{Project plan}
\label{sec:fin_man}

\emparag{General remarks}
The project ``WASA at COSY'' deals with the relocation of an existing
detector system --- the Wide Angle Shower Array (WASA) from The
Svedberg Laboratory (TSL) at Uppsala university (Sweden) to the
Research Center J\"ulich (Forschungszentrum J\"ulich, FZJ, Germany), more
specifically to an internal target position of COSY (Cooler
Synchrotron) at IKP.
WASA at COSY will be the experimental activity with the highest priority 
at the IKP in J\"ulich for the next several years.

The approach of adapting an existing detector has a number of advantages
compared to building a new one: It is faster, provided the detector
is available on short notice. Given the necessary lead-time for 
preparations at COSY, the 
date that WASA will be available
(end of 2005) is regarded as optimal.  It is
cost-effective if the detector is ``in good shape''
which is the case for WASA.

However, the movement and setting up of a detector as complex as WASA
is not free of charge --- even neglecting basic costs such as
dismounting, transport, and installation. We anticipate the following
investments for necessary repairs and changes to the detector:
Adaptation to the higher COSY energy, in particular the
forward detectors of WASA.  Exchange of electronics for which no spare
parts exist and no replacement can be purchased any longer.
Replacement of the data acquisition system to improve data taking
capabilities and to adapt the WASA-DAQ to COSY standards.  These will
allow one to carry out the initial experiments described above.

\emparag{Manpower}
Well over one hundred scientists have committed themselves to actively 
persue the physics program laid out in this proposal. This includes 
members of the current WASA at CELSIUS collaboration as well as a 
wide representation from the COSY users.  As the host institute, IKP
will have a special role in providing the scientific and technical 
infrastructure for WASA at COSY. Consequently, IKP will redirect 
significant manpower (about 10 F.T.E.) to this project, which implies 
a reduction of the other activities at IKP.

\emparag{Finances}
The financing of this project is not yet finalized. In view of the high 
priority WASA at COSY has for IKP about one half of the needed investments 
summarized below could be contributed from the running budget of IKP.
This would require a one-time reduction of the number of COSY running hours by 
50\% for one year in either 2005 or 2006. The remaining investment must be 
provided by other sources in either Sweden, the EU, the German BMBF, or 
the Research Center J\"ulich.

\begin{table}[h]
\begin{tabular}[b]{lr@{~k}l}
Dismounting, transfer, set-up\hspace*{2cm}& 100 &\euro\\
Preparations at COSY                      & 300 &\euro\\
Changes at WASA (``adaptation'')          & 600 &\euro\\
New trigger- and readout electronics      & 1000&\euro\\
Sum:                                      & {\bf 2000} &{\bf \euro}\\ 
\end{tabular}
\end{table}

\emparag{Time schedule}

\begin{table}[h]
\begin{tabular}[b]{r@{~200}ll}
June  &5\hspace*{1cm}&Termination of CELSIUS operations at TSL\\
Oct.\ &5             &Finish dismounting of WASA at CELSIUS\\
Nov.\ &5             &Transfer of WASA to J\"ulich\\
June  &6             &Finish set-up of WASA at COSY\\
July  &6             &Start of commissioning\\
Jan.\ &7             &Start of experiments
\end{tabular}
\end{table}

\clearpage

\bibliographystyle{prsty}
\bibliography{proposal_lit}

\clearpage

\end{document}